\documentclass[12pt,a4paper]{article}
\pdfoutput=1

\usepackage{ifthen} 
\usepackage{booktabs} 
\newboolean{pdflatex}
\setboolean{pdflatex}{true} 
\newboolean{articletitles}
\setboolean{articletitles}{true} 

\newboolean{uprightparticles}
\setboolean{uprightparticles}{false} 

\newboolean{inbibliography}
\setboolean{inbibliography}{false}

\usepackage{microtype}
\usepackage{lineno}  
\usepackage{xspace} 
\usepackage{caption} 

\usepackage{graphicx}  
\usepackage{color}
\usepackage{colortbl}
\graphicspath{{./figs/}} 

\usepackage{amsmath} 
\usepackage{amssymb}
\usepackage{amsfonts}
\usepackage{upgreek} 

\newcommand*\patchAmsMathEnvironmentForLineno[1]{%
\expandafter\let\csname old#1\expandafter\endcsname\csname #1\endcsname
\expandafter\let\csname oldend#1\expandafter\endcsname\csname
end#1\endcsname
 \renewenvironment{#1}%
   {\linenomath\csname old#1\endcsname}%
   {\csname oldend#1\endcsname\endlinenomath}%
}
\newcommand*\patchBothAmsMathEnvironmentsForLineno[1]{%
  \patchAmsMathEnvironmentForLineno{#1}%
  \patchAmsMathEnvironmentForLineno{#1*}%
}
\AtBeginDocument{%
\patchBothAmsMathEnvironmentsForLineno{equation}%
\patchBothAmsMathEnvironmentsForLineno{align}%
\patchBothAmsMathEnvironmentsForLineno{flalign}%
\patchBothAmsMathEnvironmentsForLineno{alignat}%
\patchBothAmsMathEnvironmentsForLineno{gather}%
\patchBothAmsMathEnvironmentsForLineno{multline}%
\patchBothAmsMathEnvironmentsForLineno{eqnarray}%
}

\usepackage{hyperref}    
\usepackage[all]{hypcap} 

\def\lhcb {\mbox{LHCb}\xspace}

\ifthenelse{\boolean{uprightparticles}}%
{

 \def\PDelta      {\ensuremath{\Delta}\xspace}                 
 \def\PXi      {\ensuremath{\Xi}\xspace}                 
 \def\PLambda      {\ensuremath{\Lambda}\xspace}                 
 \def\PSigma      {\ensuremath{\Sigma}\xspace}                 
 \def\POmega      {\ensuremath{\Omega}\xspace}                 
 \def\PUpsilon      {\ensuremath{\Upsilon}\xspace}                 
 
 \def\PB      {\ensuremath{\mathrm{B}}\xspace}                 
                  
 \def\PD      {\ensuremath{\mathrm{D}}\xspace}

 \def\PK      {\ensuremath{\mathrm{K}}\xspace}

 \def\Pb      {\ensuremath{\mathrm{b}}\xspace}                 
 \def\Pc      {\ensuremath{\mathrm{c}}\xspace}

 \def\Pi      {\ensuremath{\mathrm{i}}\xspace}

}
{

 \mathchardef\PDelta="7101
 \mathchardef\PXi="7104
 \mathchardef\PLambda="7103
 \mathchardef\PSigma="7106
 \mathchardef\POmega="710A
 \mathchardef\PUpsilon="7107
                  
 \def\PB      {\ensuremath{B}\xspace}                 
                  
 \def\PD      {\ensuremath{D}\xspace}

 \def\PK      {\ensuremath{K}\xspace}

 \def\Pb      {\ensuremath{b}\xspace}                 
 \def\Pc      {\ensuremath{c}\xspace}

 \def\Pi      {\ensuremath{i}\xspace}

}

\makeatletter
\ifcase \@ptsize \relax
  \newcommand{\miniscule}{\@setfontsize\miniscule{4}{5}}
\or
  \newcommand{\miniscule}{\@setfontsize\miniscule{5}{6}}
\or
  \newcommand{\miniscule}{\@setfontsize\miniscule{5}{6}}
\fi
\makeatother

\DeclareRobustCommand{\optbar}[1]{\shortstack{{\miniscule (\rule[.5ex]{1.25em}{.18mm})}
  \\ [-.7ex] $#1$}}

\def\cquark    {{\ensuremath{\Pc}}\xspace}

\def\bquark    {{\ensuremath{\Pb}}\xspace}

\def\kaon    {{\ensuremath{\PK}}\xspace}
  \def\Kbar    {{\kern 0.2em\overline{\kern -0.2em \PK}{}}\xspace}

\def\KorKbar    {\kern 0.18em\optbar{\kern -0.18em K}{}\xspace}

\def\KS      {{\ensuremath{\kaon^0_{\rm\scriptscriptstyle S}}}\xspace}

  \def\Dbar    {{\kern 0.2em\overline{\kern -0.2em \PD}{}}\xspace}

\def\DorDbar    {\kern 0.18em\optbar{\kern -0.18em D}{}\xspace}

\def\B       {{\ensuremath{\PB}}\xspace}
\def\Bbar    {{\ensuremath{\kern 0.18em\overline{\kern -0.18em \PB}{}}}\xspace}

\def\BorBbar    {\kern 0.18em\optbar{\kern -0.18em B}{}\xspace}

\def\Bd      {{\ensuremath{\B^0}}\xspace}

  \def\Y#1S{\ensuremath{\PUpsilon{(#1S)}}\xspace}

\def\Lbar        {{\ensuremath{\kern 0.1em\overline{\kern -0.1em\PLambda}}}\xspace}
\def\LorLbar    {\kern 0.18em\optbar{\kern -0.18em \PLambda}{}\xspace}

\def\to                 {\ensuremath{\rightarrow}\xspace}

\def\AT#1     {\ensuremath{A_{\mathrm{T}}^{#1}}\xspace}           

\def\C#1      {\ensuremath{\mathcal{C}_{#1}}\xspace}                       
\def\Cp#1     {\ensuremath{\mathcal{C}_{#1}^{'}}\xspace}                    
\def\Ceff#1   {\ensuremath{\mathcal{C}_{#1}^{\mathrm{(eff)}}}\xspace}        
\def\Cpeff#1  {\ensuremath{\mathcal{C}_{#1}^{'\mathrm{(eff)}}}\xspace}       
\def\Ope#1    {\ensuremath{\mathcal{O}_{#1}}\xspace}                       
\def\Opep#1   {\ensuremath{\mathcal{O}_{#1}^{'}}\xspace}                    

\newcommand{\tev}{\ifthenelse{\boolean{inbibliography}}{\ensuremath{~T\kern -0.05em eV}\xspace}{\ensuremath{\mathrm{\,Te\kern -0.1em V}}}\xspace}
\newcommand{\gev}{\ensuremath{\mathrm{\,Ge\kern -0.1em V}}\xspace}
\newcommand{\mev}{\ensuremath{\mathrm{\,Me\kern -0.1em V}}\xspace}
\newcommand{\kev}{\ensuremath{\mathrm{\,ke\kern -0.1em V}}\xspace}
\newcommand{\ev}{\ensuremath{\mathrm{\,e\kern -0.1em V}}\xspace}
\newcommand{\gevc}{\ensuremath{{\mathrm{\,Ge\kern -0.1em V\!/}c}}\xspace}
\newcommand{\mevc}{\ensuremath{{\mathrm{\,Me\kern -0.1em V\!/}c}}\xspace}
\newcommand{\gevcc}{\ensuremath{{\mathrm{\,Ge\kern -0.1em V\!/}c^2}}\xspace}
\newcommand{\gevgevcccc}{\ensuremath{{\mathrm{\,Ge\kern -0.1em V^2\!/}c^4}}\xspace}
\newcommand{\mevcc}{\ensuremath{{\mathrm{\,Me\kern -0.1em V\!/}c^2}}\xspace}

\def\mm   {\ensuremath{\rm \,mm}\xspace}

\def\mum  {\ensuremath{{\,\upmu\rm m}}\xspace}

\def\fb   {\ensuremath{\mbox{\,fb}}\xspace}

\newcommand{\chisq}{\ensuremath{\chi^2}\xspace}

\newcommand{\chisqip}{\ensuremath{\chi^2_{\rm IP}}\xspace}

\def\gsim{{~\raise.15em\hbox{$>$}\kern-.85em
          \lower.35em\hbox{$\sim$}~}\xspace}
\def\lsim{{~\raise.15em\hbox{$<$}\kern-.85em
          \lower.35em\hbox{$\sim$}~}\xspace}

\def\ptot       {\mbox{$p$}\xspace}
\def\pt         {\mbox{$p_{\rm T}$}\xspace}

\def\evtgen     {\mbox{\textsc{EvtGen}}\xspace}

\def\geant      {\mbox{\textsc{Geant4}}\xspace}

\def\photos     {\mbox{\textsc{Photos}}\xspace}

\def\pythia     {\mbox{\textsc{Pythia}}\xspace}

\def\tell1  {TELL1\xspace}
\def\ukl1   {UKL1\xspace}

\usepackage{cite} 
\usepackage{mciteplus}

\usepackage{longtable} 
\usepackage{chngpage}

\def \light {\ensuremath{\rm light\mbox{-}parton}\xspace}
\def \xip {\ensuremath{\chi^2_{\mathrm{IP}}}\xspace}
\def \bdtbcl {\ensuremath{{\rm BDT}(bc|udsg)}\xspace}
\def \bdtbc {\ensuremath{{\rm BDT}(b|c)}\xspace}

\begin{document}

\renewcommand{\thefootnote}{\fnsymbol{footnote}}
\setcounter{footnote}{1}

\begin{titlepage}
\pagenumbering{roman}

\vspace*{-1.5cm}
\centerline{\large EUROPEAN ORGANIZATION FOR NUCLEAR RESEARCH (CERN)}
\vspace*{1.5cm}
\hspace*{-0.5cm}
\begin{tabular*}{\linewidth}{lc@{\extracolsep{\fill}}r}
\ifthenelse{\boolean{pdflatex}}
{\vspace*{-2.7cm}\mbox{\!\!\!\includegraphics[width=.14\textwidth]{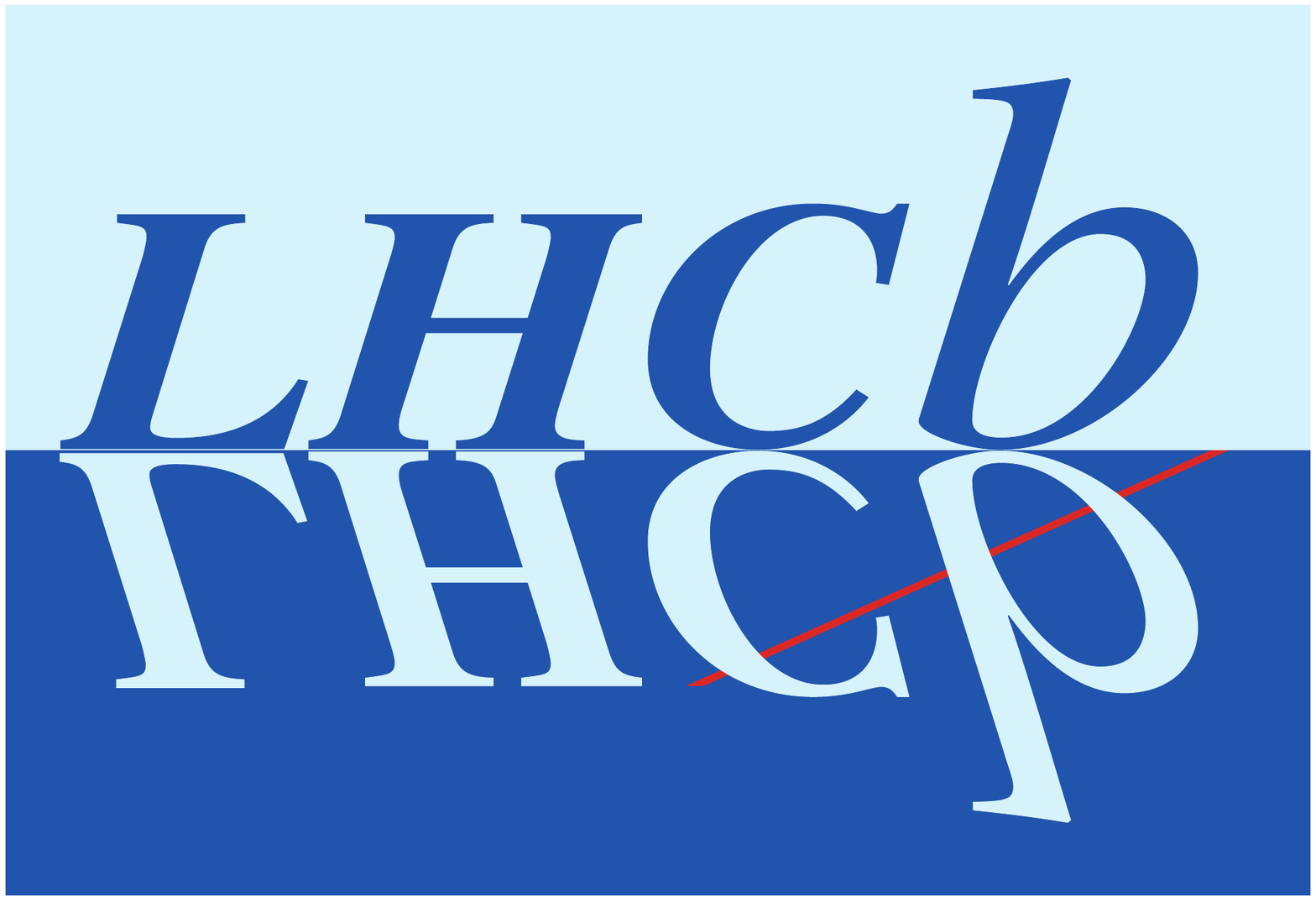}} & &}%
{\vspace*{-1.2cm}\mbox{\!\!\!\includegraphics[width=.12\textwidth]{lhcb-logo.eps}} & &}%
\\
 & & CERN-PH-EP-2015-101 \\  
 & & LHCb-PAPER-2015-016 \\  
 & & May 28, 2015 \\ 
 & & \\
\end{tabular*}

\vspace*{4.0cm}

{\bf\boldmath\huge
\begin{center}
  Identification of beauty and charm quark jets at LHCb
\end{center}
}

\vspace*{2.0cm}

\begin{center}
The LHCb collaboration\footnote{Authors are listed at the end of this paper.}
\end{center}

\vspace{\fill}

\begin{abstract}
  \noindent
  Identification of jets originating from beauty and charm quarks is important for measuring Standard Model processes and for searching for new physics.  
The performance of algorithms developed to select $b$- and $c$-quark jets
is measured using data recorded by LHCb from proton-proton collisions at $\sqrt{s}=7\tev$ in 2011 and at $\sqrt{s}=8\tev$ in 2012.
The efficiency for identifying a $b(c)$ jet is about 65\%(25\%) with a probability for misidentifying a light-parton jet of 0.3\% for jets with transverse momentum $\pt > 20\gev$ and pseudorapidity $2.2 < \eta < 4.2$.  The dependence of the performance on the \pt and $\eta$ of the jet is also measured.  
\end{abstract}

\vspace*{1.0cm}

\begin{center}
  JINST {\bf 10} (2015) P06013  
\end{center}

\vspace{\fill}

{\footnotesize 
\centerline{\copyright~CERN on behalf of the \lhcb collaboration, license \href{http://creativecommons.org/licenses/by/4.0/}{CC-BY-4.0}.}}
\vspace*{2mm}

\end{titlepage}

\newpage
\setcounter{page}{2}
\mbox{~}
\newpage

\renewcommand{\thefootnote}{\arabic{footnote}}
\setcounter{footnote}{0}

\pagestyle{plain} 
\setcounter{page}{1}
\pagenumbering{arabic}


\clearpage

\section{Introduction}

Identification of jets that originate from the hadronization of beauty ($b$) and charm ($c$) quarks is important for studying Standard Model (SM) processes and for searching for new physics.
For example, the ability to efficiently identify $b$ jets with minimal misidentification of $c$ and \light jets is crucial for the measurement of top-quark production.  The study of $t\bar{t}$ production in the forward region probes the structure of the proton~\cite{gauld} and can be used to search for physics beyond the SM~\cite{PhysRevLett.107.082003}.  
Measuring charge asymmetries in di-$b$-jet production also probes beyond the SM physics~\cite{grinstein,afc}.
Furthermore, identification of $c$ jets is important for probing the structure of the proton, {\em e.g.} in $W\!+\!c$ production.  

The signature of a $b$ or $c$ jet is the presence of a long-lived $b$ or $c$ hadron that carries a sizable fraction of the jet energy.
The LHCb detector was designed to identify $b$ and $c$ hadrons, 
and so is expected to perform well at identifying, or tagging, $b$ and $c$ jets.  
This paper describes two algorithms for identifying $b$ and $c$ jets, one designed to identify both $b$ and $c$ jets offline, and another initially designed to identify $b$-hadron decays in the trigger.
The performance of each algorithm is measured using several subsamples of the $3\fb^{-1}$ of proton-proton collision data collected at $\sqrt{s}=7\tev$ in 2011 and at 8\tev in 2012 by the LHCb detector.
The distributions of observable quantities used to discriminate between $b$, $c$ and \light jets are compared between data and simulation.

\section{The LHCb detector}

The \lhcb detector~\cite{Alves:2008zz,Aaij:2014jba} is a single-arm forward
spectrometer covering the \mbox{pseudorapidity} range $2<\eta <5$,
designed for the study of particles containing \bquark or \cquark
quarks. The detector includes a high-precision tracking system
consisting of a silicon-strip vertex detector surrounding the $pp$
interaction region~\cite{LHCbVELOGroup:2014uea}, a large-area silicon-strip detector located
upstream of a dipole magnet with a bending power of about
$4{\rm\,Tm}$, and three stations of silicon-strip detectors and straw
drift tubes~\cite{LHCb-DP-2013-003} placed downstream of the magnet.
The tracking system provides a measurement of momentum, \ptot, of charged particles with
a relative uncertainty that varies from 0.5\% at low momentum to 1.0\% at 200\gev ($c=1$ throughout this paper).
The minimum distance of a track to a primary vertex, the impact parameter, is measured with a resolution of $(15+29/\pt)\mum$,
where \pt is the component of the momentum transverse to the beam, in\,\gev.
Different types of charged hadrons are distinguished using information
from two ring-imaging Cherenkov detectors. 
Photons, electrons and hadrons are identified by a calorimeter system consisting of
scintillating-pad and preshower detectors, an electromagnetic
calorimeter and a hadronic calorimeter.
The electromagnetic and hadronic calorimeters have an energy resolution of $\sigma(E)/E = 10\%/\sqrt{E}\oplus 1\%$ and $\sigma(E)/E = 69\%/\sqrt{E}\oplus 9\%$ (with $E$ in GeV), respectively.
Muons are identified by a
system composed of alternating layers of iron and multiwire
proportional chambers~\cite{LHCb-DP-2012-002}.

The trigger~\cite{LHCb-DP-2012-004} consists of a
hardware stage, based on information from the calorimeter and muon
systems, followed by a software stage, which applies a full event
reconstruction.  
This analysis requires that either a high-\pt muon or a $(b,c)$-hadron\footnote{The notation $(b,c)$ is used to mean $b$ or $c$ throughout this paper.} candidate satisfies the trigger requirements.
Events recorded due to the presence of a high-\pt muon are required to have a muon candidate with $\pt>10\gev$.  
Events recorded due to the presence of a $(b,c)$-hadron decay
require that at least one track should have $\pt >
  1.7\gev$ and \chisqip with respect to any
  primary interaction greater than 16, where \chisqip is defined as the
  difference in \chisq of a given primary $pp$ interaction vertex~(PV) reconstructed with and
  without the considered track.
Decays of $b$ hadrons are inclusively identified by requiring a two-, three- or four-track
  secondary vertex (SV) with a large sum of \pt of
  the tracks and a significant displacement from the PV.
 A specialized boosted decision tree (BDT)~\cite{Breiman}  algorithm is used for
  the identification of SVs consistent with the decay
  of a \bquark hadron~\cite{BBDT}.
This inclusive trigger algorithm is called the topological trigger (TOPO) and is studied as a $b$-jet tagger in this paper. 
Decays of long-lived $c$ hadrons are identified either exclusively using decay modes with large branching fractions, or in $D^*(2010)^{\pm} \to D^0\pi^{\pm}$ decays where the $D^0$ is selected inclusively by the presence of a two-track SV.

In the simulation, $pp$ collisions are generated using
\pythia~\cite{Sjostrand:2007gs} 
 with a specific \lhcb
configuration~\cite{LHCb-PROC-2010-056}.  Decays of hadronic particles
are described by \evtgen~\cite{Lange:2001uf}, in which final-state
radiation is generated using \photos~\cite{Golonka:2005pn}. The
interaction of the generated particles with the detector, and its
response, are implemented using the \geant
toolkit~\cite{Allison:2006ve, *Agostinelli:2002hh} as described in
Ref.~\cite{LHCb-PROC-2011-006}.

\section{Jet identification algorithms}

Jets are clustered using the anti-$k_T$ algorithm~\cite{1126-6708-2008-04-063} with a distance parameter $0.5$, as implemented in
\textsc{Fastjet}~\cite{fastjet}.
Information from all the detector
sub-systems is used to create charged and neutral particle inputs to
the jet algorithm using a particle flow approach~\cite{LHCb-PAPER-2013-058}. 
During 2011 and 2012, LHCb collected data with a mean number of $pp$ collisions per crossing of about 1.7.  To reduce contamination from multiple $pp$ interactions, charged particles reconstructed within the vertex detector may only be clustered into a jet if they are associated to the same PV. 
The identification of $(b,c)$ jets is performed using SVs from the decays of $(b,c)$ hadrons.  The choice of using SVs and not single-track or other non-SV-based jet properties, {\em e.g.} the number of particles in the jet, is driven by the need for a small misidentification probability of \light jets in the analyses performed at LHCb.  
Furthermore, the properties of SVs from $(b,c)$-hadron decays are known to be well modeled in LHCb simulation.

\subsection{The SV tagger}

The tracks used as inputs to the SV-tagger algorithm are required to have $\pt > 0.5\gev$ and $\xip > 16$.  The \xip requirement is rarely satisfied by tracks reconstructed from particles originating directly from the PV.  
Hadronic particle identification is not used and, instead, all particles are assigned the pion mass.
In contrast to many other jet-tagging algorithms, tracks are not required to have $\Delta R \equiv \sqrt{\Delta\eta^2 + \Delta\phi^2} < 0.5$, where $\Delta\eta(\Delta\phi)$ is the difference in pseudorapidity (azimuthal angle) between the track momentum and jet axis, 
 since for low \pt jets tracks outside of the jet cone help to discriminate between $c$ and $b$ jets.

All possible two-track SVs are built using pairs of the input tracks such that the distance of closest approach between the tracks is less than 0.2\mm, the vertex fit $\chi^2 < 10$ and the two-body mass is in the range $0.4\gev < M < M(B)$, where $M(B)$ is the nominal \Bd mass~\cite{PDG2014}.  Since all particles are assigned a pion mass, the upper mass requirement rarely removes SVs from any long-lived $b$ hadrons.  The lower mass requirement removes SVs from most strange-particle decays, including the $\Lambda$ baryon whose computed mass is always below 0.4\gev when the proton is assigned a pion mass.
At this stage tracks are allowed to belong to multiple SVs. 
Next, all two-track SVs with $\Delta R < 0.5$ relative to the jet axis, where the direction of flight is taken as the PV to SV vector, are collected as candidates for a so-called linking procedure.  This procedure involves merging SVs that share tracks until none of the remaining SVs with $\Delta R < 0.5$ share tracks.  The SV position is taken to be the weighted average of the 2-body SV positions using the inverse of the 2-body vertex $\chi^2$ values as the weights.

The linking procedure can produce SVs that contain any number of tracks.  The linked $n$-track SVs are required to have $\pt > 2\gev$, significant spatial separation from the PV, and to contain at most one track with $\Delta R > 0.5$ relative to the jet axis.  
If the SV has only two tracks and a mass consistent with that of the \KS~\cite{PDG2014}, the SV is rejected.  
Interactions with material, and strange-particle decays, are suppressed by requiring that the flight distance divided by the momentum of the SV is less than 1.5\mm/GeV; this quantity serves as a proxy for the hadron lifetime.
The SV position is also required to be within a restricted region consistent with that of $(b,c)$-hadron decays.

An important quantity for discriminating between hadron types is the so-called corrected mass defined as
\begin{equation}
\label{eq:mcor}
M_{\rm cor} = \sqrt{M^2 + p^2\sin^2{\theta}} + p\sin{\theta},
\end{equation}
where $M$ and $p$ are the invariant mass and momentum of the particles that form the SV and $\theta$ is the angle between the
momentum and the direction of flight of the SV.  The corrected mass is the minimum mass that the long-lived hadron
can have that is consistent with the direction of flight.
The linked $n$-track SVs are required to have $M_{\rm cor} > 0.6\gev$ to remove any remaining kaon or hyperon decays.
A few percent of jets contain multiple SVs that pass all requirements; in such cases the SV with the highest \pt is chosen. The fraction of multi-SV-tagged jets is consistent in data and simulation.

Two BDTs are used to identify $b$ and $c$ jets: \bdtbcl trained to separate $(b,c)$ jets from \light jets and \bdtbc trained to separate $b$ jets from $c$ jets.  Both BDTs are trained on simulated samples of $b$, $c$ and \light jets. The inputs to both BDTs are as follows:
\begin{itemize}
\item the SV mass $M$;
\item  the SV corrected mass $M_{\rm cor}$;
\item the transverse flight distance of the two-track SV closest to the PV;
\item the fraction of the jet \pt carried by the SV, $\pt({\rm SV})/\pt({\rm jet})$;
\item $\Delta R$ between the SV flight direction and the jet;
\item the number of tracks in the SV;
\item  the number of SV tracks with $\Delta R < 0.5$ relative to the jet axis;
\item the net charge of the tracks that form the SV;
\item the flight distance $\chi^2$;
\item the sum of all SV track \xip.
\end{itemize} 
For jets that contain an SV passing all of the requirements, the two BDT responses are used to identify the jet as either $b$, $c$ or \light.

\subsection{The topological trigger}

The topological trigger algorithm uses SVs that satisfy similar criteria to those used in the SV-tagger algorithm to build two-, three- and four-track SVs.  The TOPO SVs are required to have large \pt and significant flight distance from the PV. 
The TOPO provides an efficient trigger option for generic $b$-jet events, as the SV used by the TOPO to trigger recording of the event can also be used to tag a $b$ jet.
  The BDT used in the TOPO algorithm uses the following inputs:
\begin{itemize}
\item the SV mass;
\item the SV corrected mass;
\item the sum of the \pt of the SV tracks;
\item the maximum distance of closest approach between the SV tracks;
\item the \xip of the SV formed using the momentum of the tracks that form the SV and SV position;
\item the flight distance $\chi^2$ of the SV from the PV;
\item the minimum \pt of the SV tracks.
\end{itemize}
To ensure stability during data-taking the TOPO BDT uses discretized inputs as described in detail in Ref.~\cite{BBDT}.  Further details about the TOPO algorithm and its performance on $b$-hadron decays as measured in LHCb data can be found in Ref.~\cite{LHCb-DP-2012-004}.

\subsection{Performance in simulation}

\begin{figure}[]
  \centering
  \includegraphics[width=0.32\textwidth]{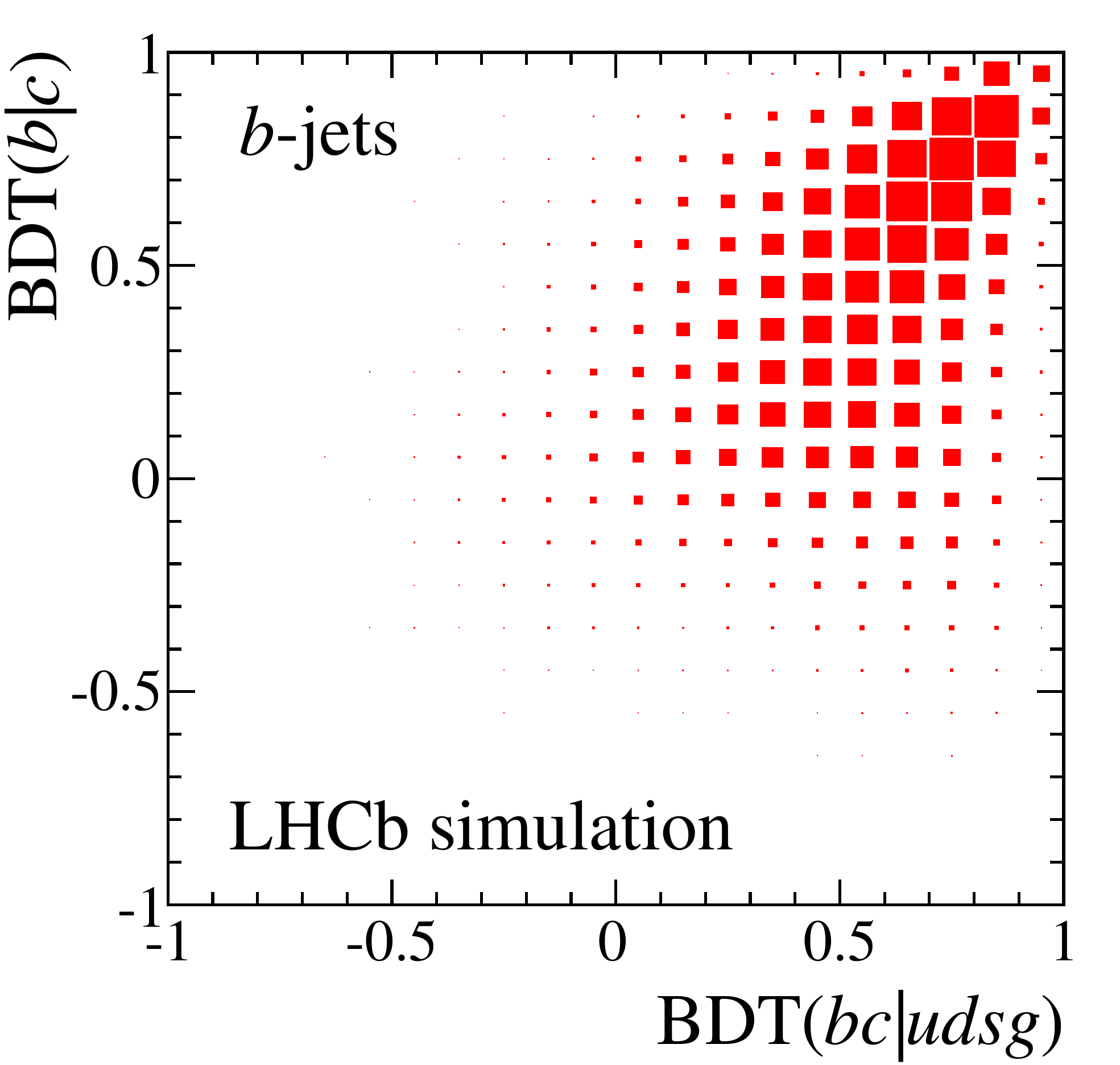}
  \includegraphics[width=0.32\textwidth]{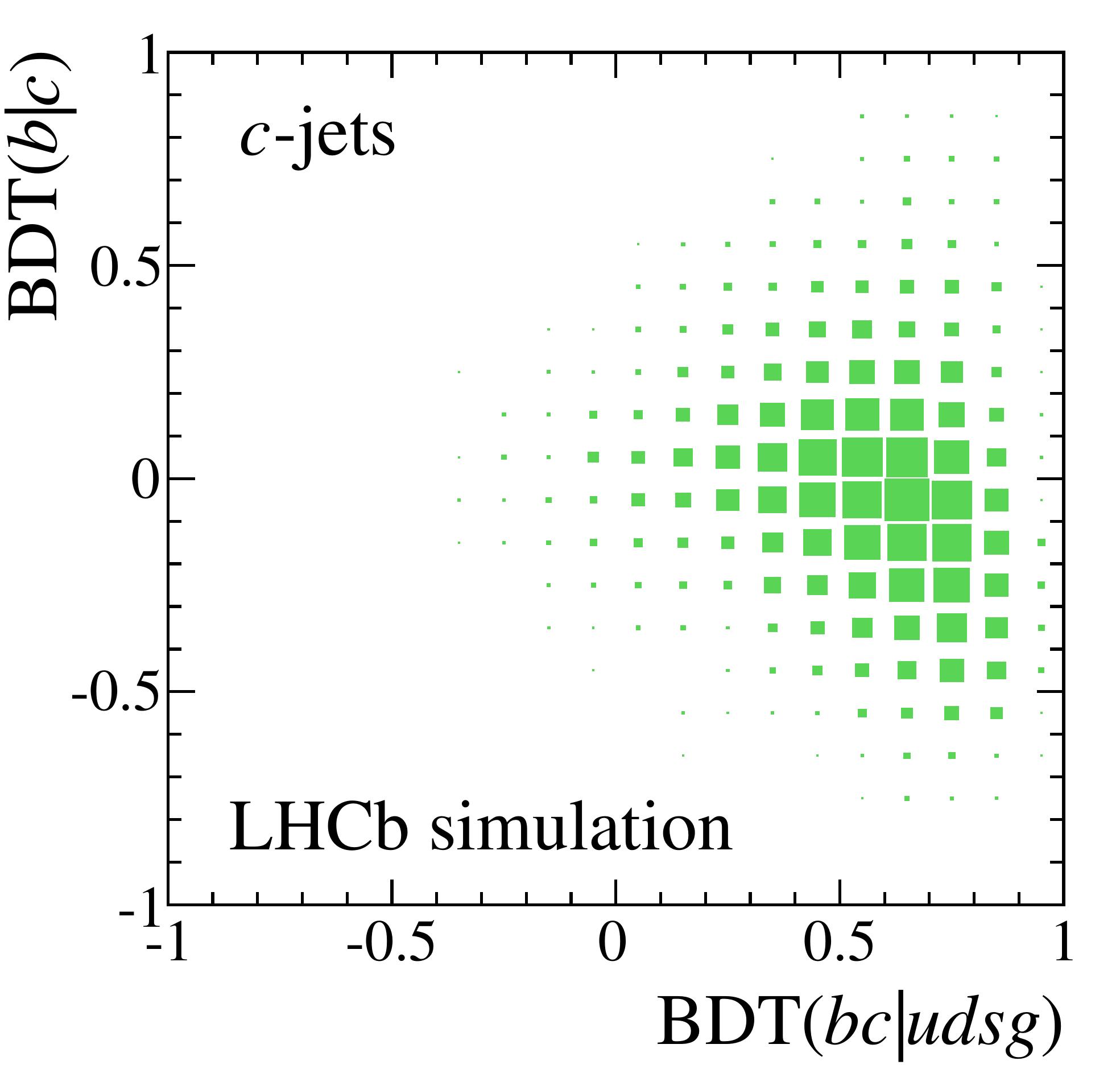}
  \includegraphics[width=0.32\textwidth]{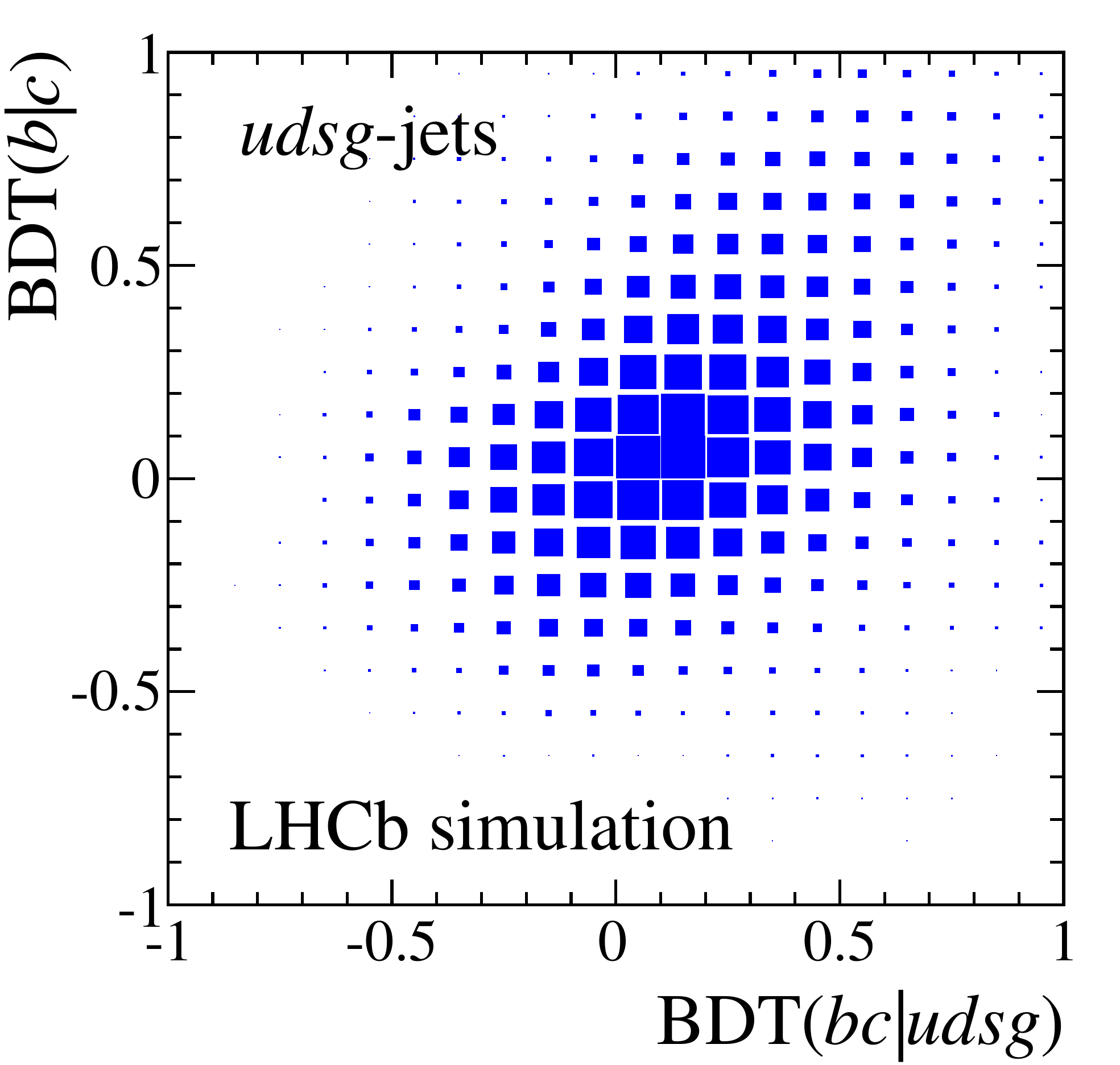}
  \caption{\label{fig:mc2d}
    SV-tagger algorithm \bdtbc versus \bdtbcl distributions obtained from simulation for (left) $b$, (middle) $c$ and (right) \light jets.
  }
\end{figure}

\begin{figure}[]
  \centering
  \includegraphics[width=0.49\textwidth]{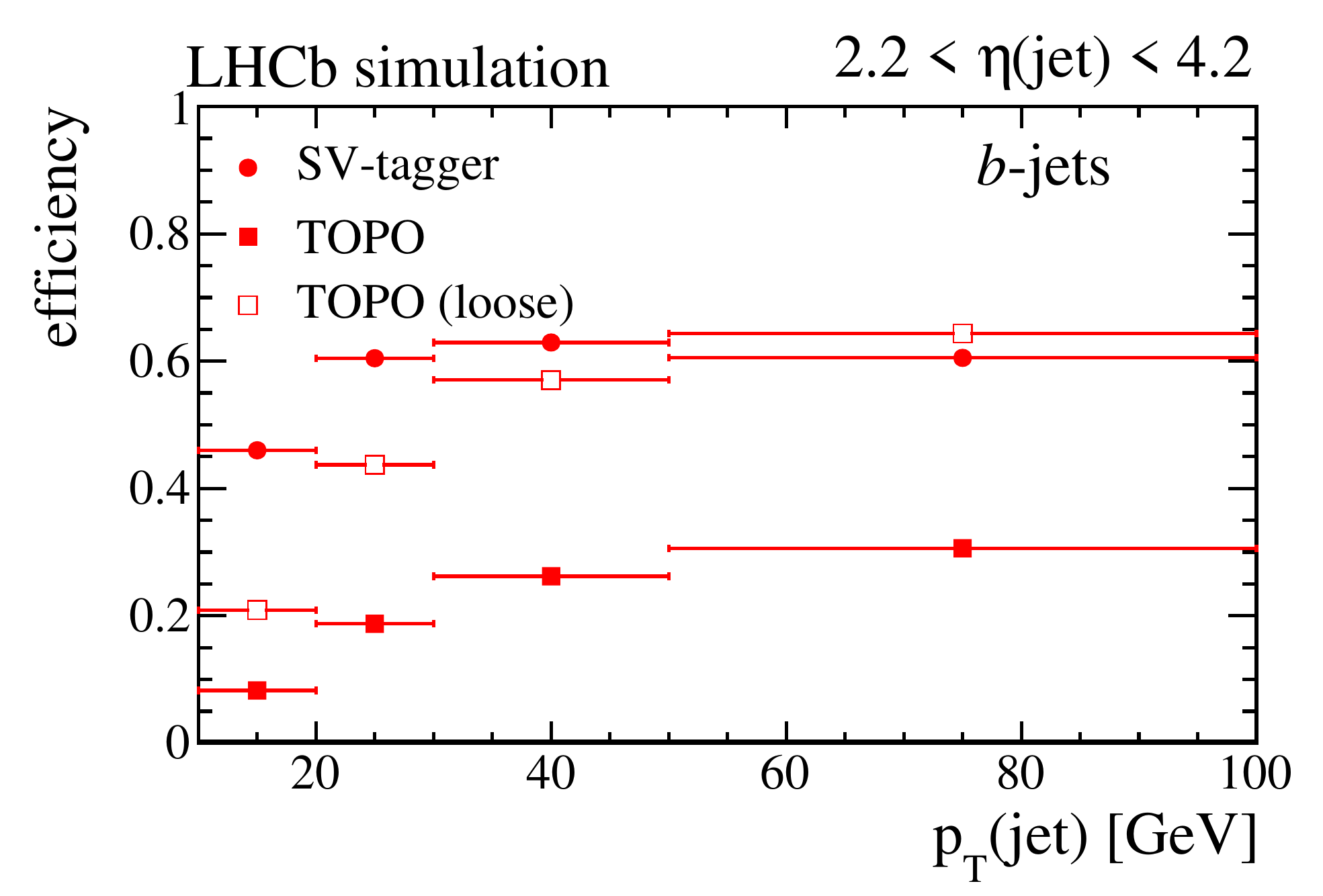}
  \includegraphics[width=0.49\textwidth]{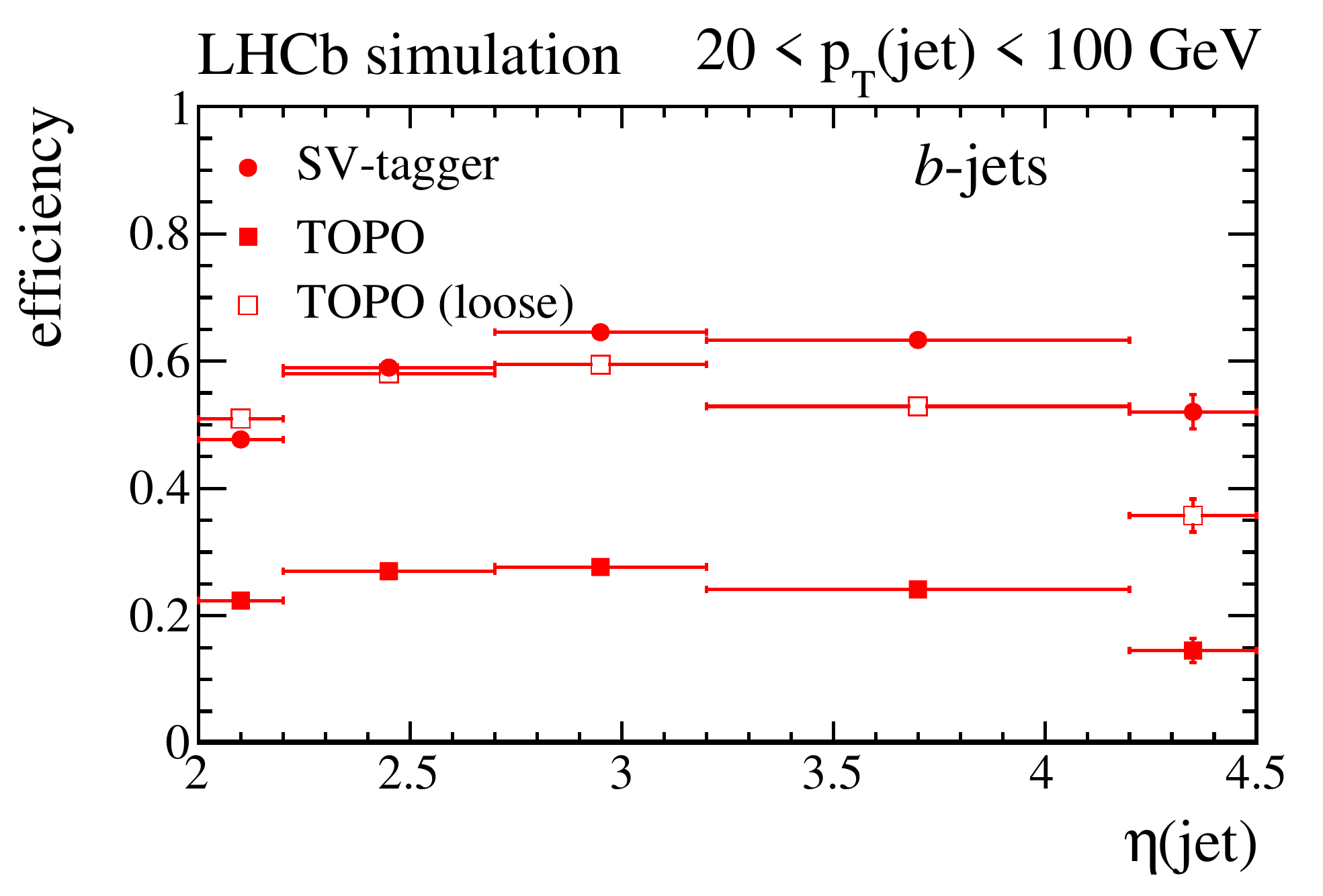}\\
  \includegraphics[width=0.49\textwidth]{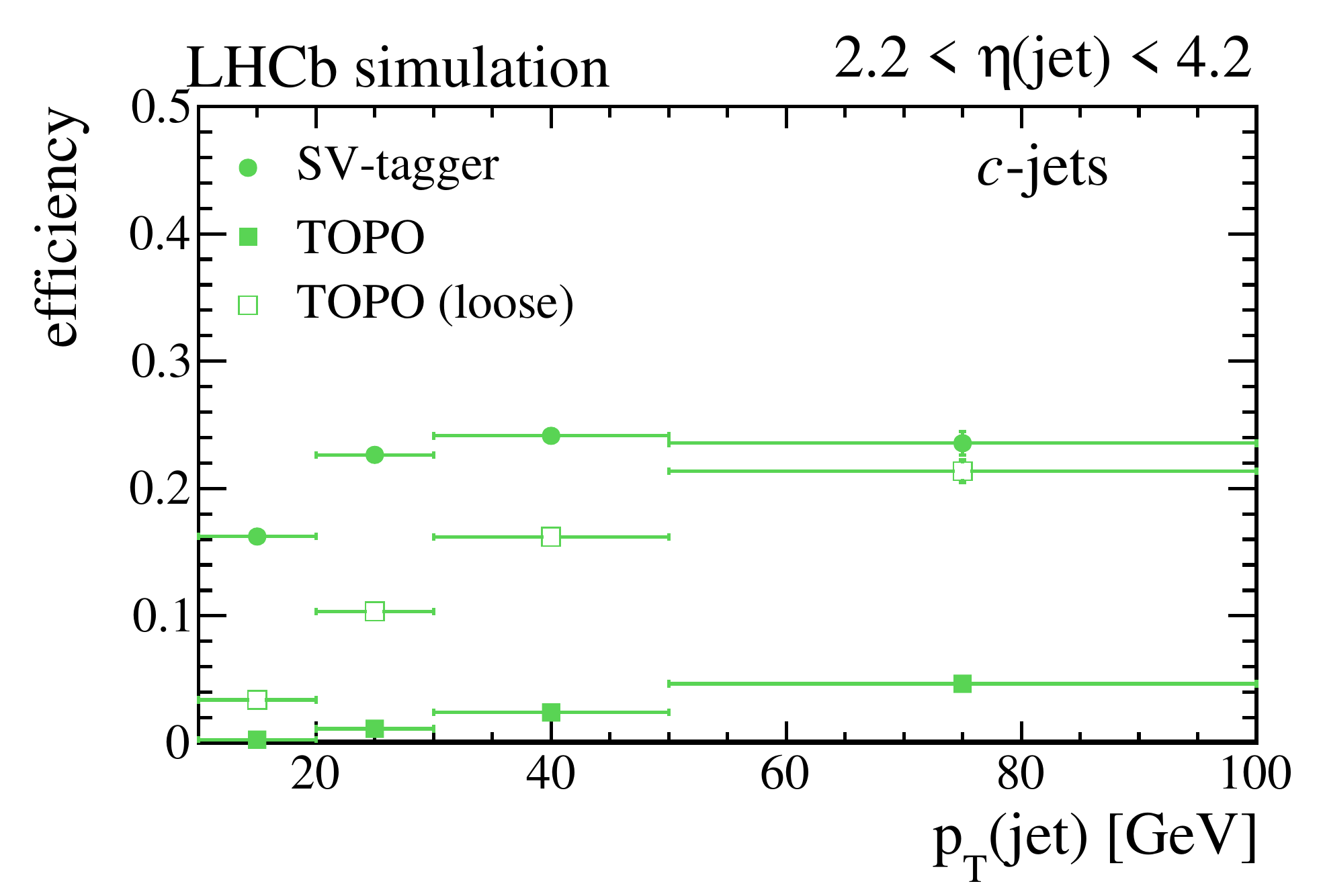}
  \includegraphics[width=0.49\textwidth]{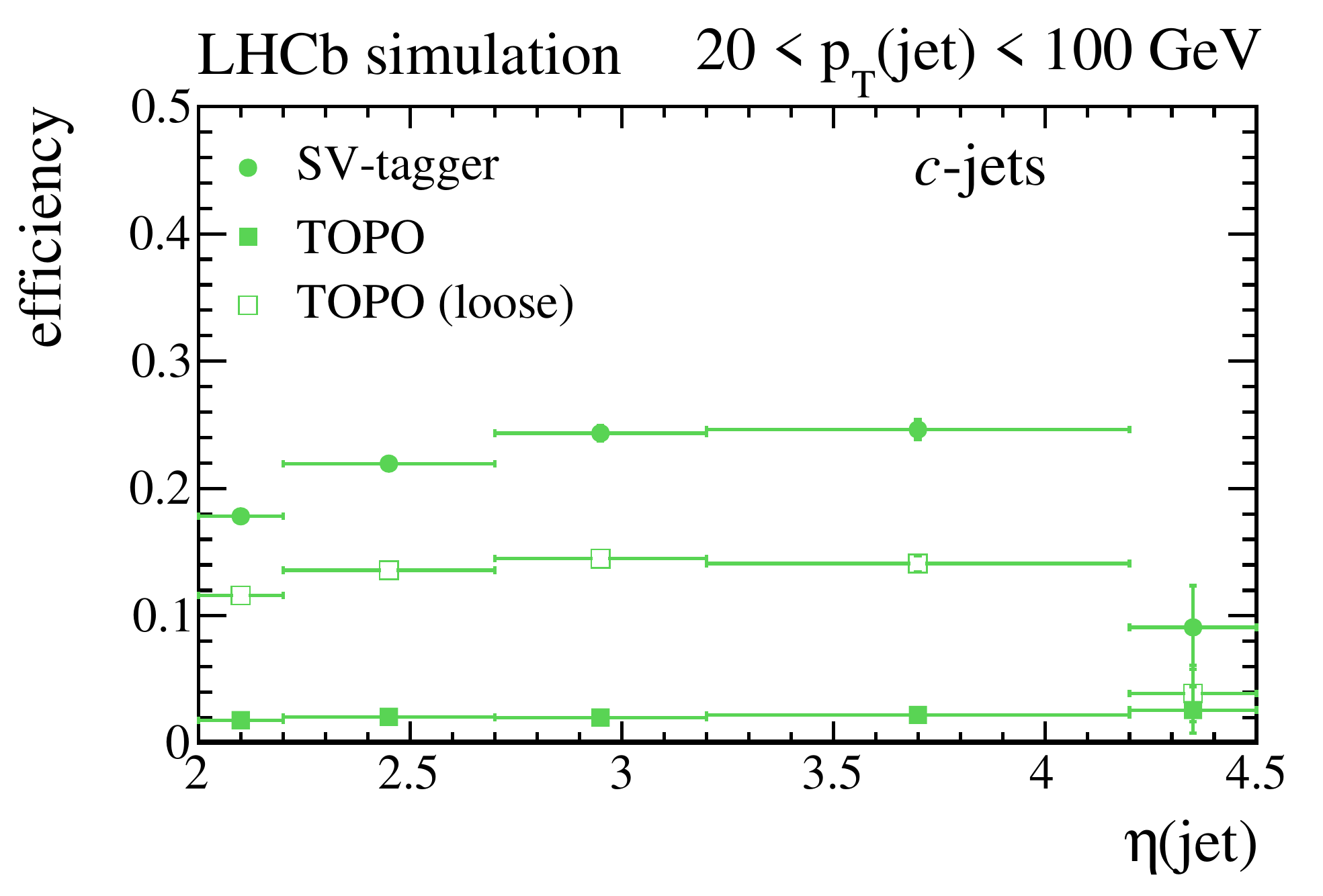}\\
  \includegraphics[width=0.49\textwidth]{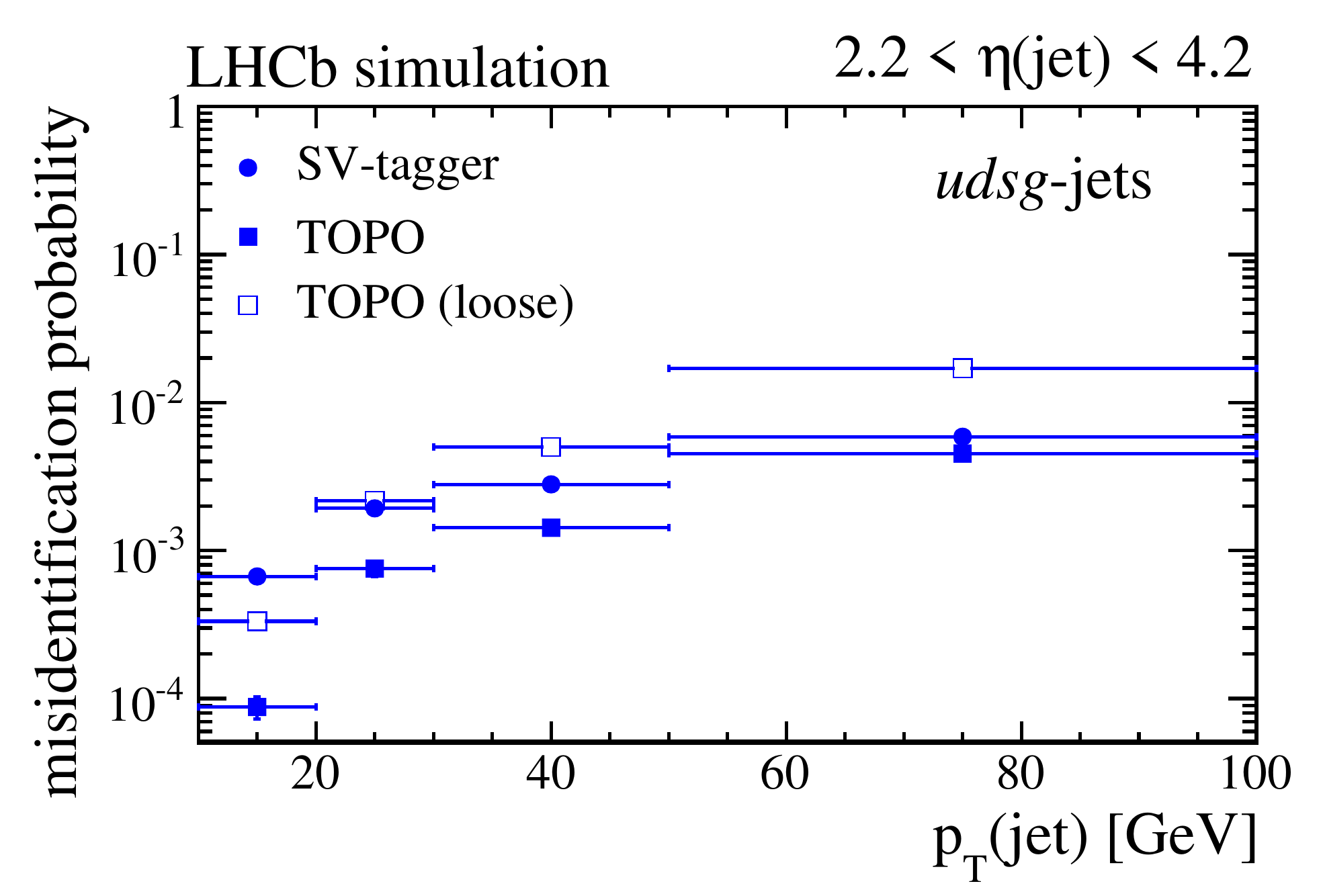}
  \includegraphics[width=0.49\textwidth]{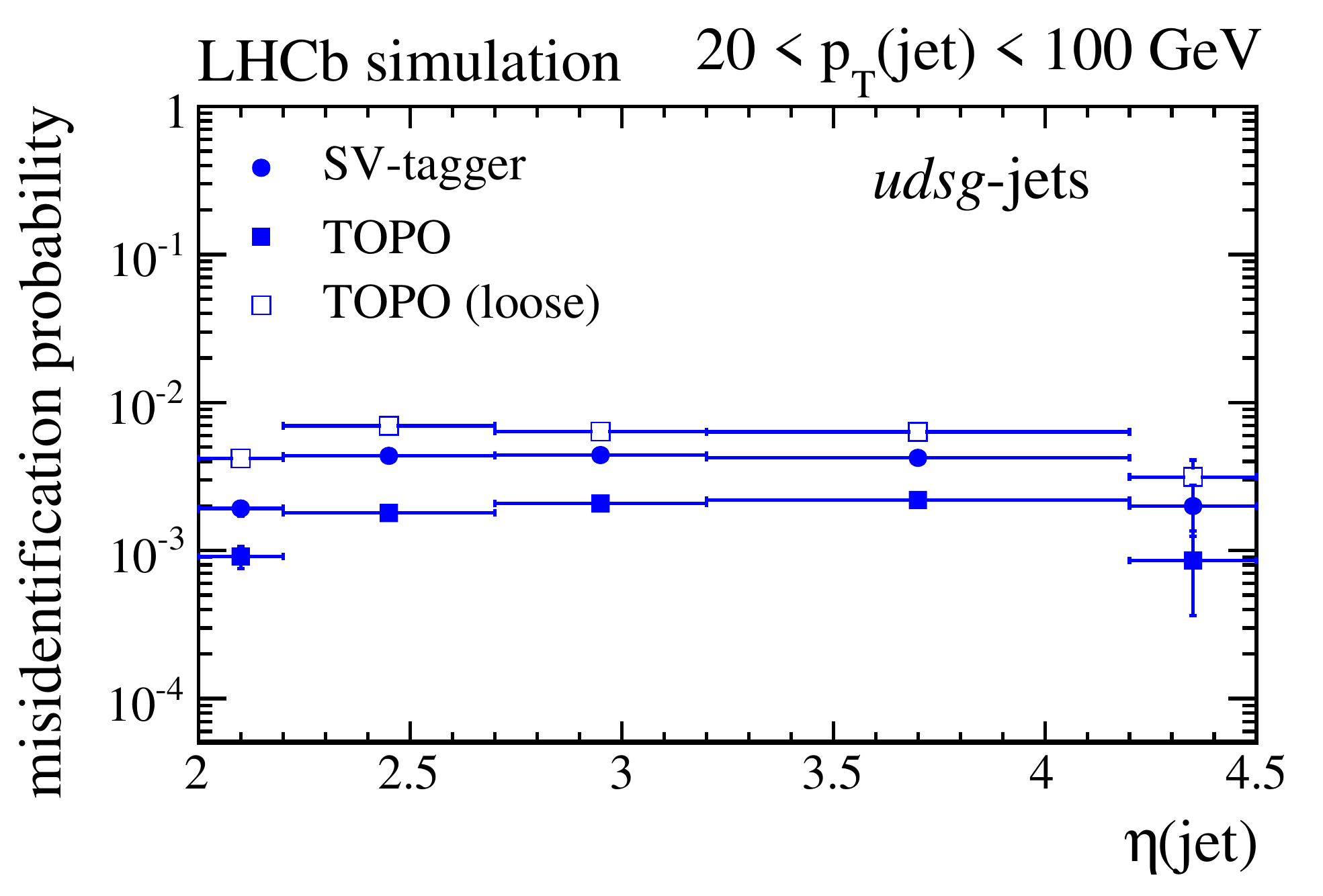}
  \caption{\label{fig:mc}
    Efficiencies and mistag probabilities obtained from simulation for the SV-tagger and TOPO algorithms for (top) $b$, (middle) $c$  and (bottom) \light jets.  The left plots show the dependence on \pt for $2.2 < \eta < 4.2$, while the right plots show the dependence on $\eta$ for  $\pt > 20\gev$ (see text for details). The ``loose'' label for the TOPO refers to the BDT requirement used in the trigger for SVs that contain muon candidates.
  }
\end{figure}

\begin{figure}[]
  \centering
  \includegraphics[width=0.55\textwidth]{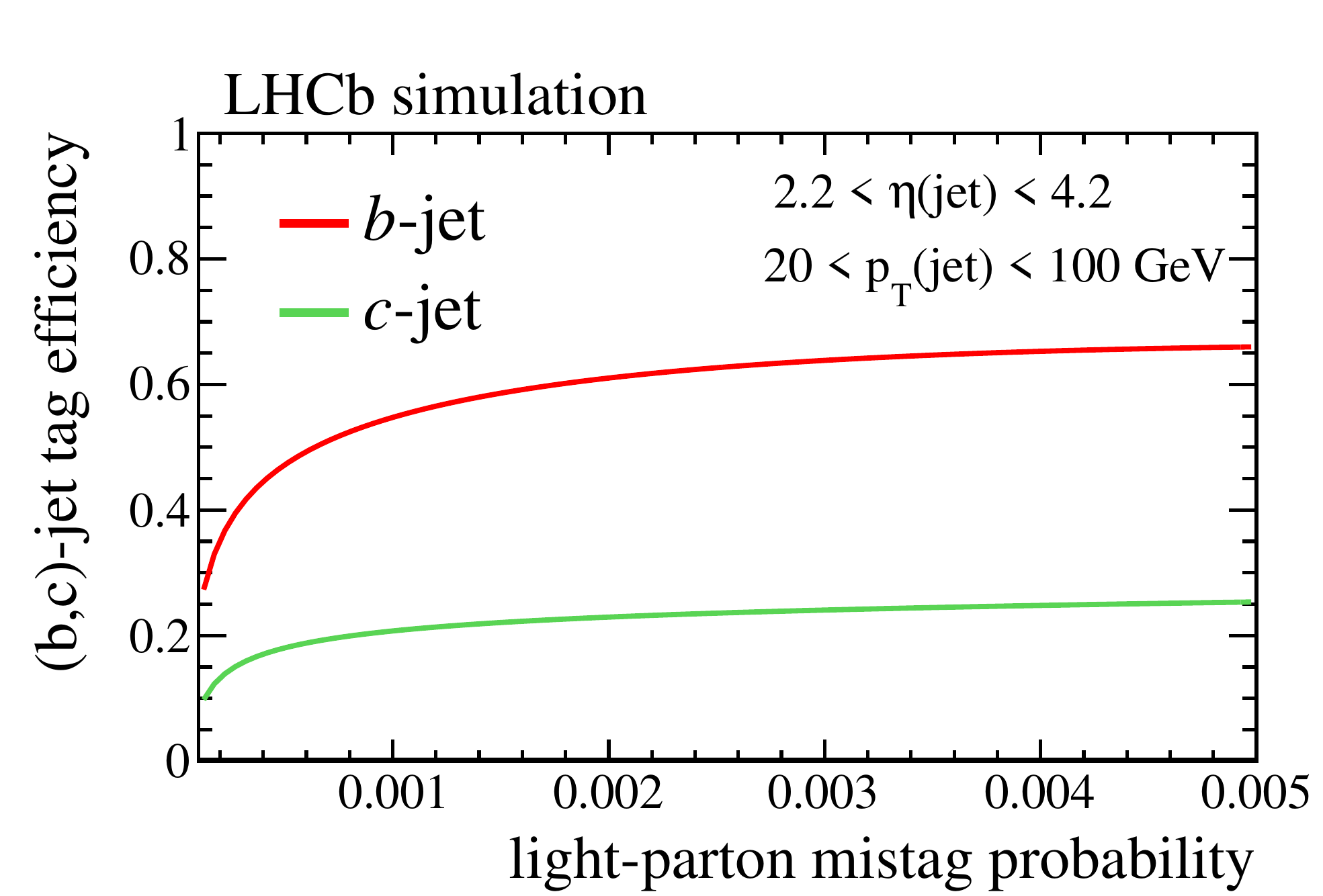}
  \caption{\label{fig:mc2}
    Efficiencies for SV-tagging a $(b,c)$-jet versus mistag probability for a \light jet from simulation.  The curves are obtained by varying the \bdtbcl requirement.
  }
\end{figure}

Figure~\ref{fig:mc2d} shows the SV-tagger BDT distributions obtained from simulated $W\!+$jet events for each jet type.  The distributions in the two-dimensional BDT plane of SV-tagged $b$, $c$, and \light jets are clearly distinguishable. 
The full two-dimensional distribution is fitted in data to determine the jet flavor content. 
However, to aid in comparison to other jet-tagging algorithms, a requirement of  $\bdtbcl > 0.2$ is applied to display the performance obtained from simulated events in Fig.~\ref{fig:mc}.
This requirement is about 90\% efficient on SV-tagged $(b,c)$ jets and highly suppresses \light jets.
The $(b,c)$-jet efficiencies are nearly uniform for jet $\pt > 20\gev$ and for $2.2 < \eta < 4.2$, but are lower for low-\pt jets and for jets near the edges of the detector.  
The misidentification probability of \light jets is less than 0.1\% for low-\pt jets and increases to about 1\% at 100\gev.  
Figure~\ref{fig:mc2} shows the $(b,c)$-jet efficiencies versus the mistag probability of \light jets obtained by increasing the \bdtbcl cut.

For the TOPO algorithm, in the trigger a BDT requirement is always applied; the requirement is looser when the SV contains a muon.  
In the LHCb measurement of the charge asymmetry in $b\bar{b}$ production~\cite{LHCb-PAPER-2014-023}, this same looser BDT requirement was applied to tag a second jet in the event. 
Figure~\ref{fig:mc} shows the performance of the TOPO algorithm, obtained from simulated events, for both the nominal and loose BDT requirements.  
The nominal trigger BDT requirement strongly suppresses $c$ and \light jets, with the misidentification probability of \light jets being 0.01\% for low-\pt jets.  
Such a strong suppression is required during online running due to output rate limitations.

The jet-tagging performance is measured in simulated events with one $pp$ collision and two or more $pp$ collisions and found to be consistent.
The tagging performance is also studied in simulation using different event types, {\em e.g.} top-quark and QCD di-jet events, with only small changes in the tagging efficiencies and BDT templates observed for $(b,c)$ jets.  
The mistag probability of \light jets is found to be higher for high-\pt jets in events that also contain $(b,c)$ jets.  This is discussed in detail in Sec.~\ref{sec:light}.

\section{Efficiency measurements in data}

The tagging efficiencies for $b$ and $c$ jets are measured in data and compared with expectations from simulation.  
To measure the tagging efficiencies in a given data sample, both the number of tagged $(b,c)$ jets and the total number of $(b,c)$ jets must be determined.
The tagged $(b,c)$ yields are obtained by fitting the SV-tagger or TOPO BDT distributions in the subsample of jets that are tagged by an SV.  
The total number of $(b,c)$ jets is determined by fitting the \xip distribution of the highest-\pt track in the jet.  
The $(b,c)$-tagging efficiency is the ratio of the tagged over total $(b,c)$-jet yields.  

An alternative approach employed by other experiments  (see, {\em e.g.} Ref.~\cite{Chatrchyan:2012jua}) is to measure the efficiency using the subsample of jets that contain a muon.  This approach has the advantage that the $(b,c)$-jet content is enhanced due to the presence of muons from the semileptonic decays of $(b,c)$ hadrons; however, the disadvantage is that this method assumes that mismodeling of the tagging performance is the same for semileptonic and inclusive decays.
Both the highest-\pt track and muon-jet methods are used in this analysis to study the jet-tagging performance.

Combined fits of several data samples enriched in $(b,c)$ jets are performed to obtain the tagging efficiencies.
It is important to include the systematic uncertainties on both the tagged and total $(b,c)$-jet yields for each data sample in the combined fits.

This section is arranged as follows: the data samples used are described in Sec.~4.1; the BDT fits used to obtain the tagged $(b,c)$-jet yields are given in Sec.~4.2; the highest-\pt-track \xip fits used to obtain the total $(b,c)$-jet yields are described in Sec.~4.3;
the muon-jet subsample method is discussed in Sec.~4.4; the systematic uncertainties on the tagged and total $(b,c)$-jet yields are presented in Sec.~4.5; and the $(b,c)$-tagging efficiency results are given in Sec.~4.6.

\subsection{Data samples}

Events that contain either a high-\pt muon or a fully reconstructed $(b,c)$ hadron, referred to here as an event-tag,   are used to measure the jet-tagging efficiencies in data.
The highest-\pt jet in the event that does not have any overlap with the event-tag is chosen as the test jet.
Each event-tag is required to have satisfied specific trigger requirements and to have $\Delta\phi > 2.5$ relative to the test-jet axis to reduce the possibility of contamination of the jet from the event-tag\footnote{The event-tag samples are highly pure; however, when the event-tag is not properly reconstructed the non-overlap requirements are not guaranteed to hold.  Requiring that the event-tag and test jet are back-to-back in the transverse plane greatly reduces the probability that a particle originating from the event-tag decay but not reconstructed in the event-tag is reconstructed as part of the test jet.}.
Therefore, all events used to measure the $(b,c)$-tagging efficiency have passed the trigger independently of the presence of the test jet, which ensures that the trigger does not bias the efficiency measurement.   The following event-tags are used (labeled by the data-set identifier):
\begin{itemize}
\item ($B+$jet) a fully reconstructed $b$-hadron decay which enriches the $b$-jet content of the test-jet sample;
\item ($D+$jet) a fully reconstructed $c$-hadron decay which enriches the $c$-jet and $b$-jet content of the test-jet sample (due to $b\to c$ decays);
\item ($\mu(b,c)+$jet) a displaced high-\pt muon which enriches the $c$-jet and $b$-jet content of the test-jet sample;
\item ($W\!+$jet) a prompt isolated high-\pt muon indicative of $W\!+$jet events that consists of about 95\% \light jets.
\end{itemize}
The first three samples are used to measure the $(b,c)$-jet identification efficiencies and properties.  The final sample is used to study misidentification of \light jets.  In all samples the event-tag and test jet are required to originate from the same PV.  
The range $10 < \pt({\rm jet}) < 100\gev$ is considered since there are no large enough data samples to measure the efficiency for jet $\pt > 100\gev$.

\subsection{Tagged-jet yields}

The presence of an SV and its kinematic properties are used to discriminate between $b$, $c$ and \light jets.  As described in Section~3, the SV-tagger algorithm uses two BDTs while the TOPO uses one BDT for each SV.  
The tagged yields for each algorithm are obtained by fitting to data BDT templates obtained from simulation for $b$, $c$ and \light jets.  In all fits the template shapes are fixed and only the yields of each jet type are free to vary.

\begin{figure}[] 
  \centering 
  \includegraphics[width=0.45\textwidth]{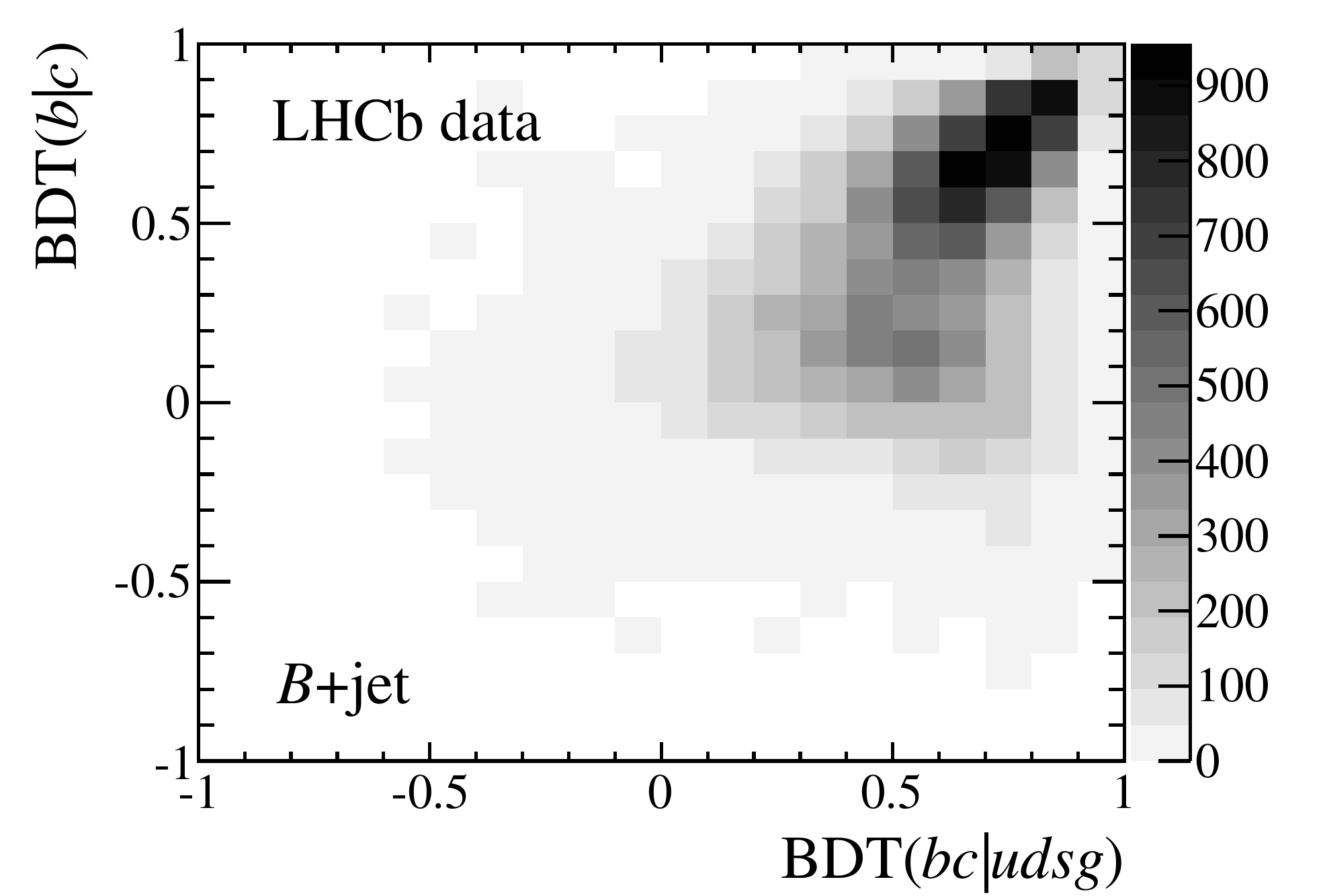}
  \includegraphics[width=0.45\textwidth]{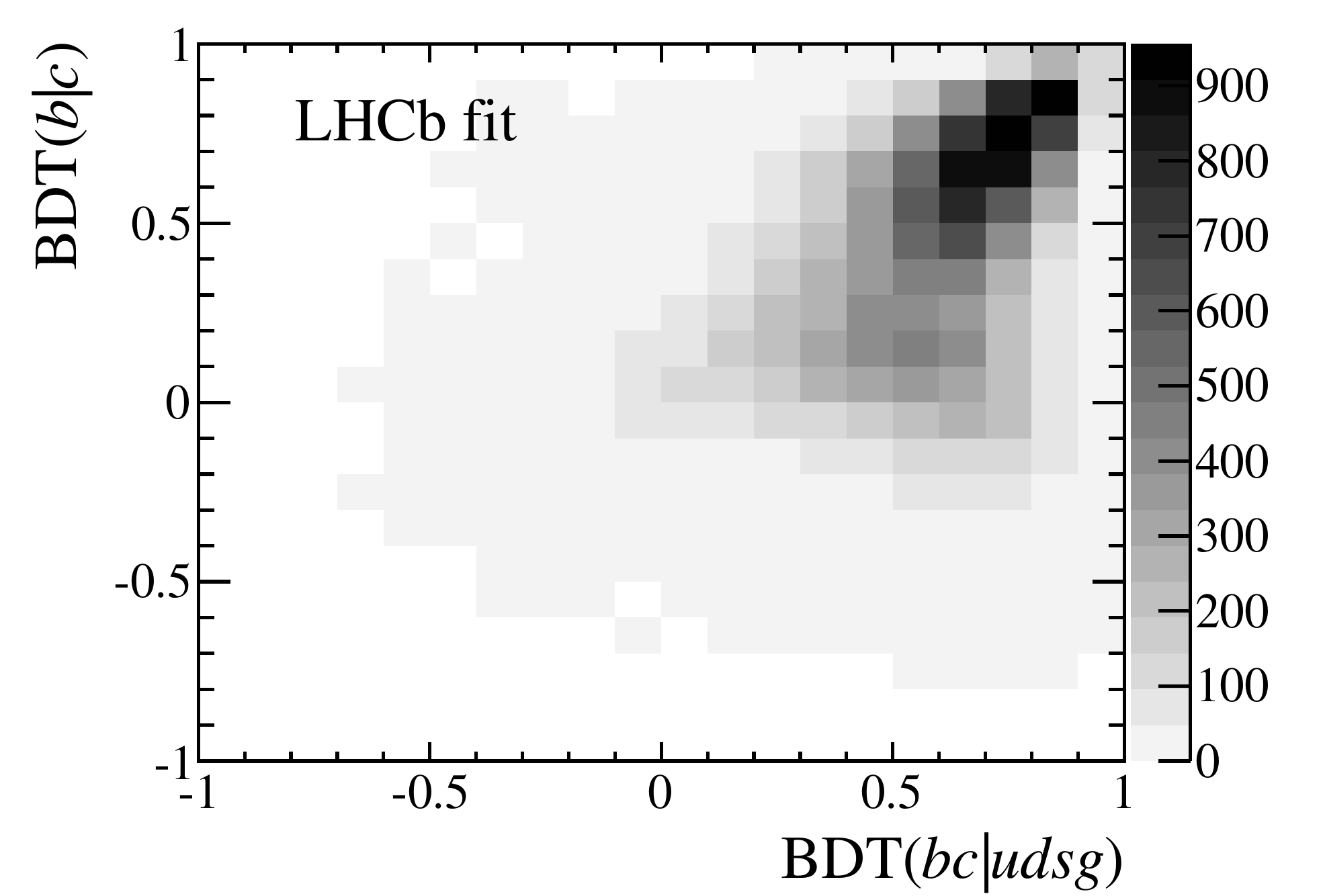}\\
  \includegraphics[width=0.45\textwidth]{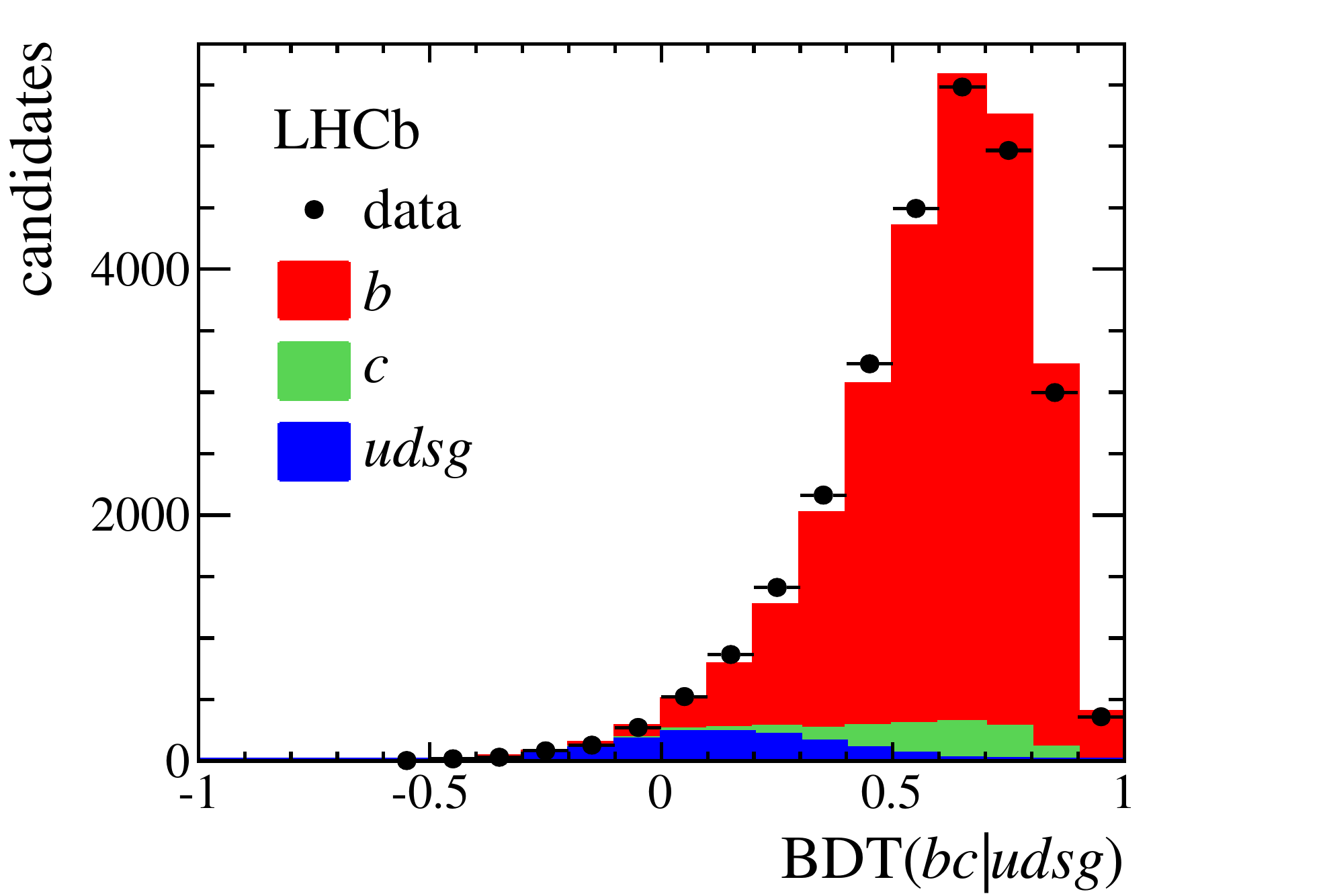}
  \includegraphics[width=0.45\textwidth]{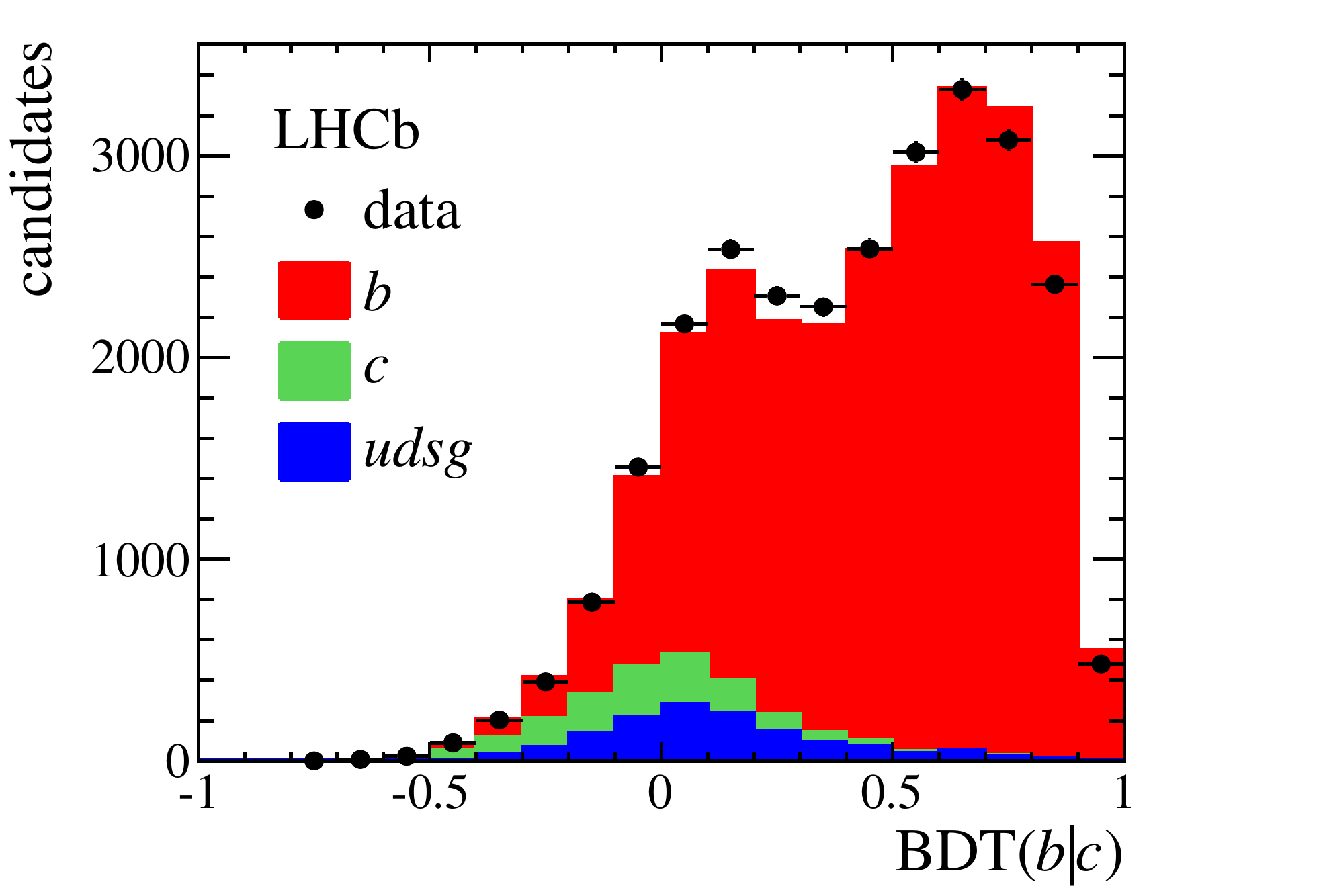}
  \caption{\label{fig:bdtfit_bhad} SV-tagger BDT fit results for the $B+$jet data sample with $10 < \pt({\rm jet}) < 100\gev$: (top left) distribution in data; (top right) two-dimensional template-fit result; and (bottom) projections of the fit result with the $b$, $c$, and \light contributions shown as stacked histograms. 
}
\end{figure}

\begin{figure}[] 
  \centering 
  \includegraphics[width=0.45\textwidth]{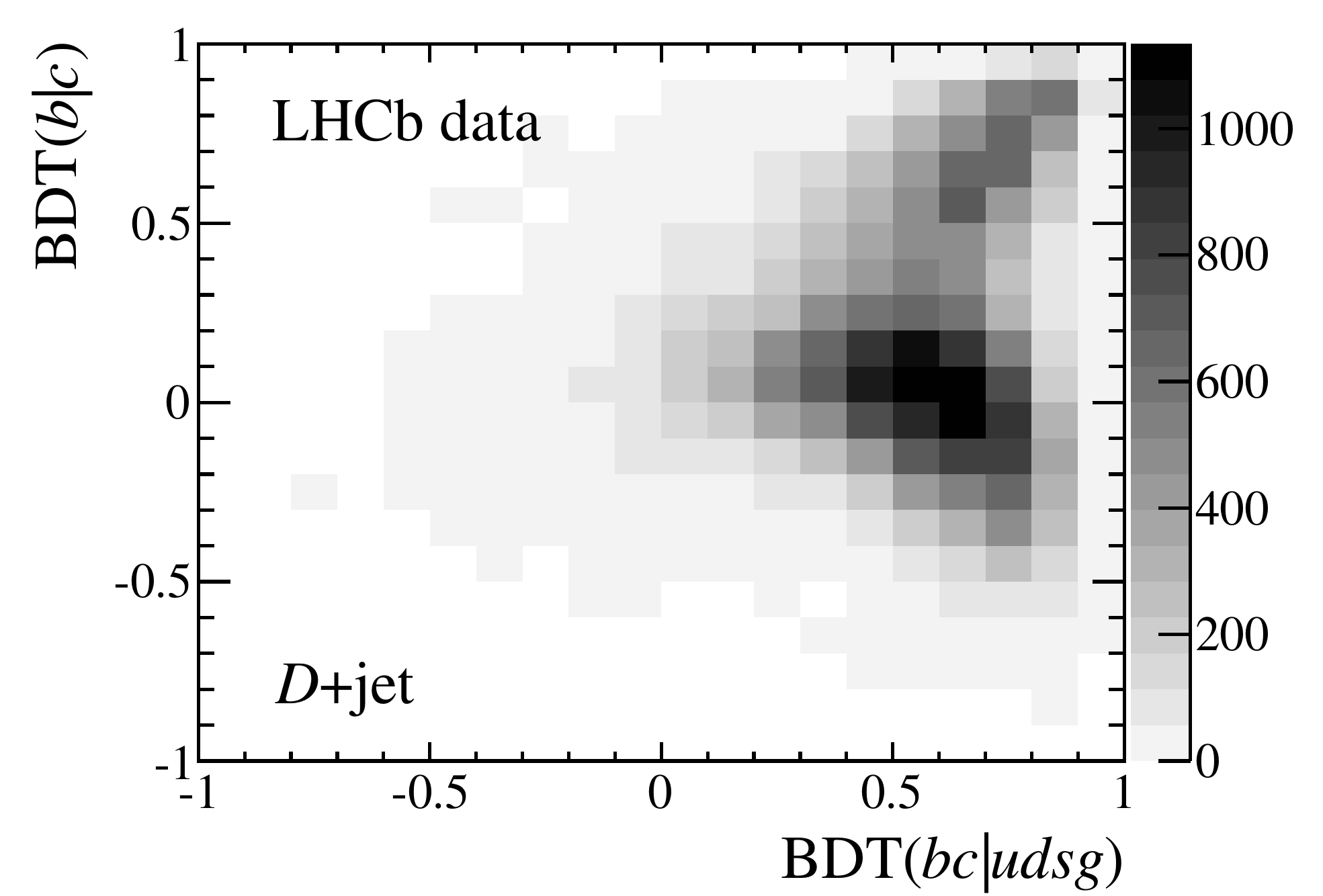}
  \includegraphics[width=0.45\textwidth]{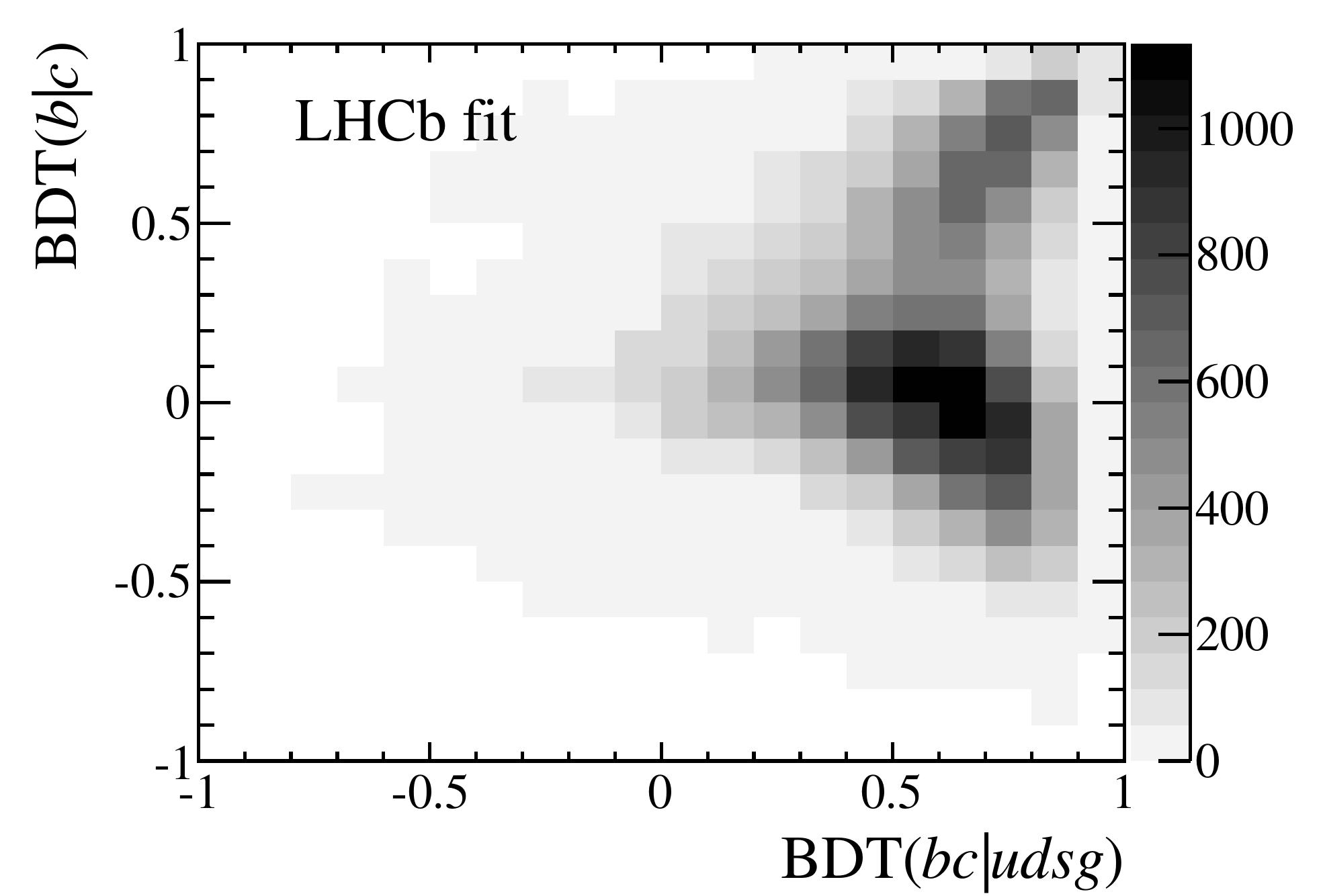}\\
  \includegraphics[width=0.45\textwidth]{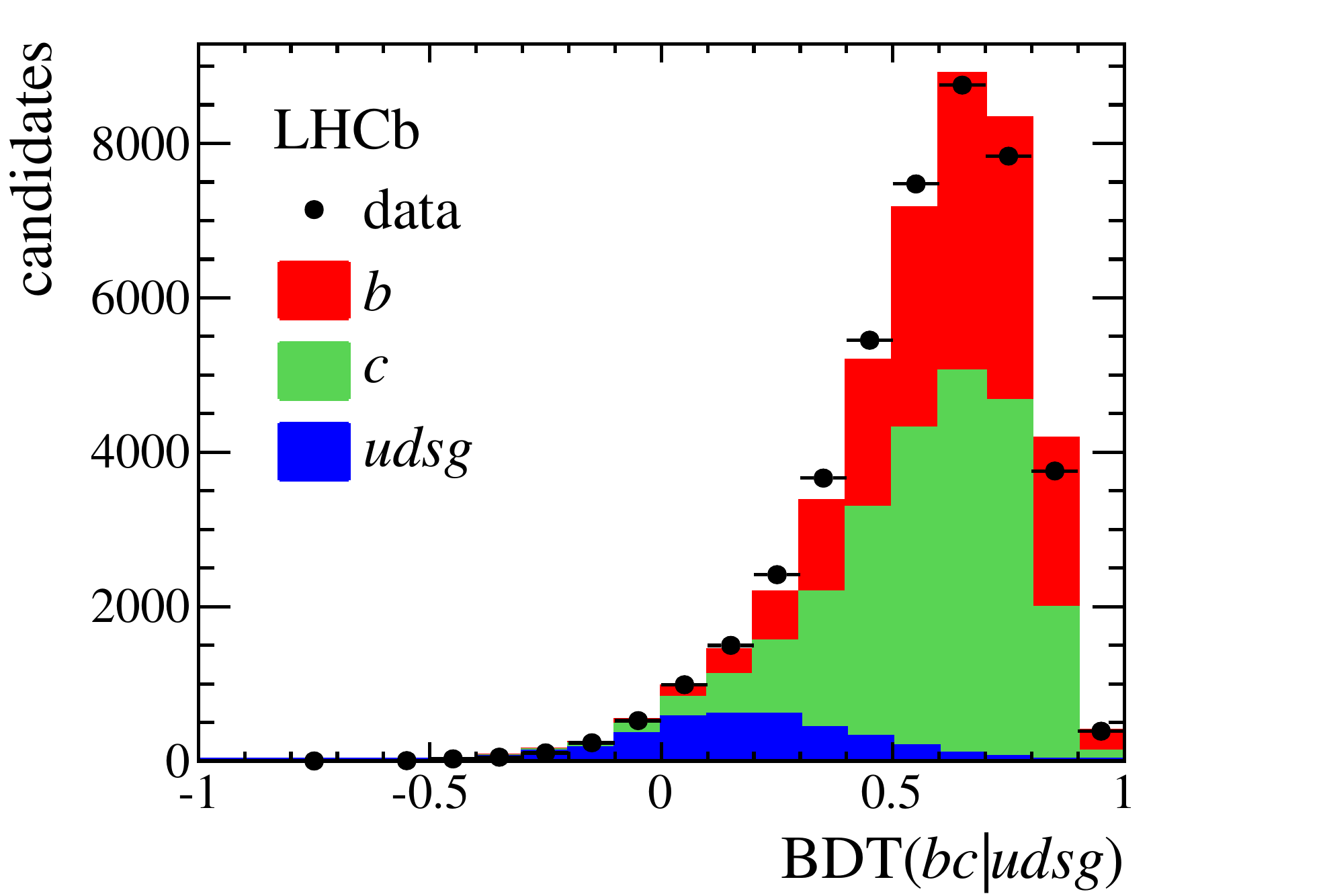}
  \includegraphics[width=0.45\textwidth]{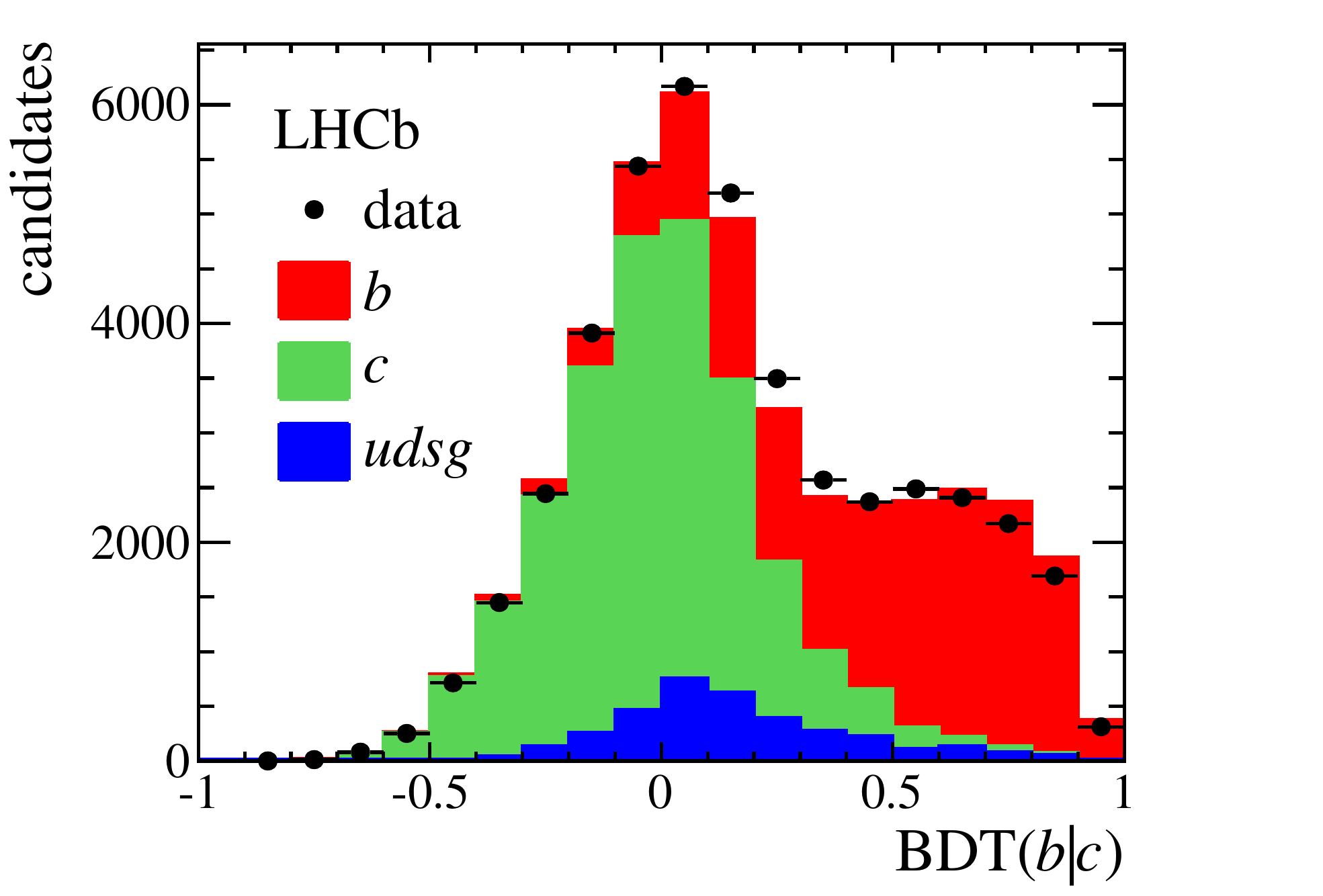}
  \caption{\label{fig:bdtfit_chad} Same as Fig.~\ref{fig:bdtfit_bhad} for the $D+$jet data sample.
}
\end{figure}

\begin{figure}[] 
  \centering 
  \includegraphics[width=0.45\textwidth]{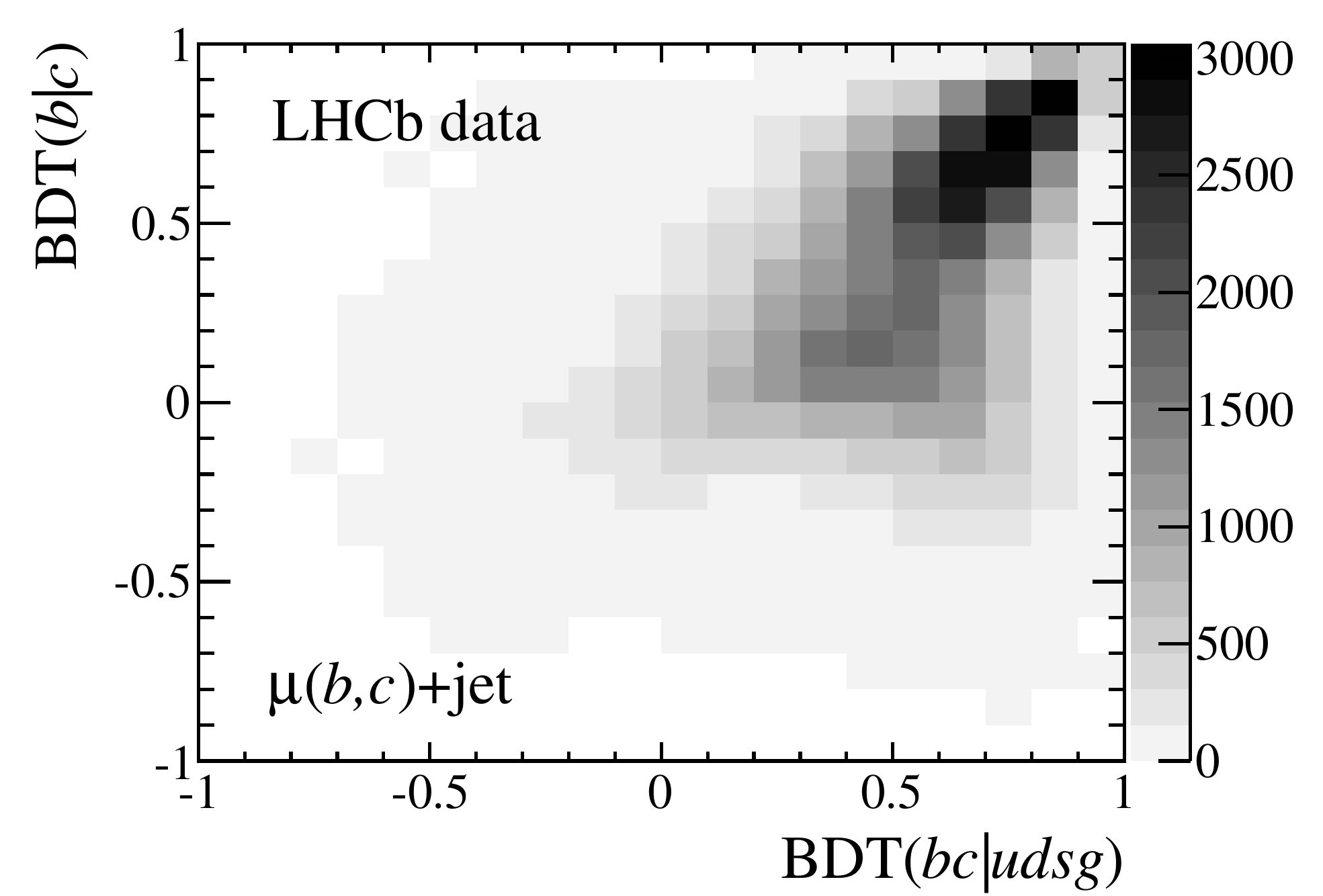}
  \includegraphics[width=0.45\textwidth]{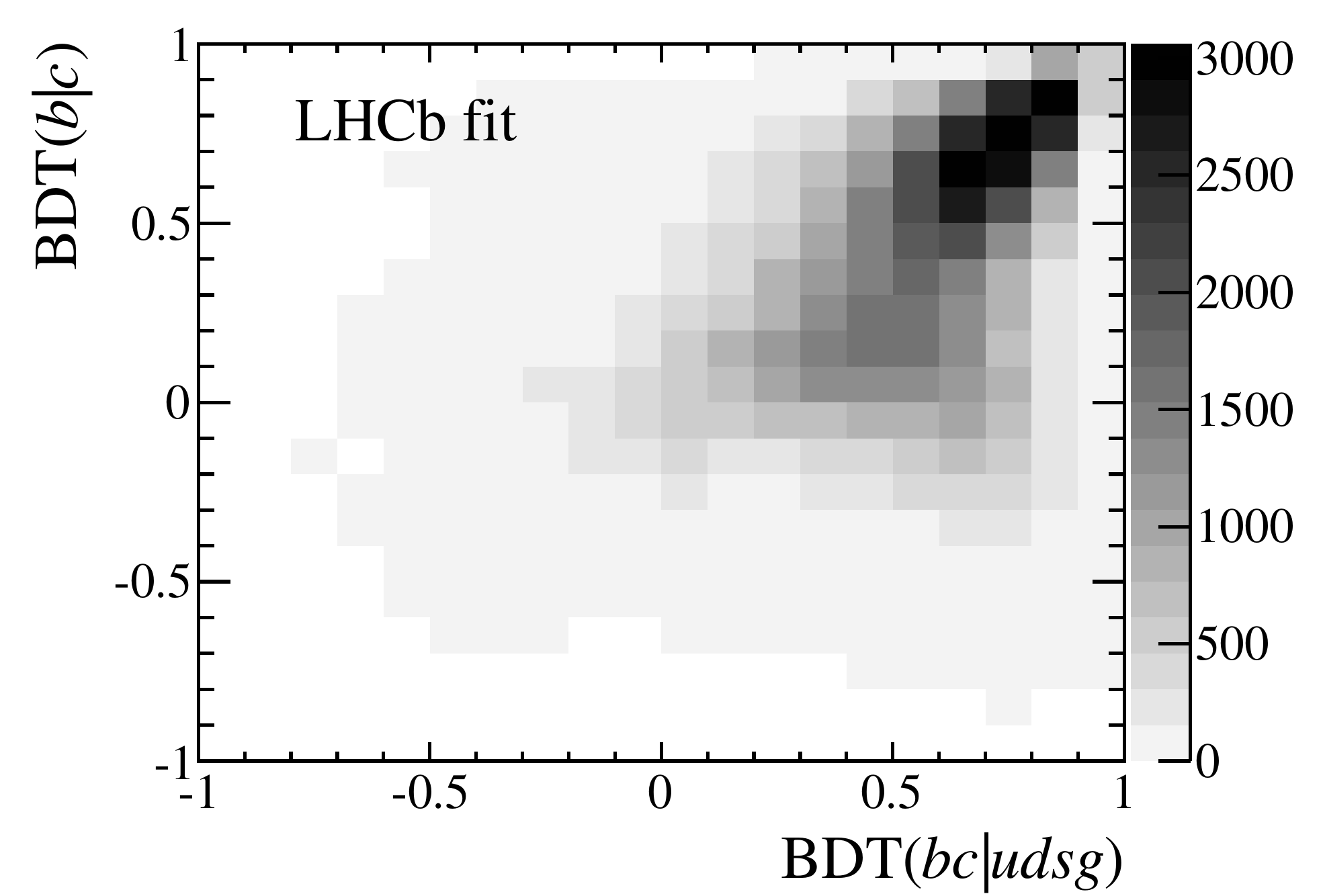}\\
  \includegraphics[width=0.45\textwidth]{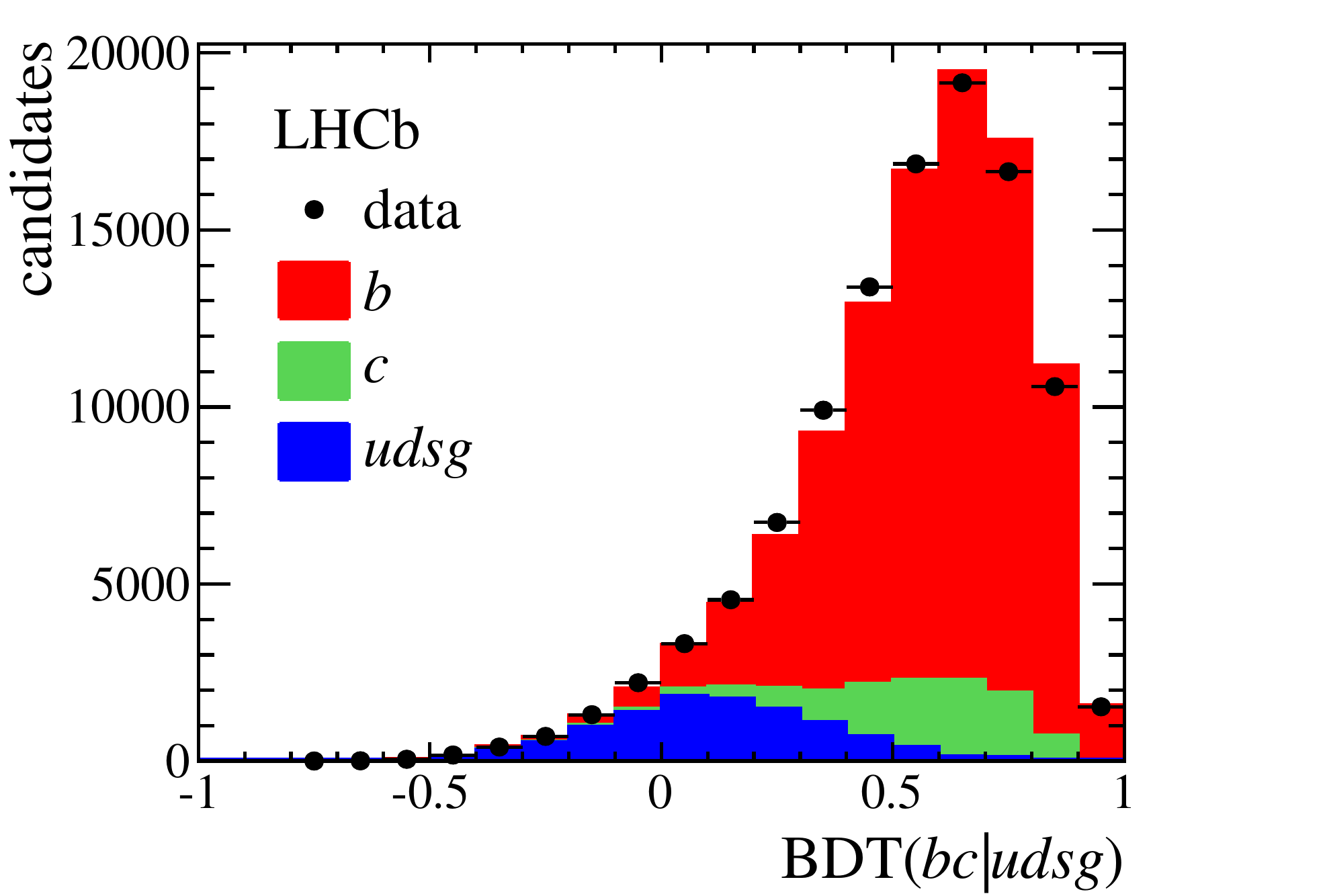}
  \includegraphics[width=0.45\textwidth]{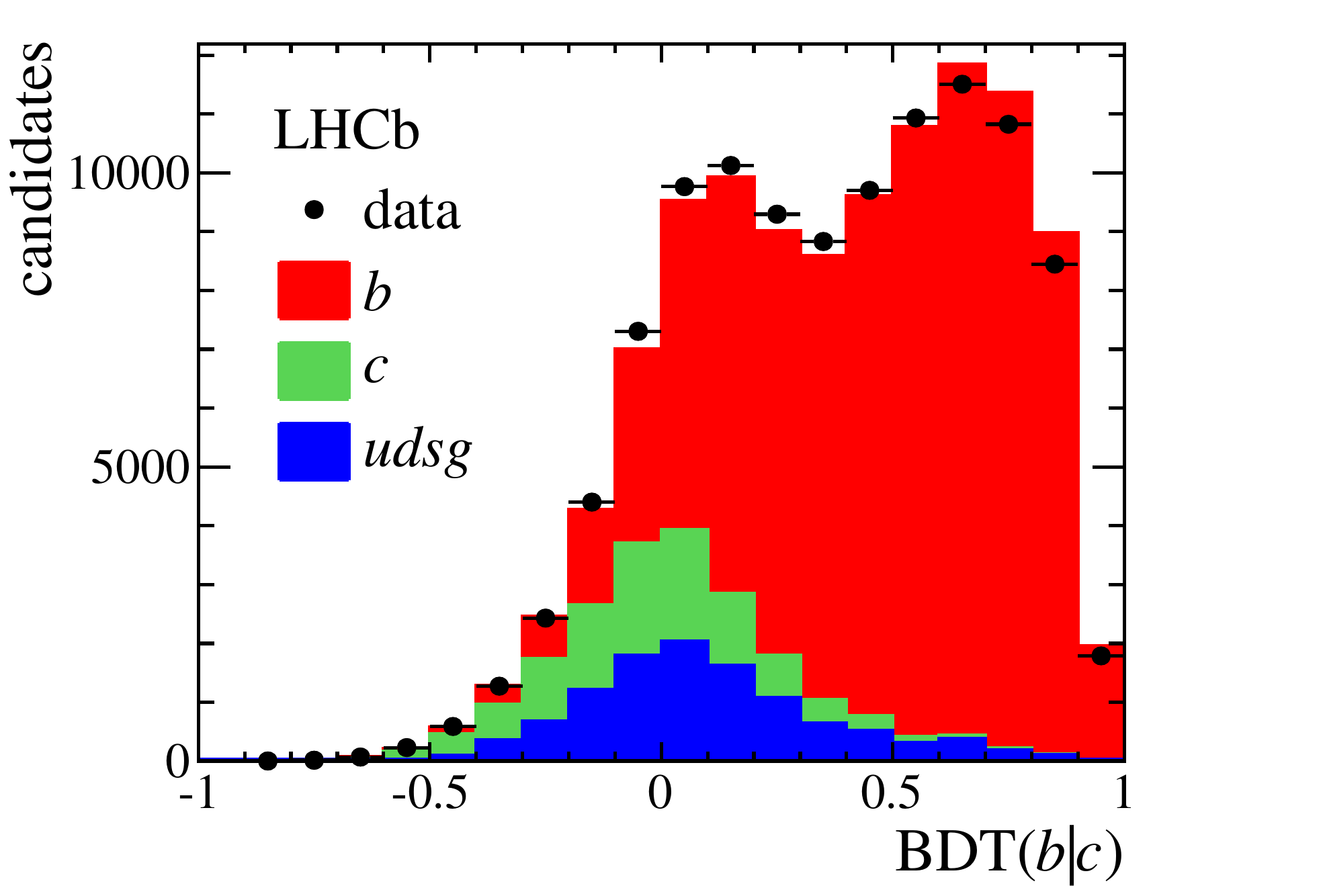}
  \caption{\label{fig:bdtfit_bmu} Same as Fig.~\ref{fig:bdtfit_bhad} for the $\mu(b,c)+$jet data sample.
}
\end{figure}

Figures~\ref{fig:bdtfit_bhad}--\ref{fig:bdtfit_bmu} show the results of fits performed to the two-dimensional SV-tagger BDT distributions in the $B+$jet, $D+$jet and $\mu(b,c)+$jet data samples.  
The $b$ and $c$ jets are clearly distinguishable in the two-dimensional BDT distributions:  $b$ jets are mostly found in the upper right corner, while $c$ jets are found in the center-right and lower-right regions.  The \light jets cluster near the origin but are difficult to see due to the low SV-tag probability of \light jets.
The BDT templates for $b$, $c$ and \light jets describe the data well.  
A dedicated study of the modeling of the \light-jet BDT distributions is discussed in Sec.~5.

\begin{figure}[] 
  \centering 
  \includegraphics[width=0.45\textwidth]{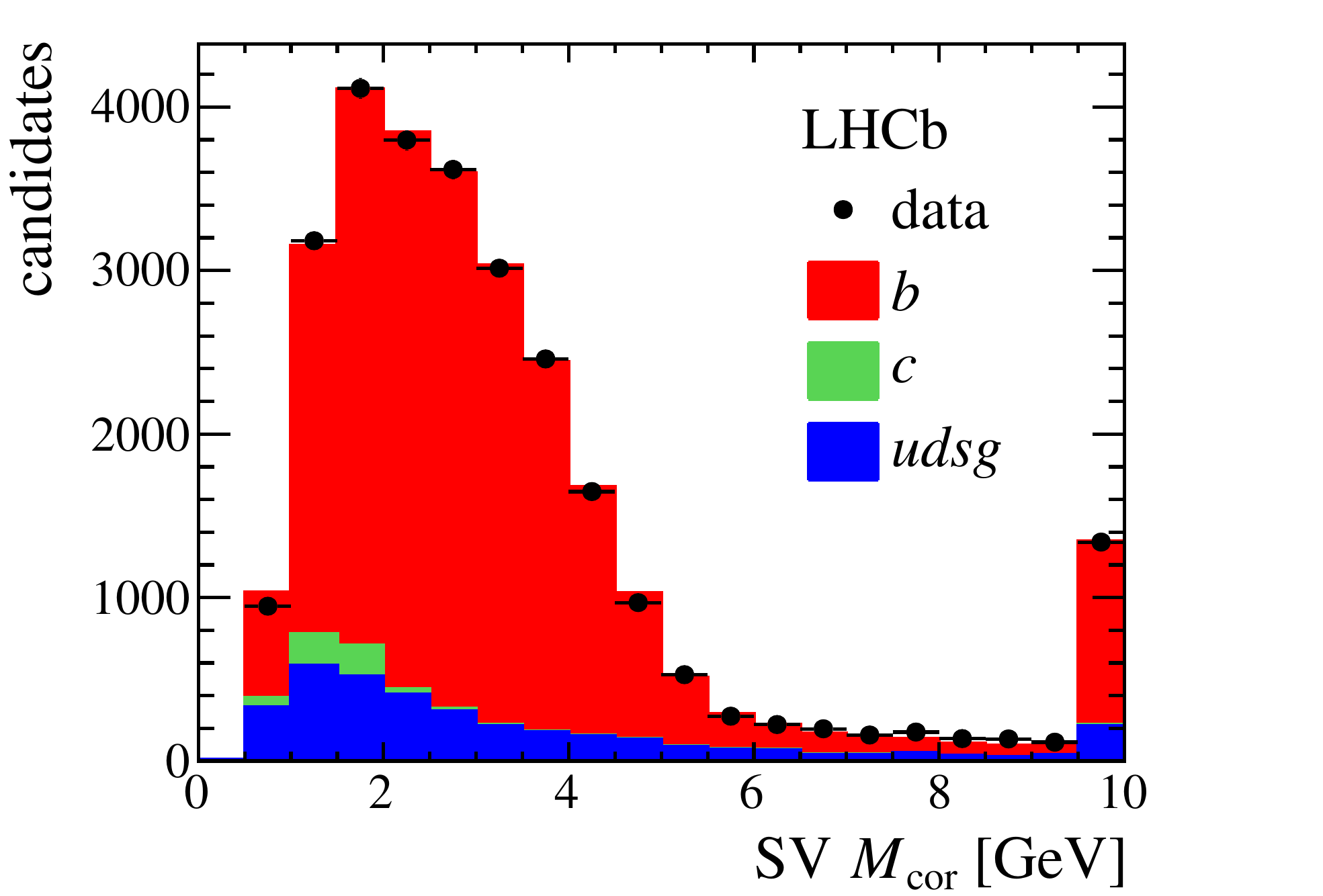}
  \includegraphics[width=0.45\textwidth]{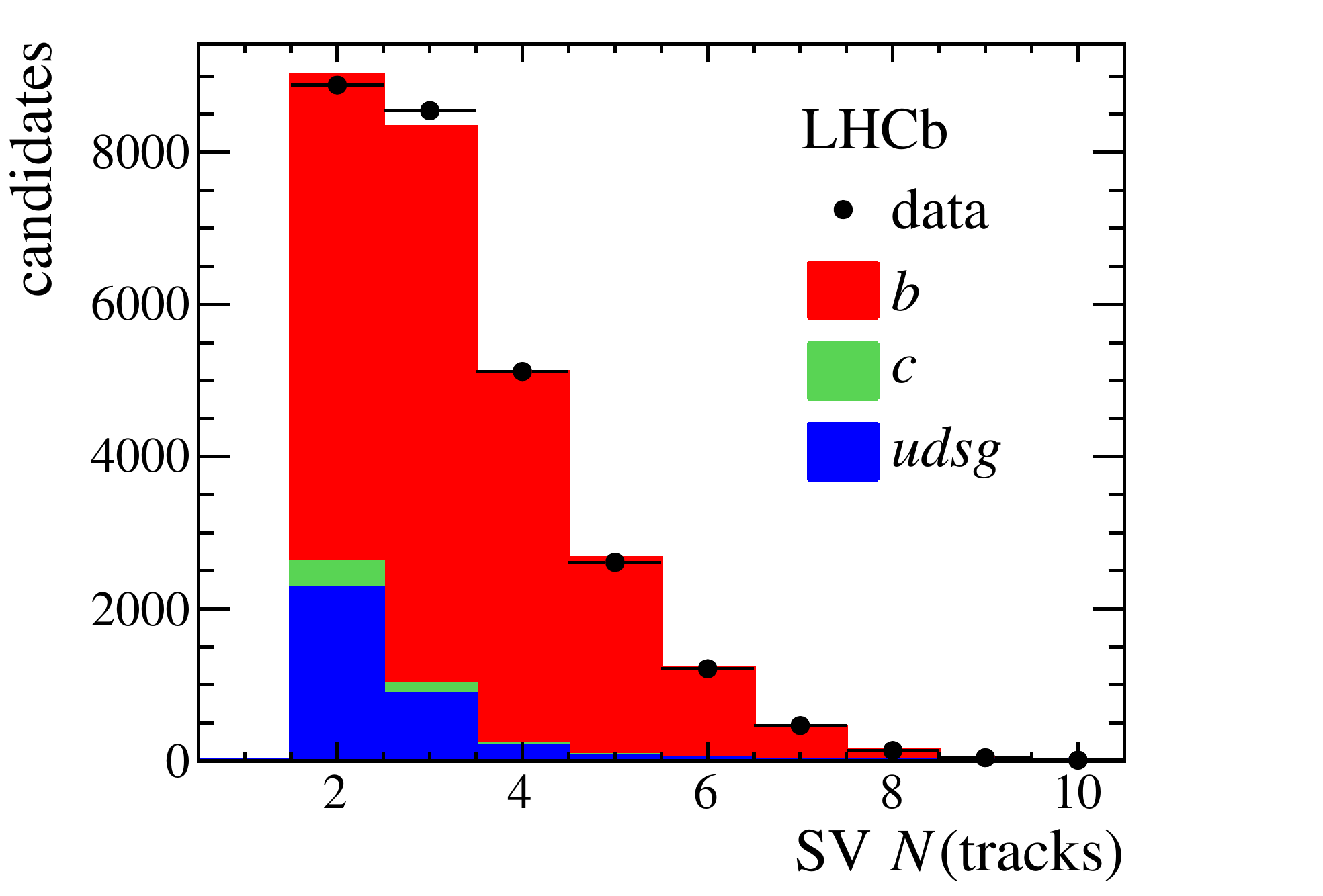}\\
  \includegraphics[width=0.45\textwidth]{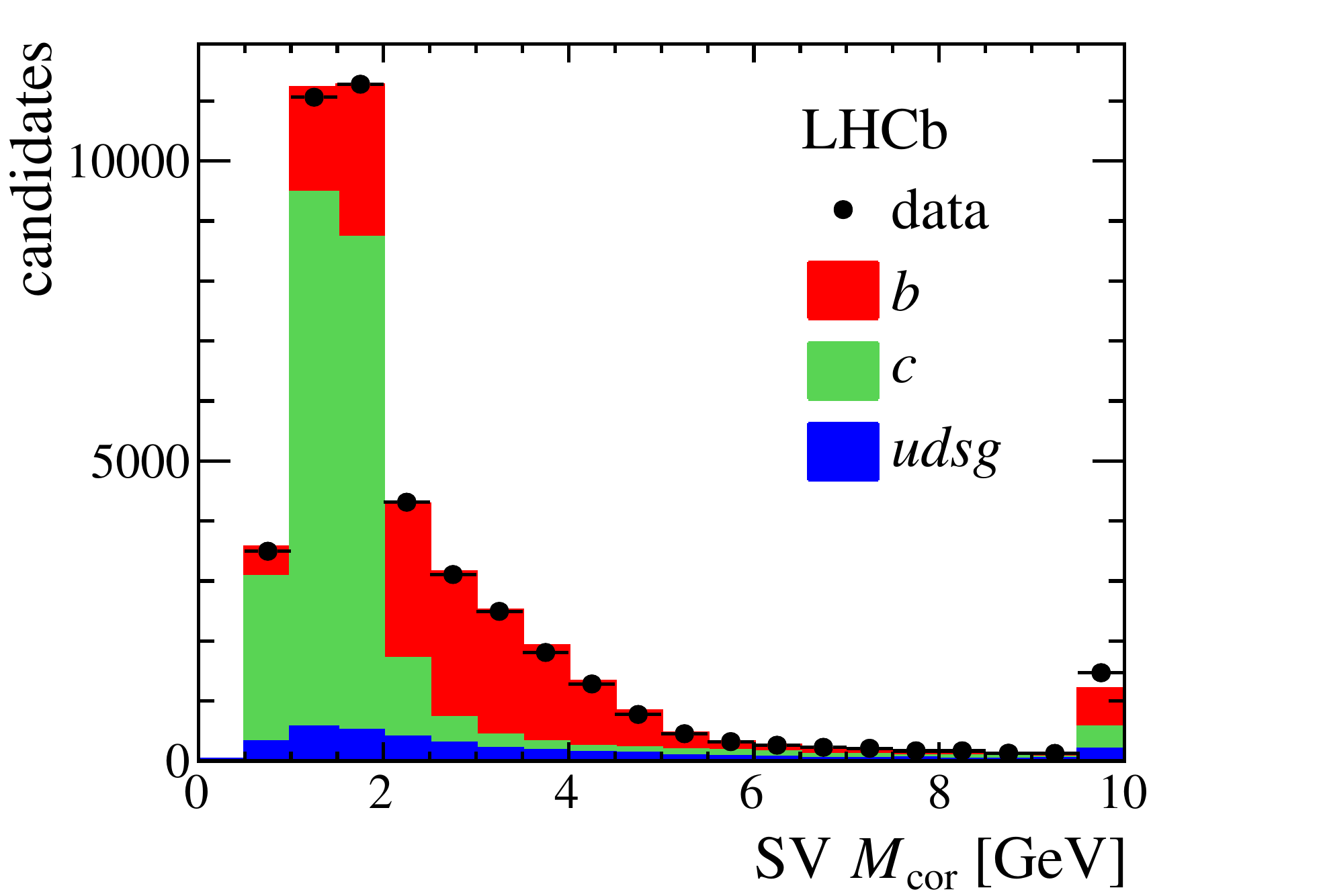}
  \includegraphics[width=0.45\textwidth]{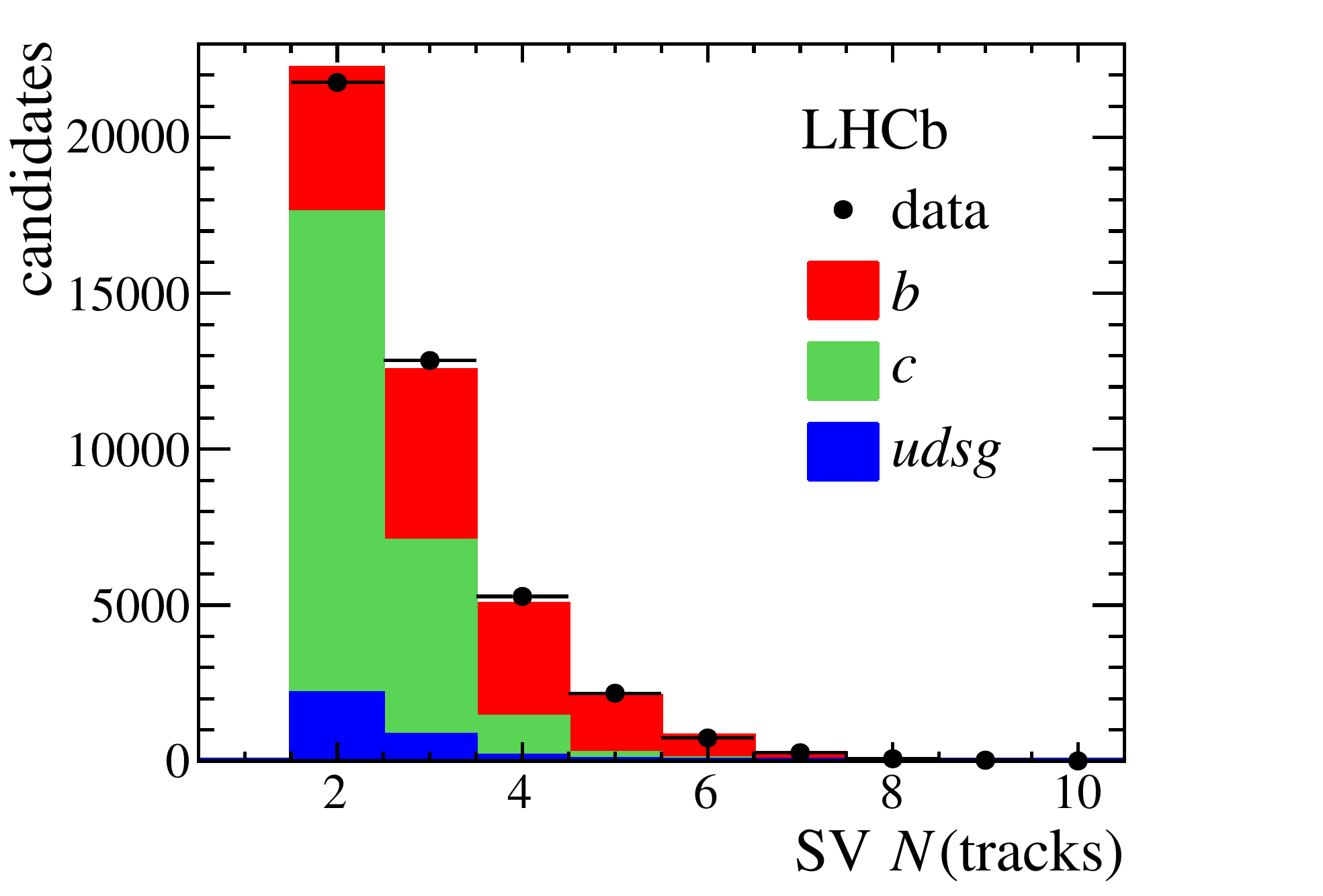}\\
  \includegraphics[width=0.45\textwidth]{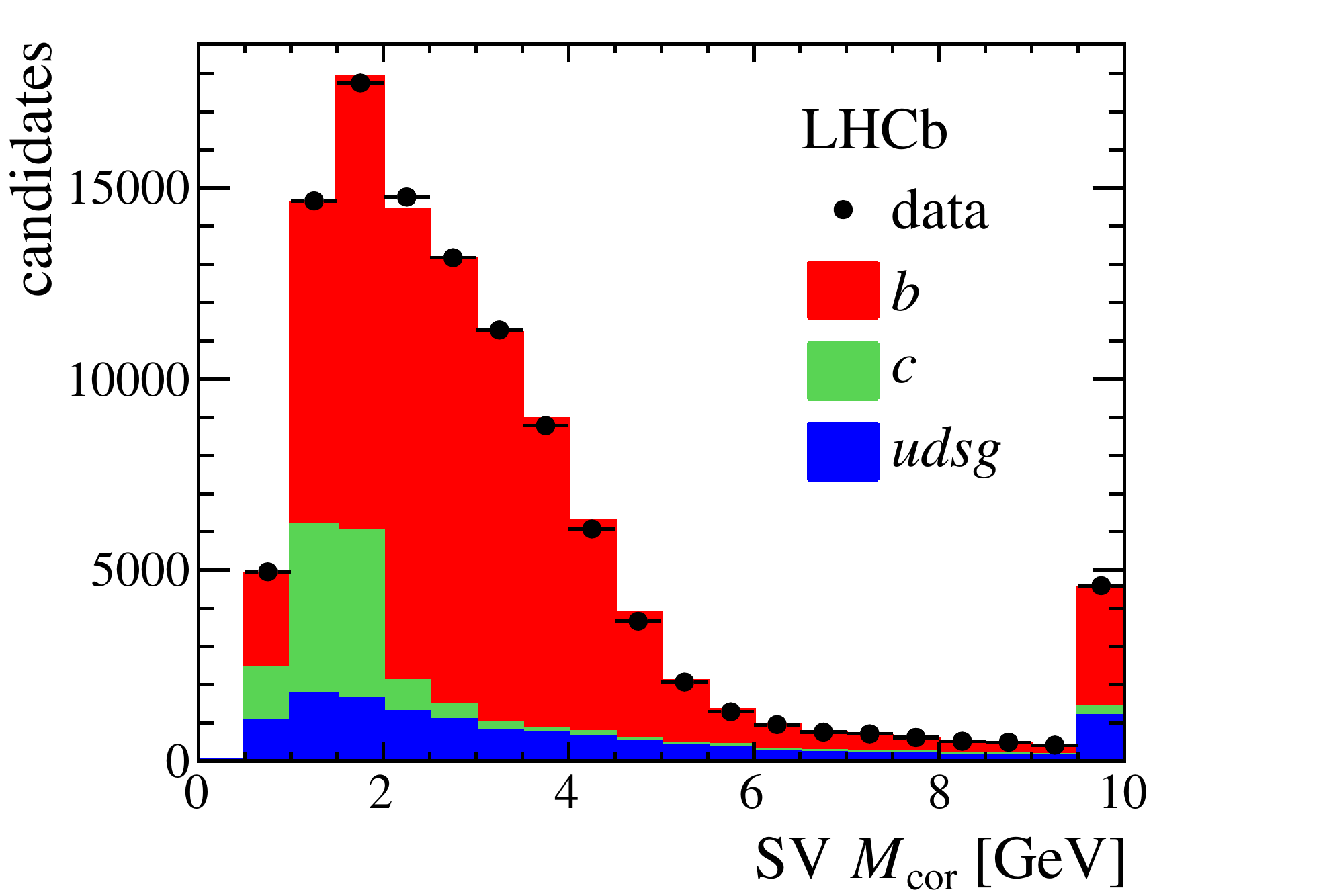}
  \includegraphics[width=0.45\textwidth]{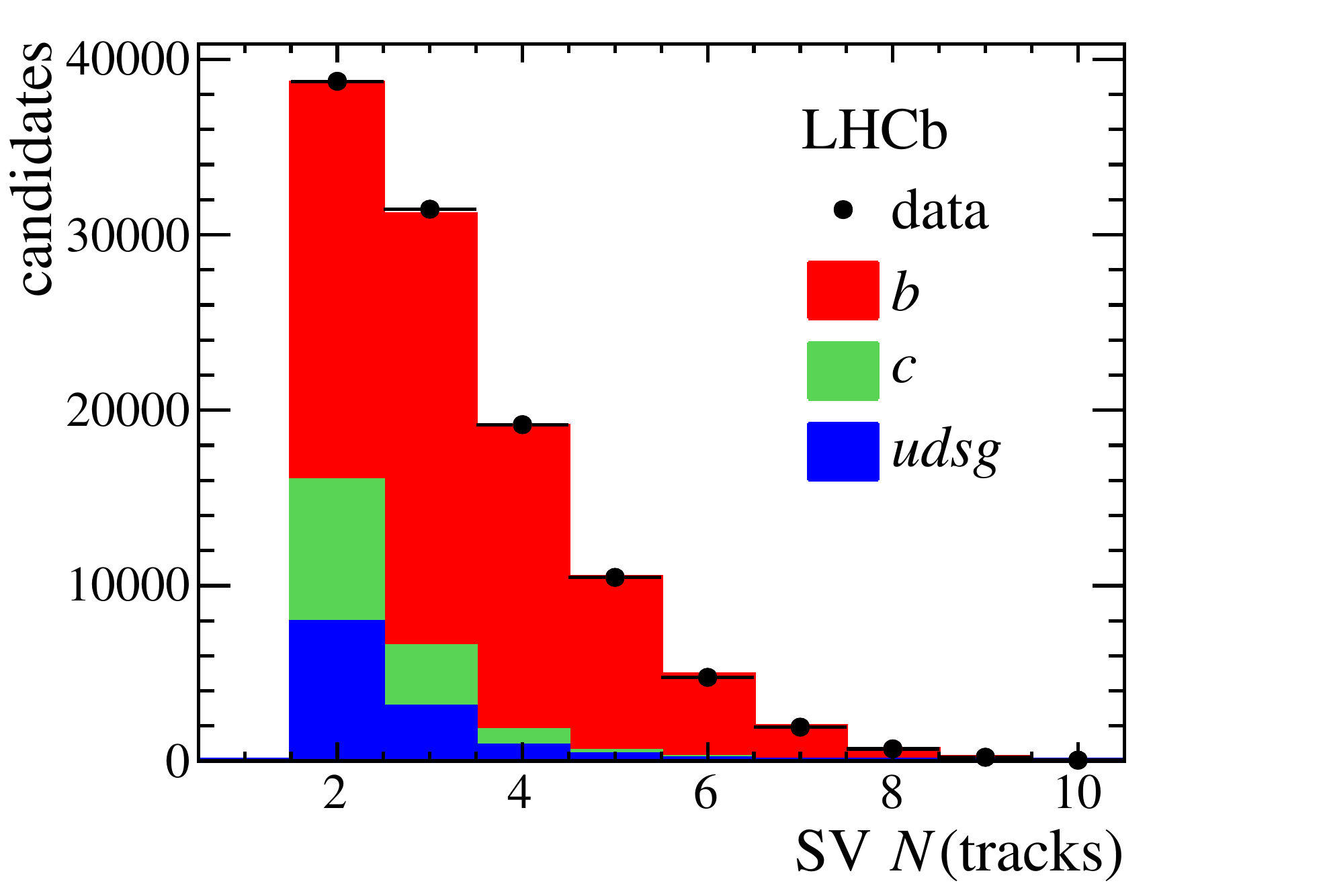}
  \caption{\label{fig:mcvntrk} Two-dimensional $M_{\rm cor}$ versus SV track multiplicity fit results for (top) $B+$jet, (middle) $D+$jet and (bottom) $\mu(b,c)+$jet data samples.  The left plots show the projection onto the $M_{\rm cor}$ axis, while the right plots show the projection onto the track multiplicity. The highest $M_{\rm cor}$ bin includes candidates with $M_{\rm cor} > 10\gev$.}
\end{figure}

A simple cross-check on the $b$, $c$ and \light yields is performed by fitting only two of the BDT inputs: the corrected mass defined in Eq.~\ref{eq:mcor} and the number of tracks in the SV.  The corrected mass provides the best discrimination between $c$ jets and other jet types due to the fact that $M_{\rm cor}$ peaks near the $D$ meson mass for $c$ jets\footnote{This is true for all long-lived $c$ hadrons when all tracks are assigned a pion mass.}.  The number of tracks in the SV identifies $b$ jets well since $b$-hadron decays often produce many displaced tracks. 
Figure~\ref{fig:mcvntrk} shows the results of a two-dimensional fit to these two SV properties.  The absolute fractions of $b$, $c$ and \light SV tags agree with the BDT fit results to within 1--2\%.  
The corrected mass has been previously used in LHCb jet analyses for determining the $c$-jet yield~\cite{LHCb-PAPER-2014-023} and for extracting the $b$-jet yield~\cite{LHCb-PAPER-2014-055}.  The clear peaking structure for $c$ jets, which relies on the excellent vertex resolution of the LHCb detector, makes them easily identifiable.

\begin{figure}[] 
  \centering 
  \includegraphics[width=0.45\textwidth]{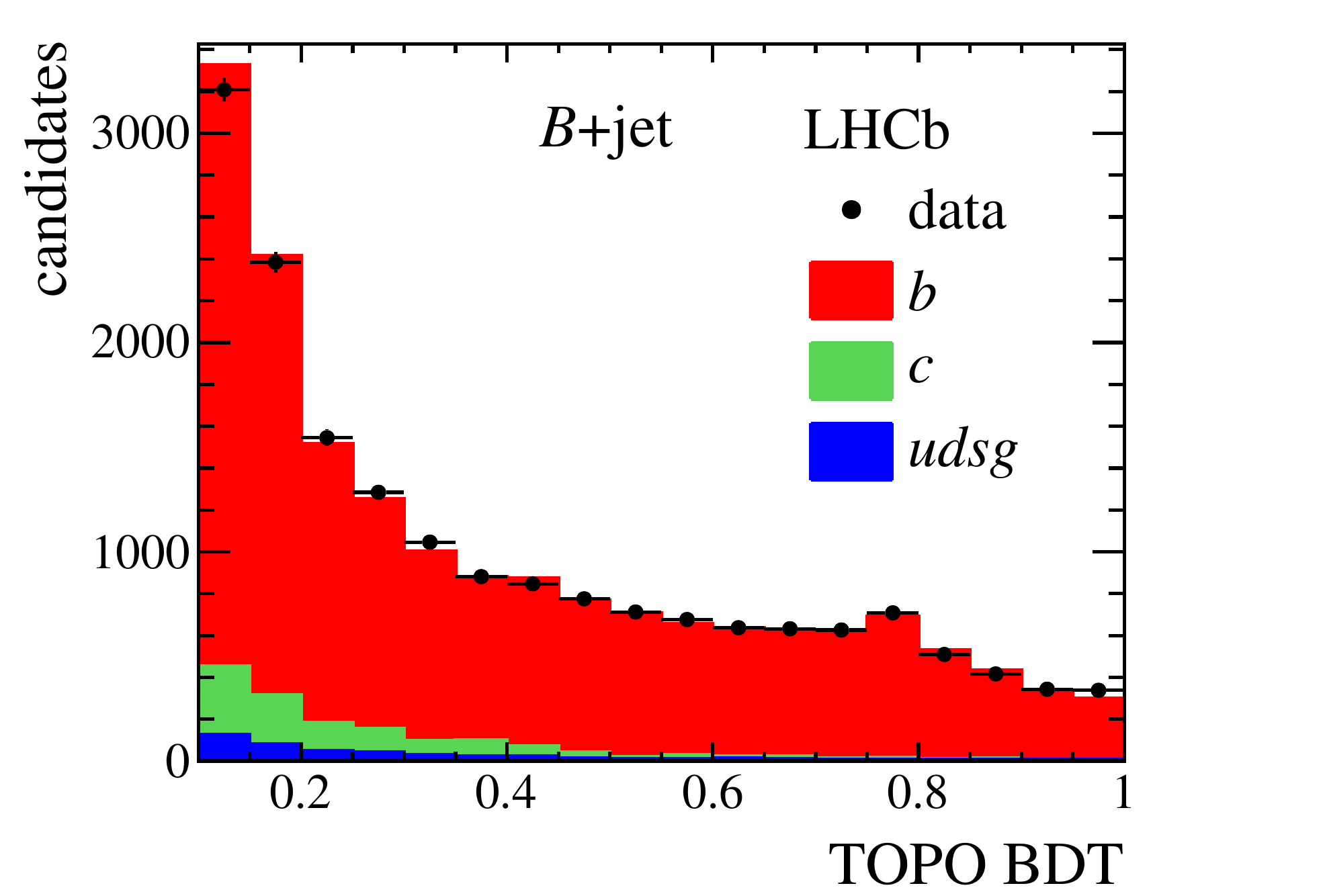}
  \includegraphics[width=0.45\textwidth]{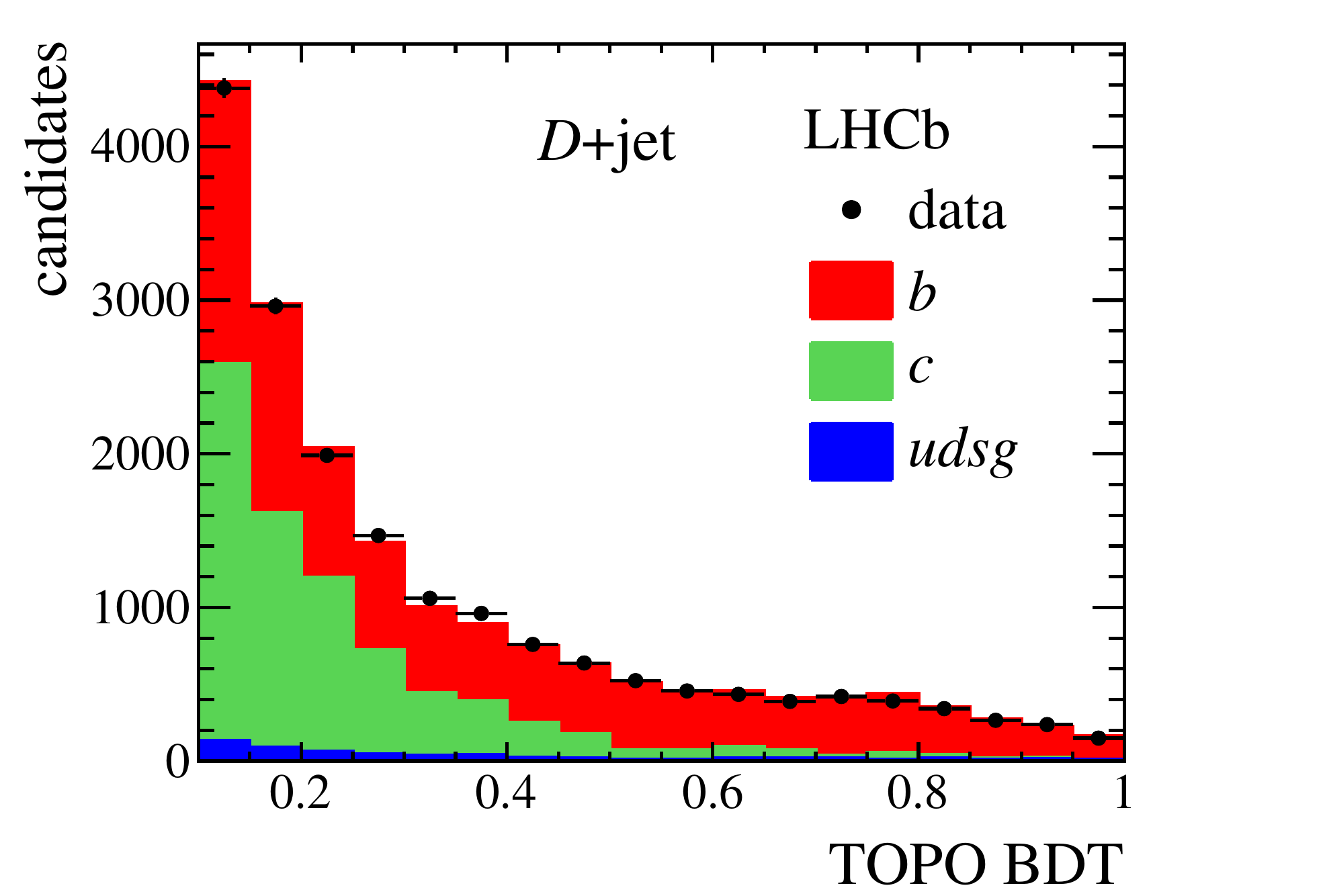}\\
  \includegraphics[width=0.45\textwidth]{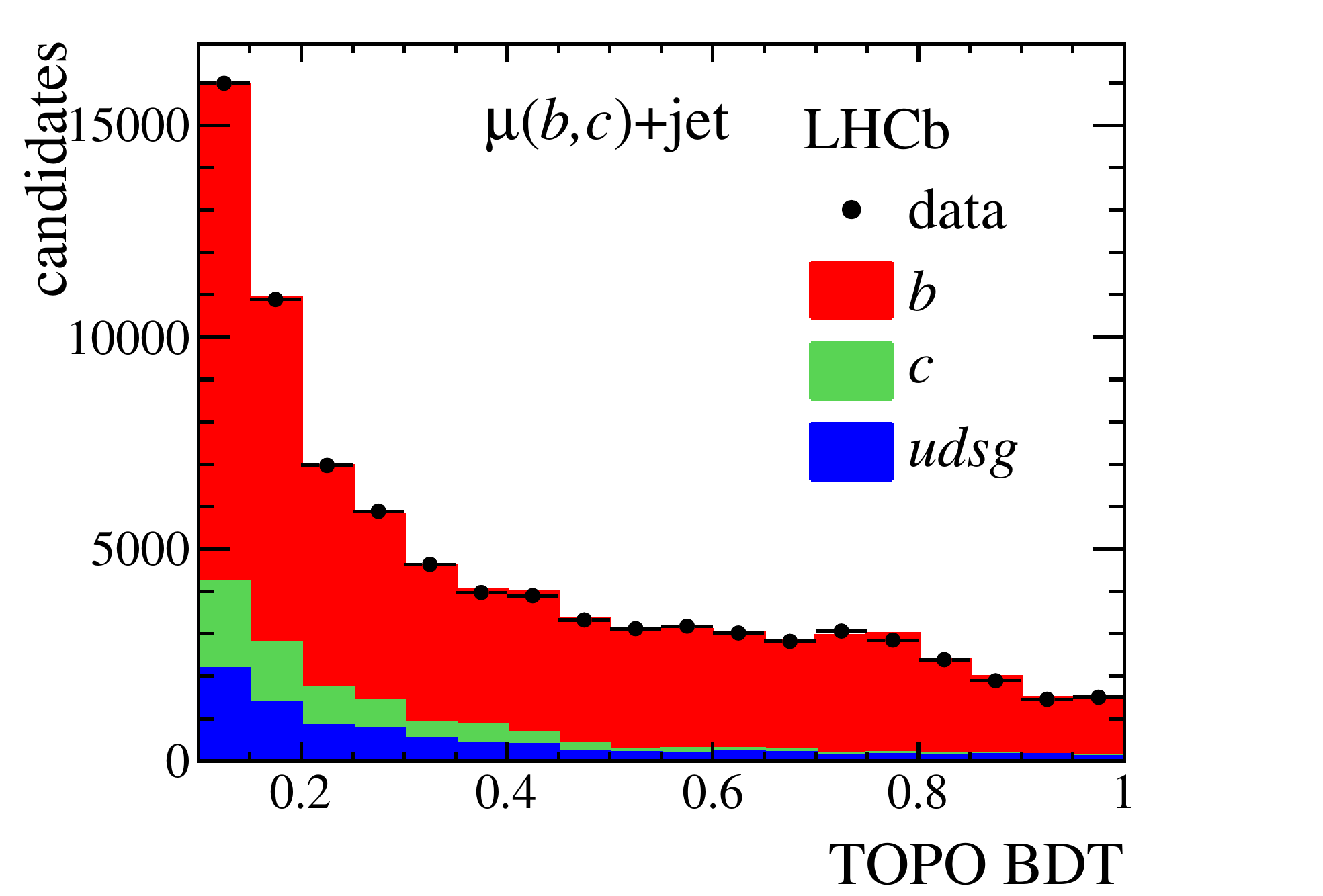}
  \caption{\label{fig:topobdtfits} 
    Fits to the TOPO BDT distribution in (left) $B+$jet, (middle) $D+$jet and (right) $\mu(b,c)+$jet data samples with $10 < \pt({\rm jet}) < 100\gev$.  
  }
\end{figure}

Figure~\ref{fig:topobdtfits} shows the results of fitting the TOPO BDT distributions in the various data samples using $b$, $c$ and \light jet template shapes obtained from simulation.  
The ratios of SV-tagger to TOPO SV-tagged $b$, $c$ and \light jets are each consistent with expectations from simulation.  
Modeling of both the SV-tagger and TOPO SV properties are sufficient to allow the SV-tagged content to be accurately determined. 

\subsection{Efficiency measurement using highest-{\boldmath $p_{\rm T}$} tracks}

To determine the jet-tagging efficiency, the jet composition prior to applying the SV tag must be determined.  This is necessarily more difficult than determining the SV-tagged composition.  The \xip distribution of the highest-\pt track in the jet is used for this task.  For \light jets the highest-\pt track will mostly originate from the PV, while for $(b,c)$ jets the highest-\pt track will often originate from the decay of the $(b,c)$ hadron.  To avoid possible issues with modeling of soft radiation, only the subset of jets for which the highest-\pt track satisfies $\pt($track$)/\pt($jet$) > 10\%$, which is about 95\% of all jets, is used.

\begin{figure}[] 
  \centering 
  \includegraphics[width=0.45\textwidth]{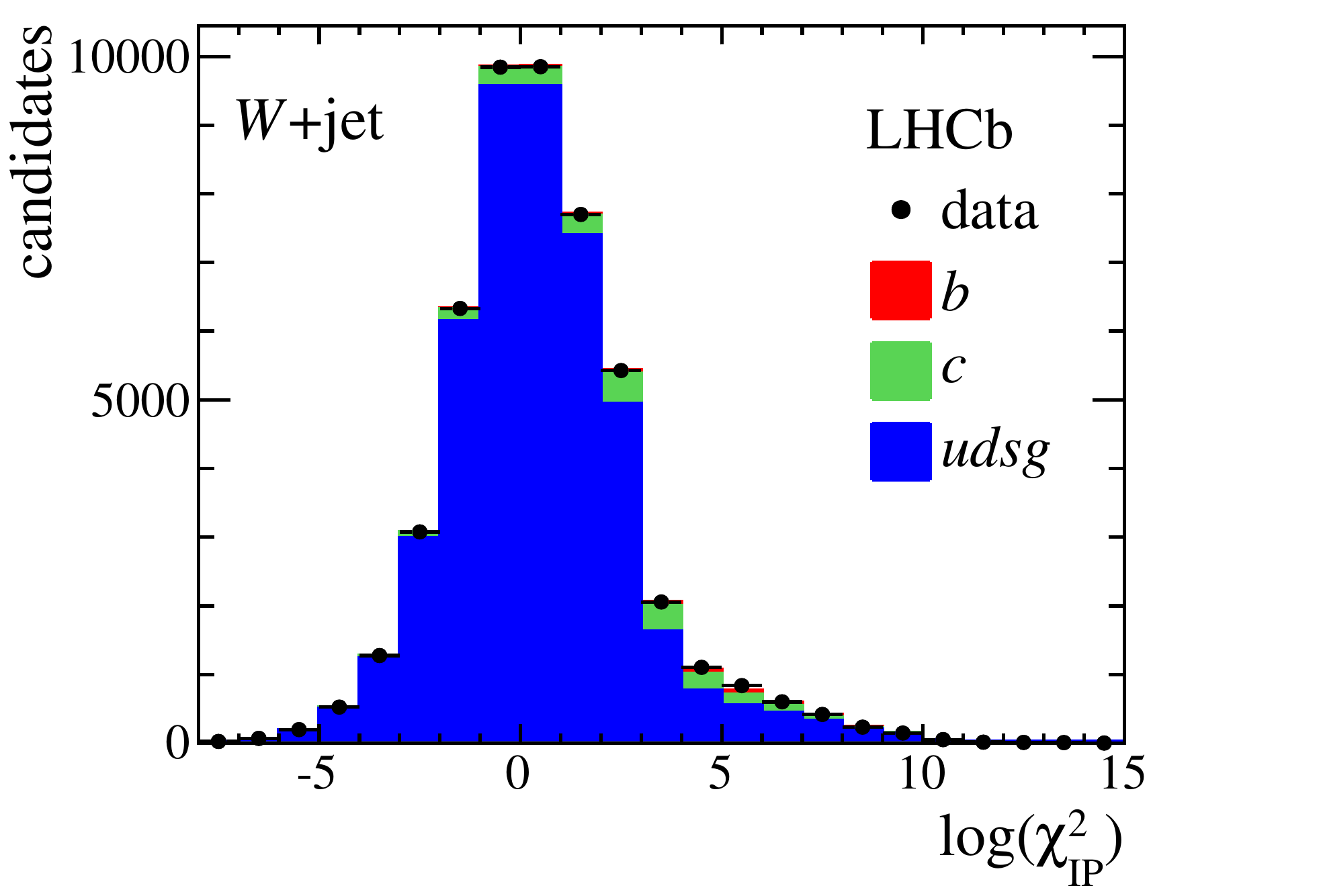}
  \caption{\label{fig:ipcalib} 
    Results of \xip calibration using $W\!+$jet data for $10 < \pt({\rm jet}) < 100\gev$. The tail out to large \xip values in the \light-jet sample is largely due to strange particle decays.
  }
\end{figure}

Since the $W\!+$jet sample is dominantly composed of \light jets, the \xip detector response can be obtained in a data-driven way using this data sample.
First, the two-dimensional SV-tagger BDT response is fitted to determine the SV-tagged $b$, $c$ and \light jet yields.  The tagging efficiencies obtained in simulation for $b$ and $c$ jets are used to estimate the total number of $b$ and $c$ jets in the $W\!+$jet data sample.  
Since the $b$ and $c$ jets combined make up only 5\% of the total data sample, any mismodeling of the SV-tagging efficiency will have negligible impact on this study.  
The IP resolution is studied by comparing the observed \xip distributions in data with templates obtained from simulation in bins of jet \pt.  
The resolution in data is found to be about 10\% worse than in the simulation which is consistent with previous LHCb studies of the IP resolution~\cite{LHCbVELOGroup:2014uea}. 
Figure~\ref{fig:ipcalib} shows that the data-driven templates describe the data well.  
The difference in the detector response between data and simulation is assumed to be universal and is applied to correct the \xip templates for $b$ and $c$ jets.

\begin{figure}[] 
  \centering 
 \includegraphics[width=0.45\textwidth]{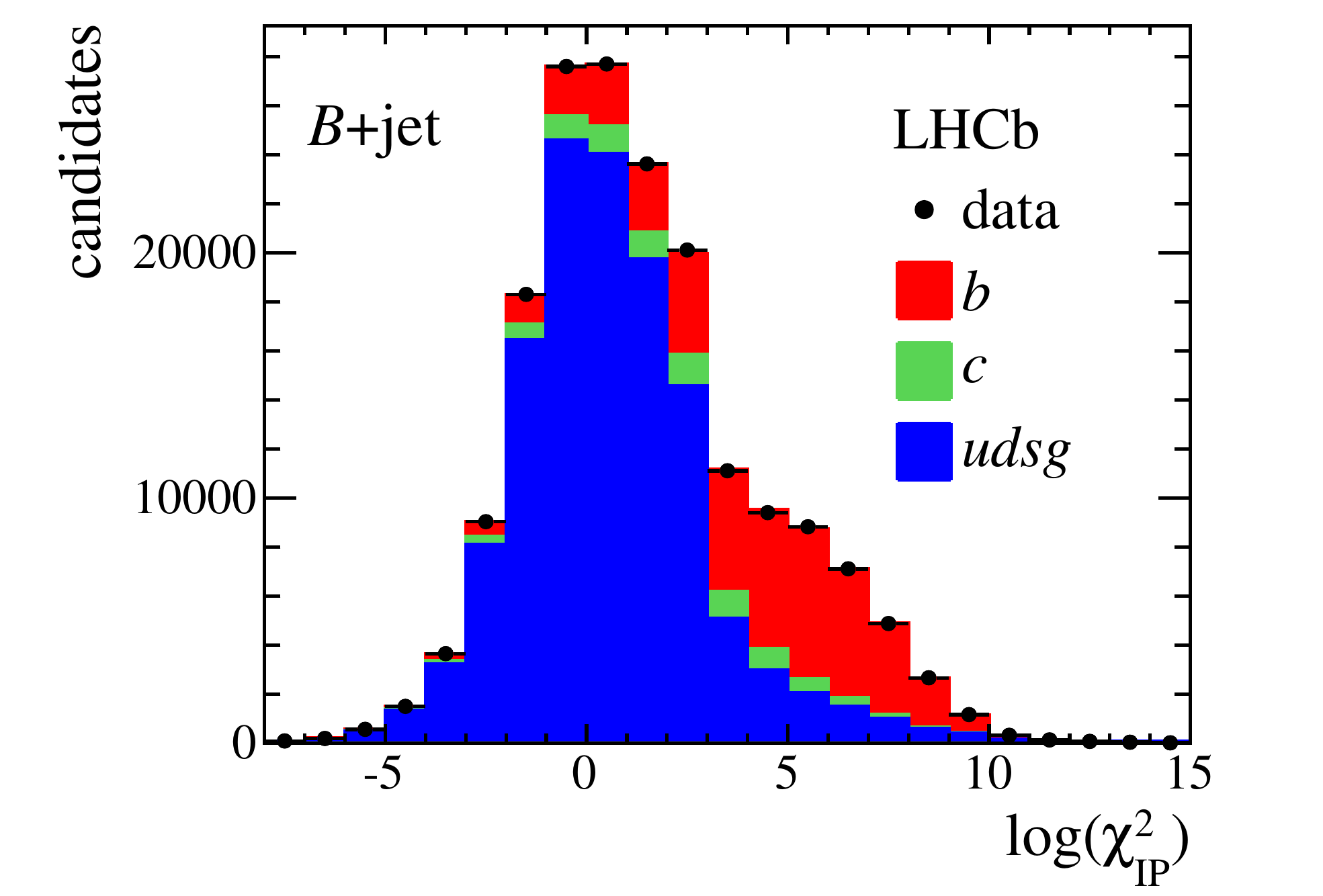}
 \includegraphics[width=0.45\textwidth]{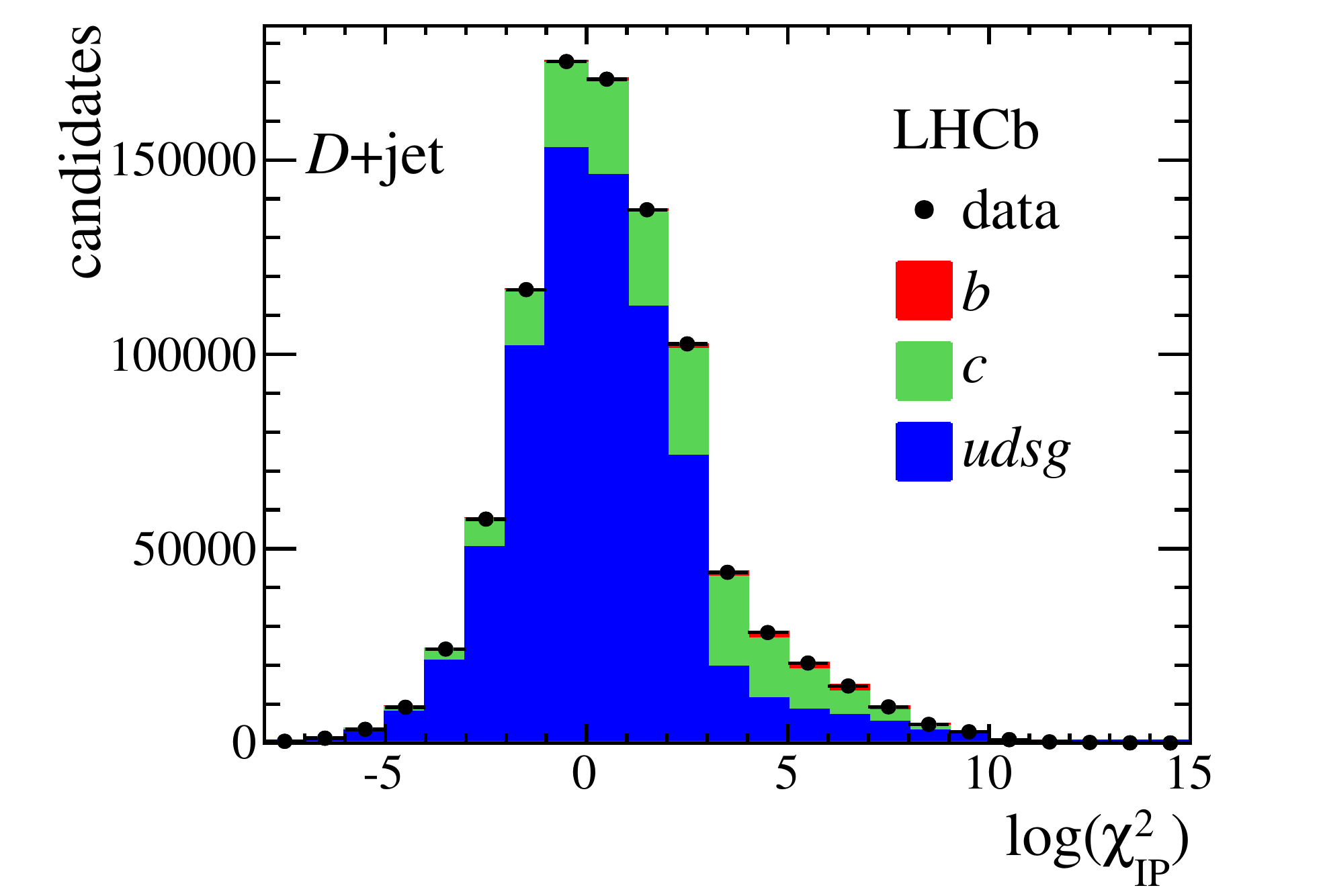}\\
 \includegraphics[width=0.45\textwidth]{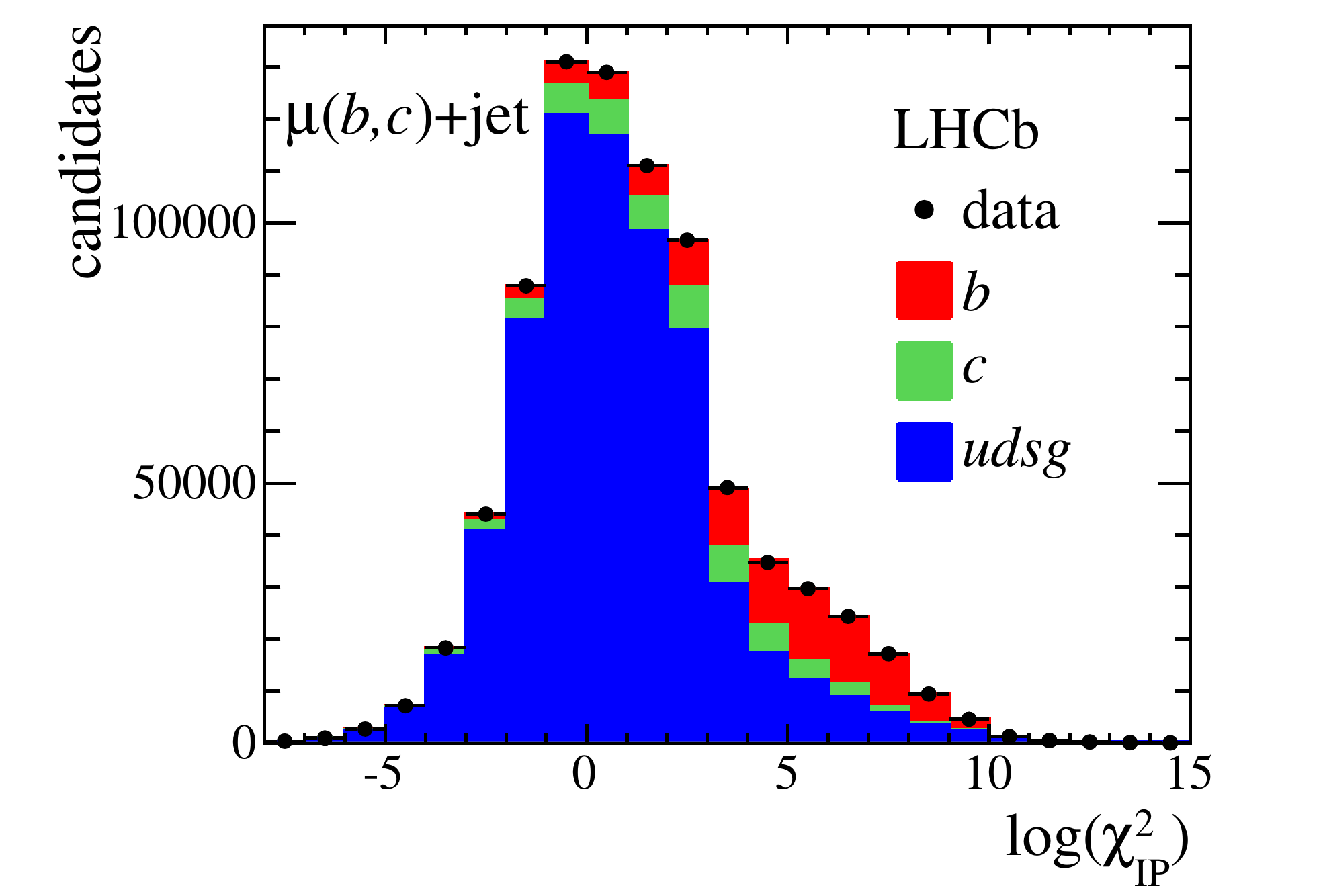}
  \caption{\label{fig:iphrd} 
    Fits to the \xip distribution in (top left) $B+$jet, (top right) $D+$jet and (bottom) $\mu(b,c)+$jet data samples.  
  }
\end{figure}

Figure~\ref{fig:iphrd} shows the results of fitting the \xip distributions in the $B+$jet, $D+$jet and $\mu(b,c)+$jet data samples.  Each sample consists of mostly \light jets prior to applying an SV tag.
While these data samples require that an event-tag containing a $(b,c)$ quark is reconstructed, the associated $(b,c)$ quarks produced in hard scattering processes are often not produced within the LHCb acceptance.
Furthermore, for the $(B,D)+$jet samples, the event-tags often have low \pt so that the associated $(b,c)$ quarks may be within the LHCb acceptance but do not form a high-\pt jet.
The \light-jet \xip template has a long tail out to large values which arises due to hyperon and kaon decays.  
In the \xip fits, the $\log{\xip} > 3$ component of the \light template is allowed to vary independently to allow for different $s$-quark content from the $W\!+$jet calibration sample.  Apart from this, all \xip templates are fixed in shape.  
The efficiency for tagging a jet originated by a quark of type $q$ is determined as
\begin{equation}
\label{eq:eff}
\epsilon_q = N_q({\rm SV})/N_q(\xip),
\end{equation}
{\em i.e.} it is the ratio of the yield determined from fits to the SV-tagged BDT distributions, either for the SV-tagger or TOPO algorithm, to the yield obtained from fits to the \xip distributions.  

\subsection{Efficiency measurement using muon jets}

\begin{figure}[] 
  \centering 
  \includegraphics[width=0.45\textwidth]{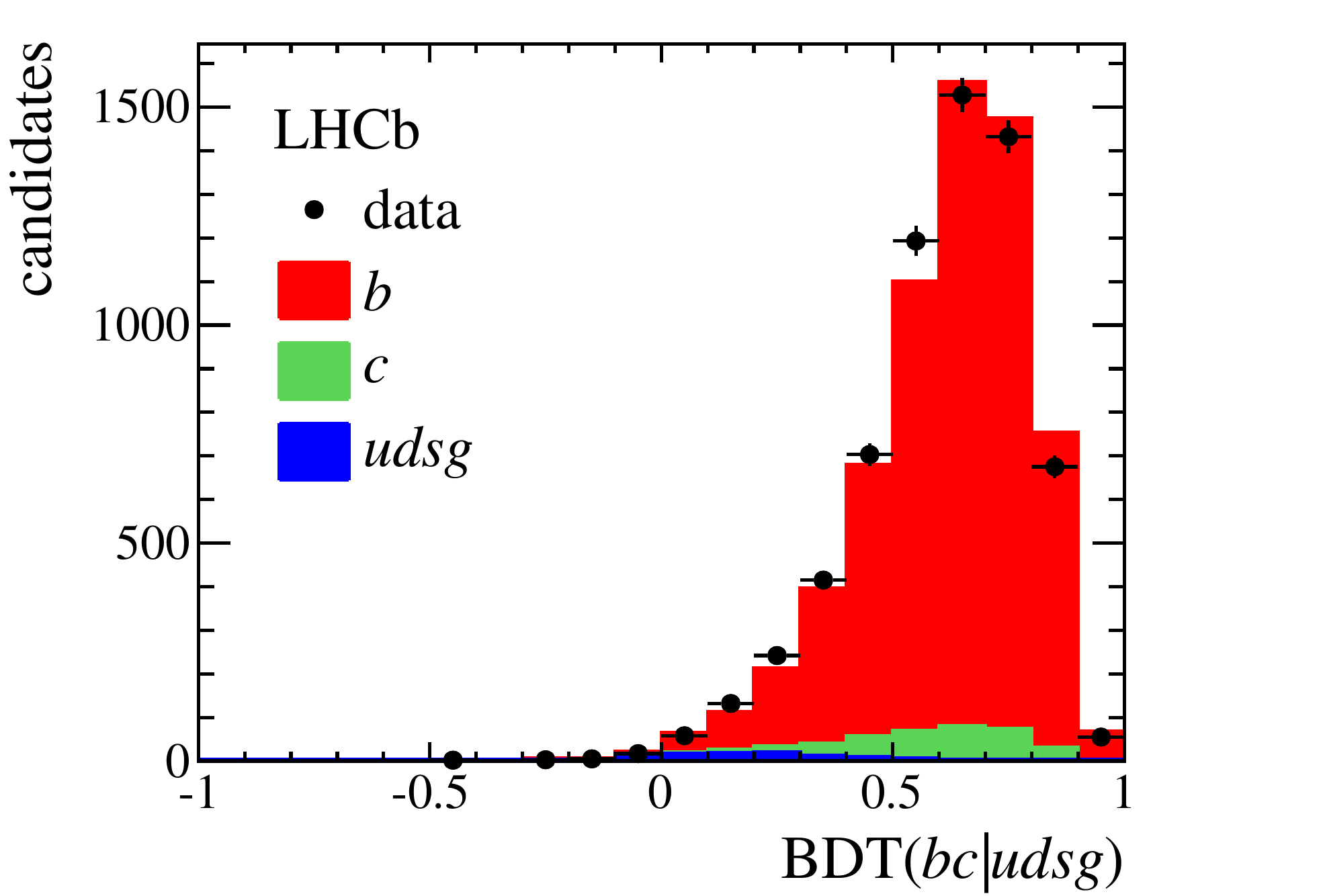}
  \includegraphics[width=0.45\textwidth]{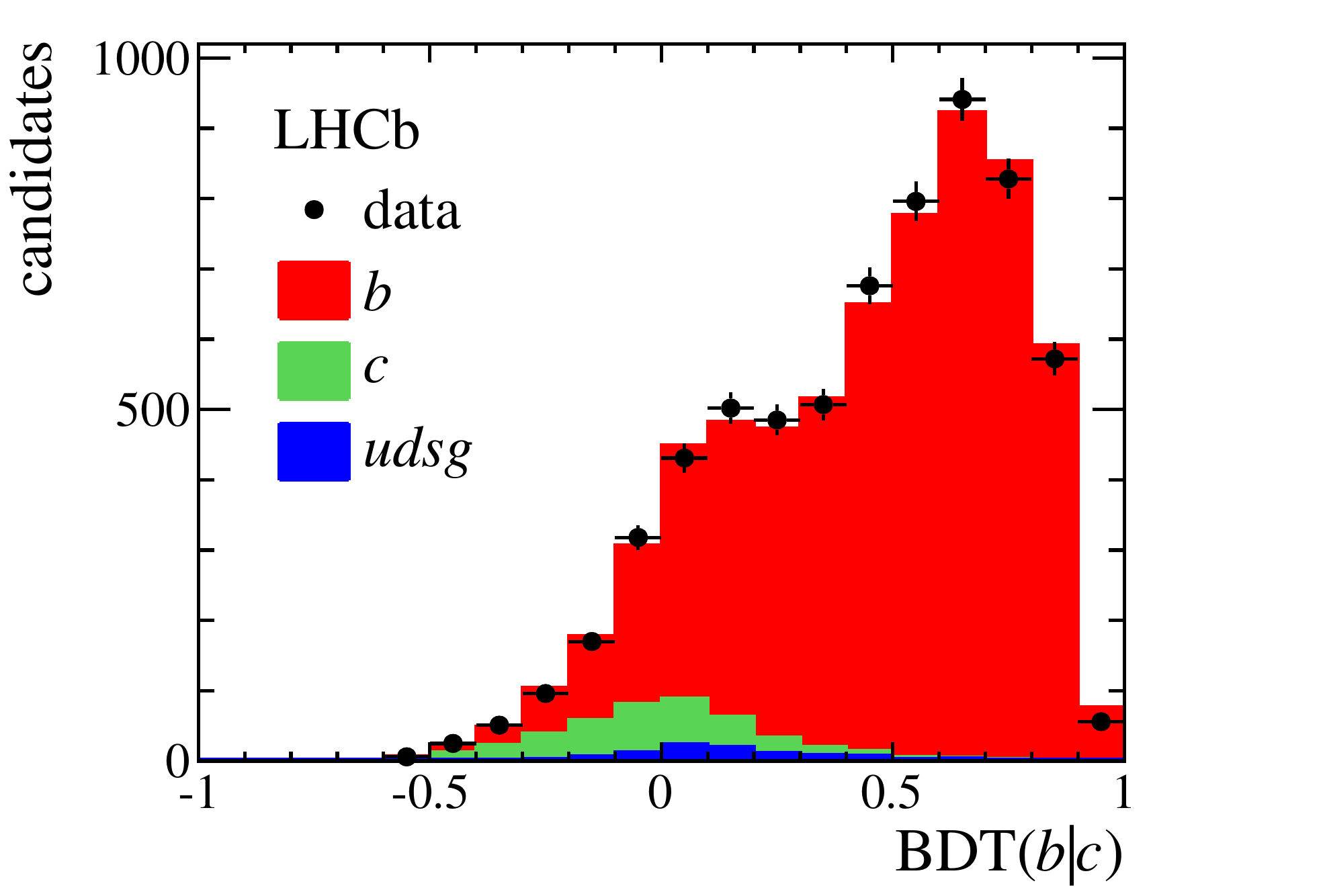}\\
  \includegraphics[width=0.45\textwidth]{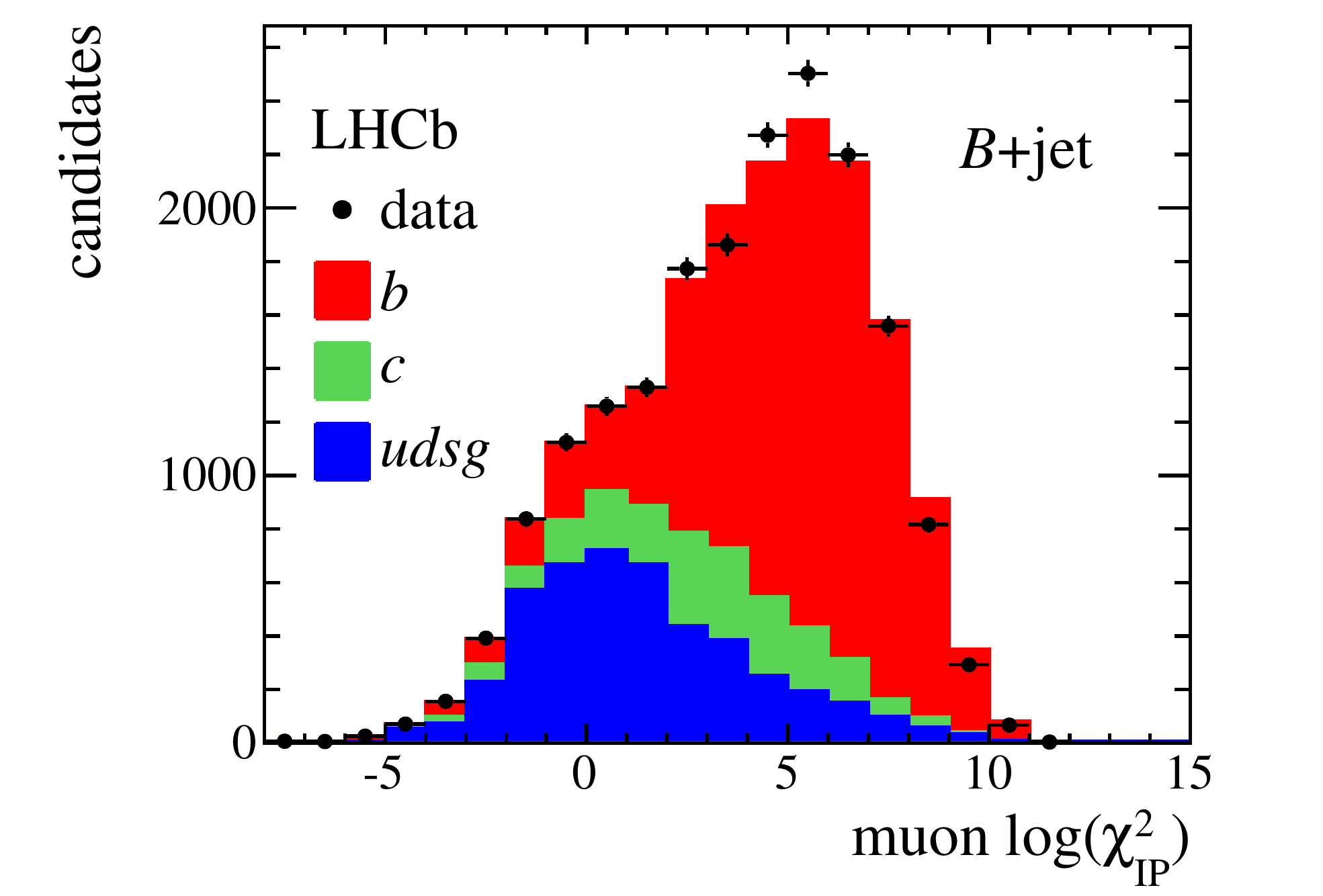}
  \caption{\label{fig:efffitmu_bhad}  (top) SV-tagger two dimensional BDT fit results projected onto the (left) \bdtbcl and (right) \bdtbc axes and (bottom) \xip fit results for the $B+$muon-jet  subsample with $10 < \pt({\rm jet}) < 100\gev$.
}
\end{figure}

\begin{figure}[] 
  \centering 
  \includegraphics[width=0.45\textwidth]{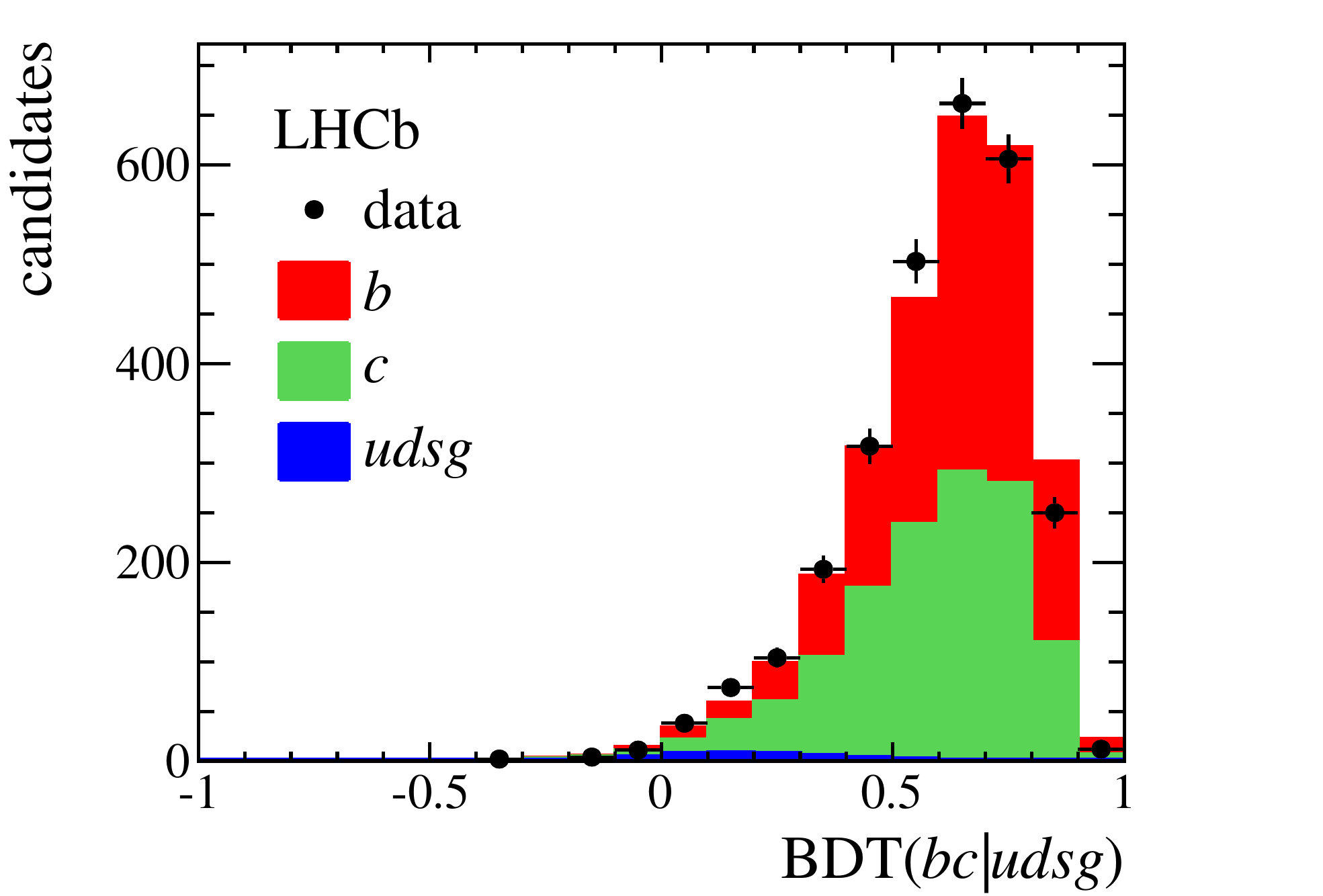}
  \includegraphics[width=0.45\textwidth]{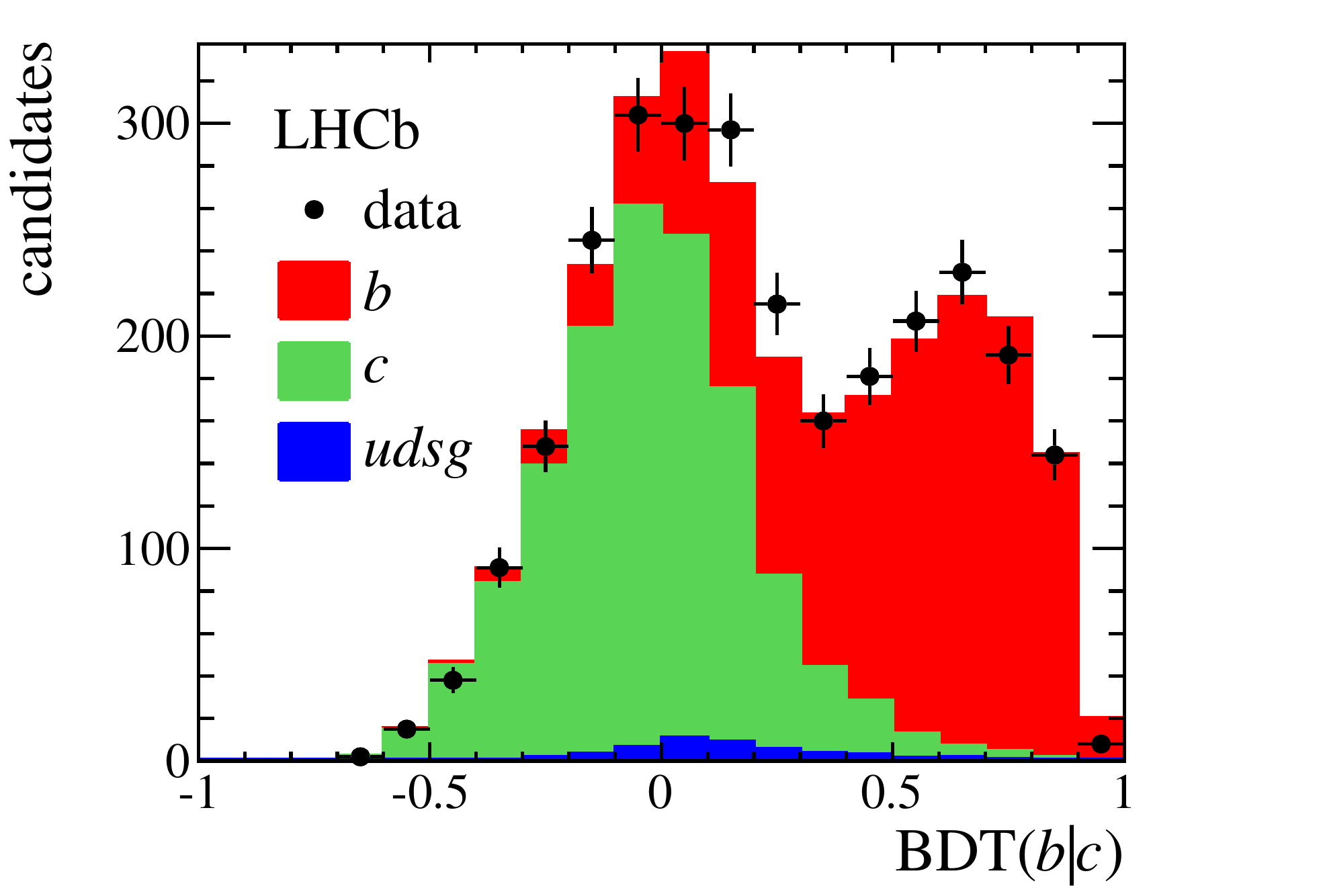}\\
  \includegraphics[width=0.45\textwidth]{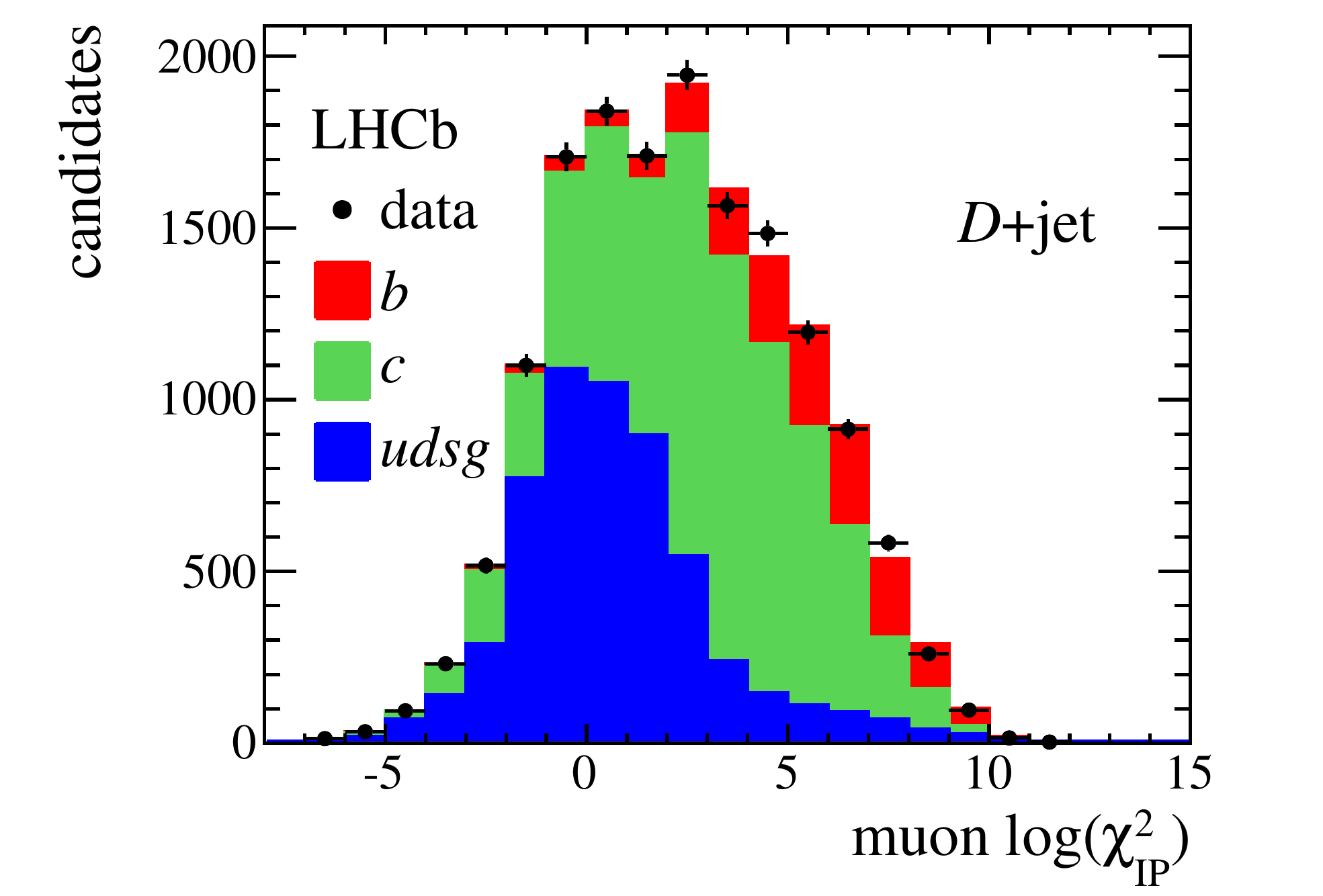}
  \caption{\label{fig:efffitmu_chad}
    Same as Fig.~\ref{fig:efffitmu_bhad} but for the $D+$muon-jet data sample.}
\end{figure}

\begin{figure}[] 
  \centering 
  \includegraphics[width=0.45\textwidth]{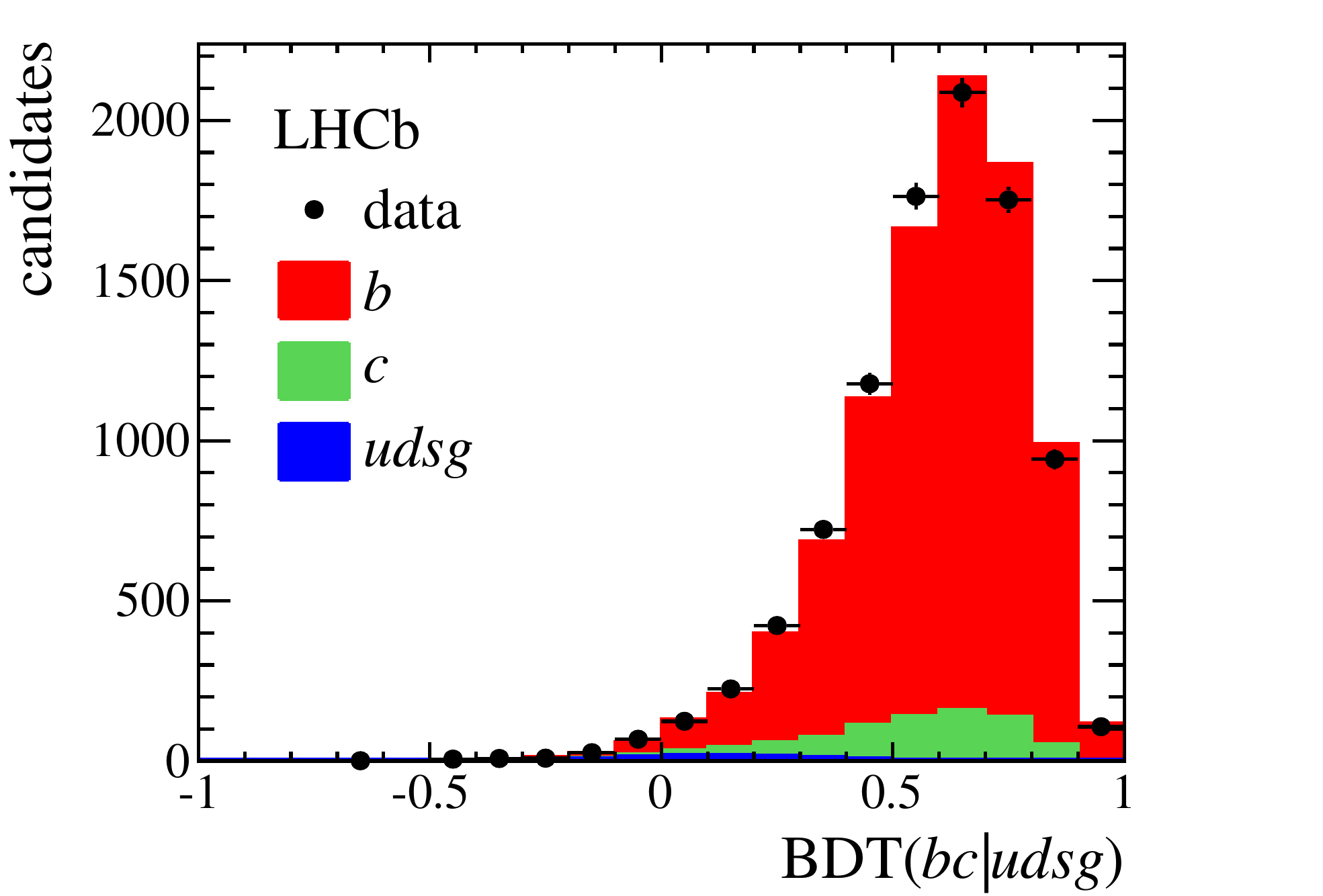}
  \includegraphics[width=0.45\textwidth]{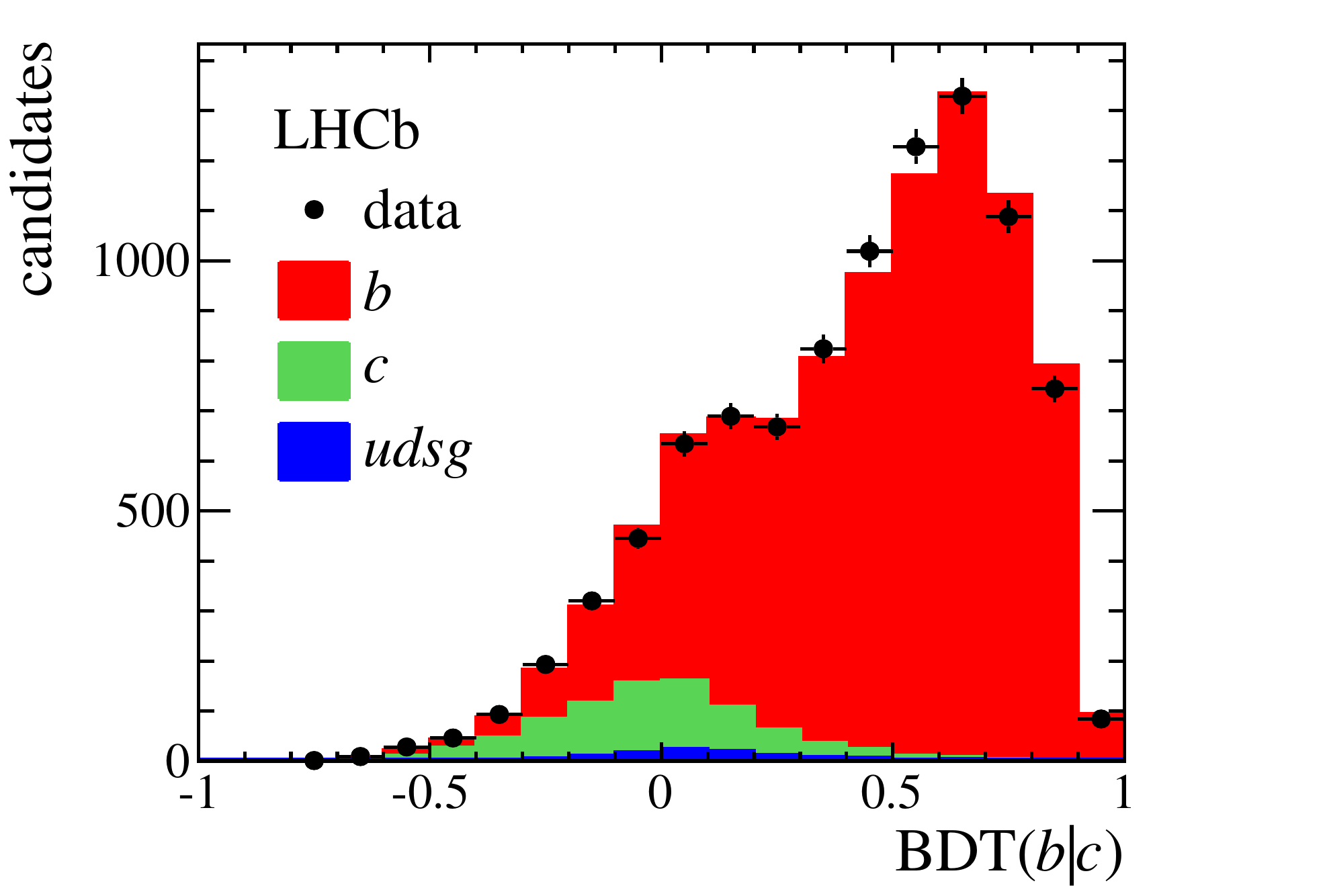}\\
  \includegraphics[width=0.45\textwidth]{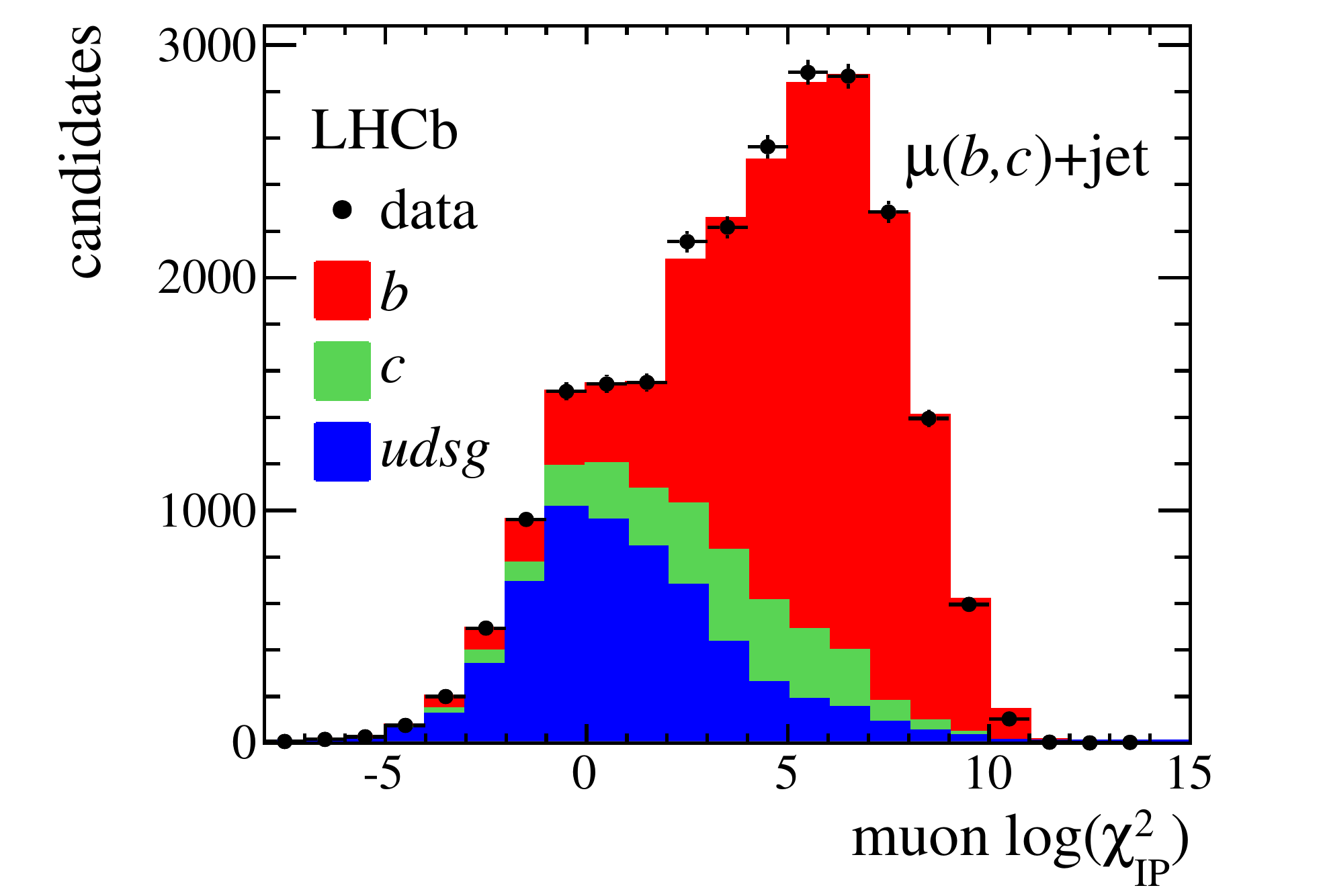}
  \caption{\label{fig:efffitmu_bmu} Same as Fig.~\ref{fig:efffitmu_bhad} but for the $\mu(b,c)+$muon-jet data sample.}
\end{figure}

The approach described in the previous subsection has the advantage that it involves measuring the efficiency on almost all of the jets in the data sample; however, its disadvantage is the large \light-jet content, which results in 10--20\% uncertainties on the pre-SV-tag jet content.  
An approach used by other experiments is to measure the efficiency on the subset of jets that contain muons.  
The tagging efficiencies are also obtained using Eq.~\ref{eq:eff} for the muon-jet subsamples. 
Figures~\ref{fig:efffitmu_bhad}--\ref{fig:efffitmu_bmu} show the SV-tagger BDT and \xip fit results for the muon-jet subsample of each data set.
In these subsamples the \xip is that of the highest-\pt muon in the jet.  The muon is required to satisfy $\pt(\mu)/\pt({\rm jet}) > 10\%$.  
The initial \light-jet content is greatly reduced in these data subsamples; however, this approach only uses about $10\%$ of the jets and it is possible that mismodeling of the jet-tagging performance in semileptonic decays is not the same as for other decays.

\subsection{Systematic uncertainties}

The systematic uncertainty on $N_{(b,c)}({\rm SV})$ is estimated using the difference between the $(b,c)$ SV-tagged yields obtained from two different fits: the fit to the BDT distributions and the fit to the $M_{\rm cor}$ versus track multiplicity distributions.
The latter approach removes jet quantities such as jet \pt from the yield determination.  While the absolute uncertainty on the SV-tagged quark content as determined by the difference in these two methods is only a few percent, the relative uncertainty is large for cases where a given jet type makes up a small fraction of the SV-tagged data sample.  For example, the relative uncertainty on the $c$-jet yield in the $B+$jet data sample is large.  
As a further cross-check the $(B,D)+$jet data samples are used to obtain data-driven BDT templates.  The difference in $(b,c)$ yields obtained by fitting the $W\!+$jet data sample using the data-driven and simulation templates is found to be negligible.

The systematic uncertainty on $N_{(b,c)}(\xip)$ has several components.
The nominal \xip fits allow the large-IP component of the \light-jet template to vary.  The \xip fits are repeated fixing this component to that observed in $W\!+$jet data, with the difference in $(b,c)$-jet yields assigned as a systematic uncertainty.  This uncertainty is sizable for the case of high-\pt $c$ jets whose \xip template is less distinct from that of \light jets which has a variable large-IP component in the fit.  
Possible dependence of the mismodeling of the IP resolution on the origin point of the particle is studied and found to be negligible.

For the case of muon jets, the misidentification probability of hadrons as muons and the jet track multiplicity must be modeled properly to obtain an accurate \xip distribution.  
Mismodeling of these properties does not lead to large uncertainty on $N_b(\xip)$, since the vast majority of reconstructed muons in $b$ jets are truly muons that arise due to semileptonic decays.  
For $c$ jets, however, mismodeling of these properties can produce sizable shifts in $N_c(\xip)$ due to the smaller fraction of $c$ jets that contain muons from semileptonic decays. 
A comparison between $W\!+$jet data and simulation of the jet fraction that satisfies the muon-jet requirements, in bins of jet \pt, is used to obtain an estimate of the probability of misidentifying a jet as a muon jet. 
Based on this study a 5\% relative uncertainty is assigned to $N_{b}(\xip)$ and 20\% to $N_{c}(\xip)$ for muon jets.    
Another possible way of misidentifying muon jets is if the semileptonic decay of a $b$ hadron outside of the jet produces a muon reconstructed as part of the jet\footnote{This can also happen for semileptonic $c$-hadron decays; however, such decays rarely produce particles with $\Delta R > 0.5$ to the jet axis due to the much lower mass of $c$ hadrons compared to that of $b$ hadrons.}.  
The $\Delta R$ distribution between the SV direction of flight and jet axis for all muons found in an SV is used to conclude that this effect is at the per mille level; it is taken to be negligible.  

Jets produced in different types of events can have different properties.  
The $b$-tag efficiency is found to agree to about 1\% in simulated $W\!+\!b$, top and QCD multi-jet events.   
The BDT shapes are studied in simulated single-jet $b$ and di-jet $b\bar{b}$ events and found to be consistent for low-\pt jets but to show small discrepancies for large jet \pt.  For example, the absolute difference in efficiency of requiring $\bdtbcl>0.2$ for $b$ jets is less than 1\% up to a jet \pt of 50\gev but reaches about 3\% at a jet \pt of 100\gev. 
In the data samples considered in this study, such effects are negligible as using BDT templates from different event types results in differences in the SV-tagged yields of less than 1\%.  

Events where multiple $b$ hadrons are produced could affect the SV BDT shapes. 
The fraction of SVs that contain a track with $\Delta R > 0.5$ relative to the jet axis is studied in data with the back-to-back requirement for the event-tag and test jet removed.  The fraction of SVs that contain such a track is found to vary by at most a few percent as a function of $\Delta R$ between the event-tag and test jet.   This could indicate percent-level cross-talk between multiple $b$ jets or could be due to changes in the jet composition.  
For the efficiency measurements presented in this paper the effect of $(b,c)$-hadron decays outside of the jet is negligible; however, such decays could have an important impact on the tagging performance in some event types, {\em e.g.} in four $b$-jet events.  

Gluon splitting to $b\bar{b}$ or $c\bar{c}$ can produce jets that contain multiple $(b,c)$ hadrons which have a higher tagging efficiency.  The requirement that a $(b,c)$-hadron-decay signature is back-to-back with the test jet suppresses gluon-splitting contributions.
The fraction of jets that contain multiple SVs in data is a few percent, which agrees to about 1\% in all bins with simulated jets that contain only a single $(b,c)$ hadron.  The systematic uncertainty due to jets that contain multiple $(b,c)$ hadrons from $g\to(b\bar{b},c\bar{c})$ is taken to be 1\%.
Finally,  there is no evidence in simulation of dependence on the number of $pp$ interactions in the event, so the uncertainty due to mismodeling of the number of $pp$ interactions is taken to be negligible.
The systematic uncertainties are summarized in Table~\ref{tab:syst}.

\begin{table}[]
  \begin{center}
    \caption{\label{tab:syst} Summary of relative systematic uncertainties ($-$ denotes negligible).
      Systematic uncertainties that dependent on jet type and \pt are marked by a $*$ (see text for details).}
    \begin{tabular}{c|cc}
      \toprule
      source  & $b$ jets & $c$ jets \\
      \midrule
      BDT templates$^*$ & $\approx 2\%$ & $\approx 2\%$ \\  
      \light-jet large IP component$^*$ &  $\approx 5\%$ & $\approx 10-30\%$ \\
      IP resolution & $-$ & $-$ \\
      hadron-as-muon probability (muon-jet subsample only) & 5\% & 20\% \\
      out-of-jet $(b,c)$-hadron decay & $-$ & $-$ \\
      gluon splitting & $1\%$ & $1\%$ \\
      number of $pp$ interactions per event & $-$ & $-$ \\
      \bottomrule
    \end{tabular}
  \end{center}
\end{table}

\subsection{Results}
 
A combined fit to the $B+$jet, $D+$jet and $\mu(b,c)+$jet data samples, including the systematic uncertainties in Table~\ref{tab:syst}, is performed to obtain the $(b,c)$-jet tagging efficiencies.
In these fits, both $N_{(b,c)}({\rm SV})$ and $N_{(b,c)}(\xip)$ are determined simultaneously under the constraint that the $(b,c)$-tagging efficiency in a given jet \pt and $\eta$ region must be the same in each data sample.
The highest-\pt track and muon-jet subsamples are fitted independently since the scale factors between data and simulation could be different for semileptonic and inclusive decays.  The scale factors for $b$ and $c$ jets are allowed to vary independently since these may be different 
for different jet types.  
The misidentification probability of \light jets is allowed to vary freely in each data sample, although the results obtained are all consistent and agree with simulation.

\begin{figure}[] 
  \centering 
  \includegraphics[width=0.45\textwidth]{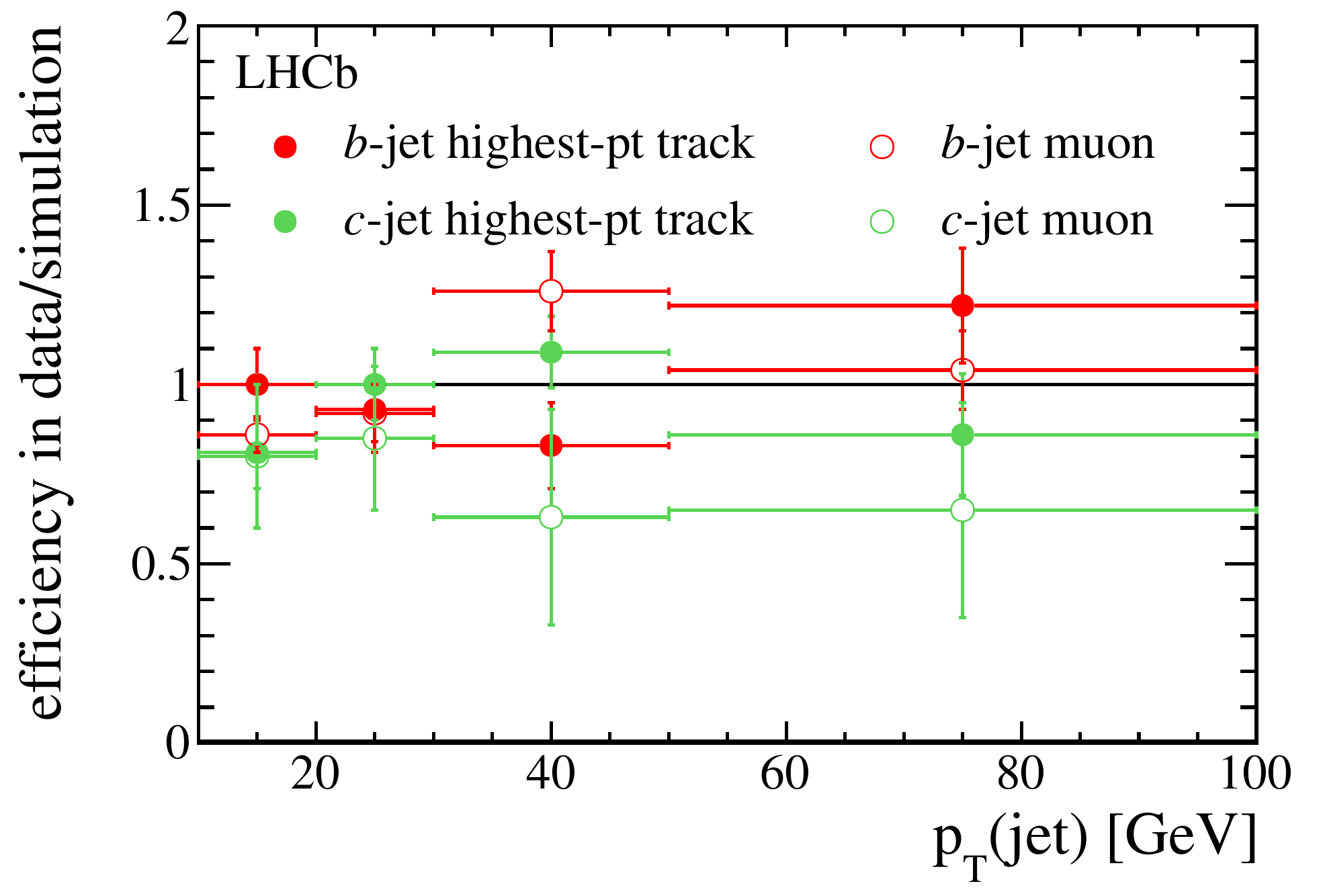}
  \includegraphics[width=0.45\textwidth]{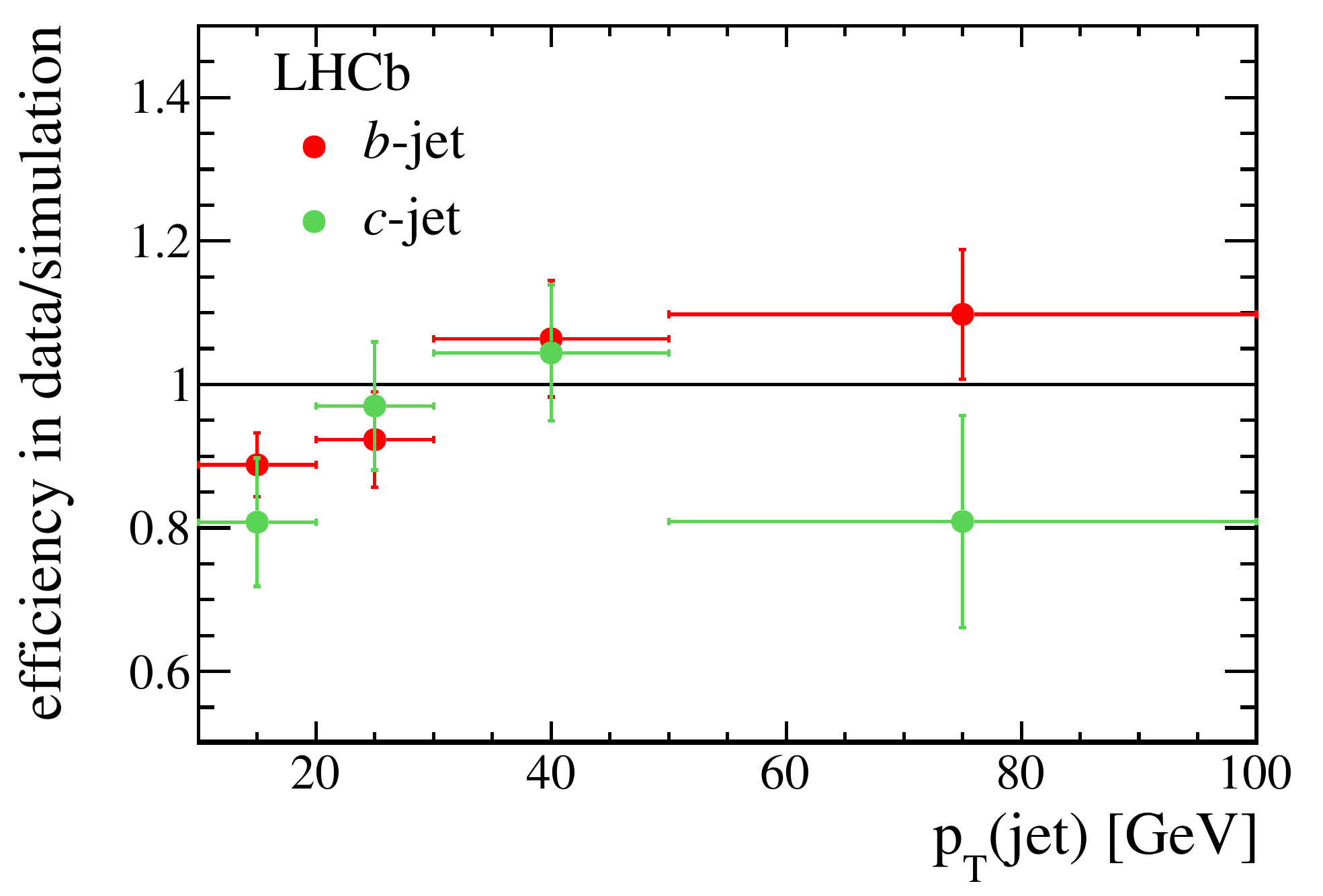}\\
  \includegraphics[width=0.45\textwidth]{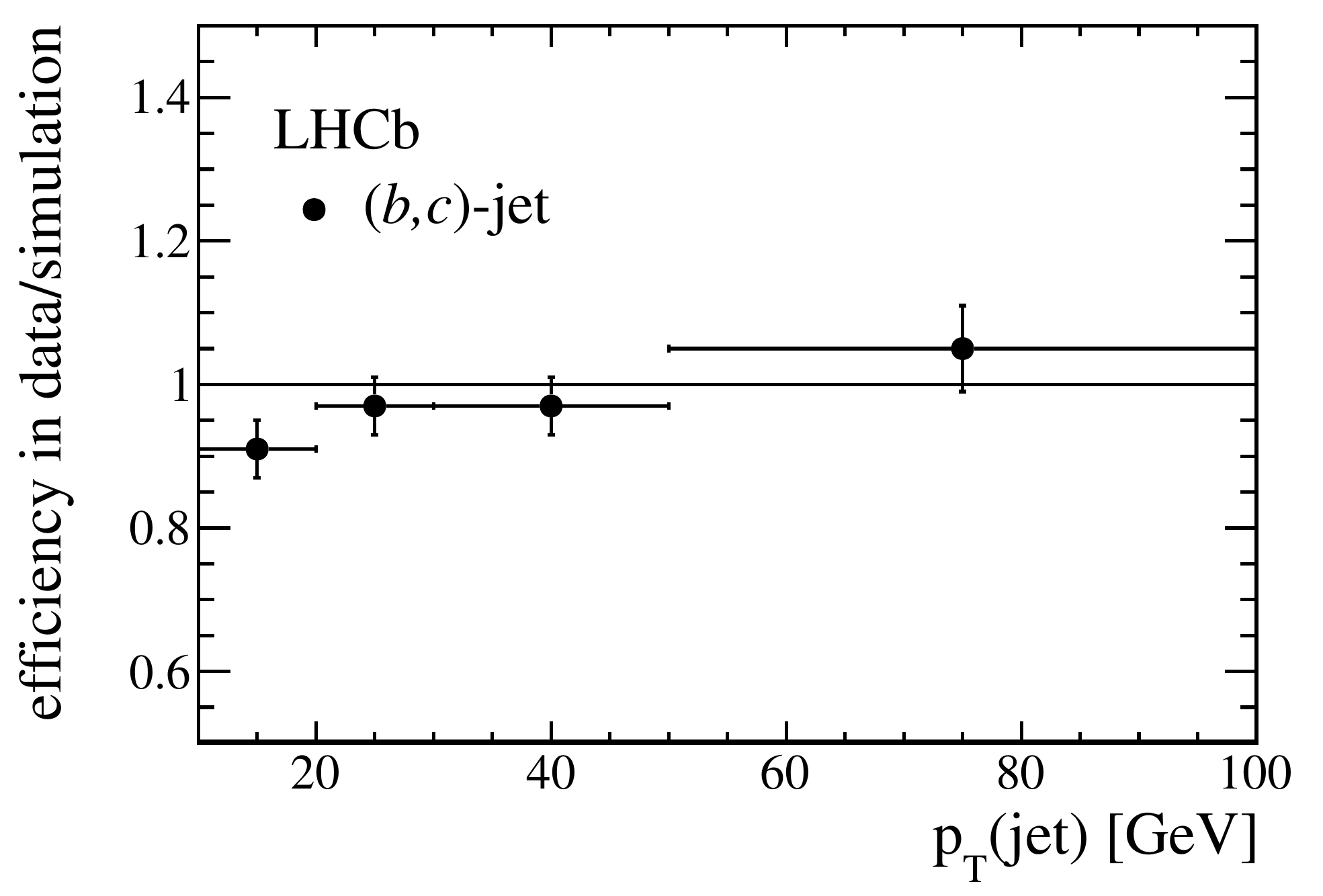}
  \includegraphics[width=0.45\textwidth]{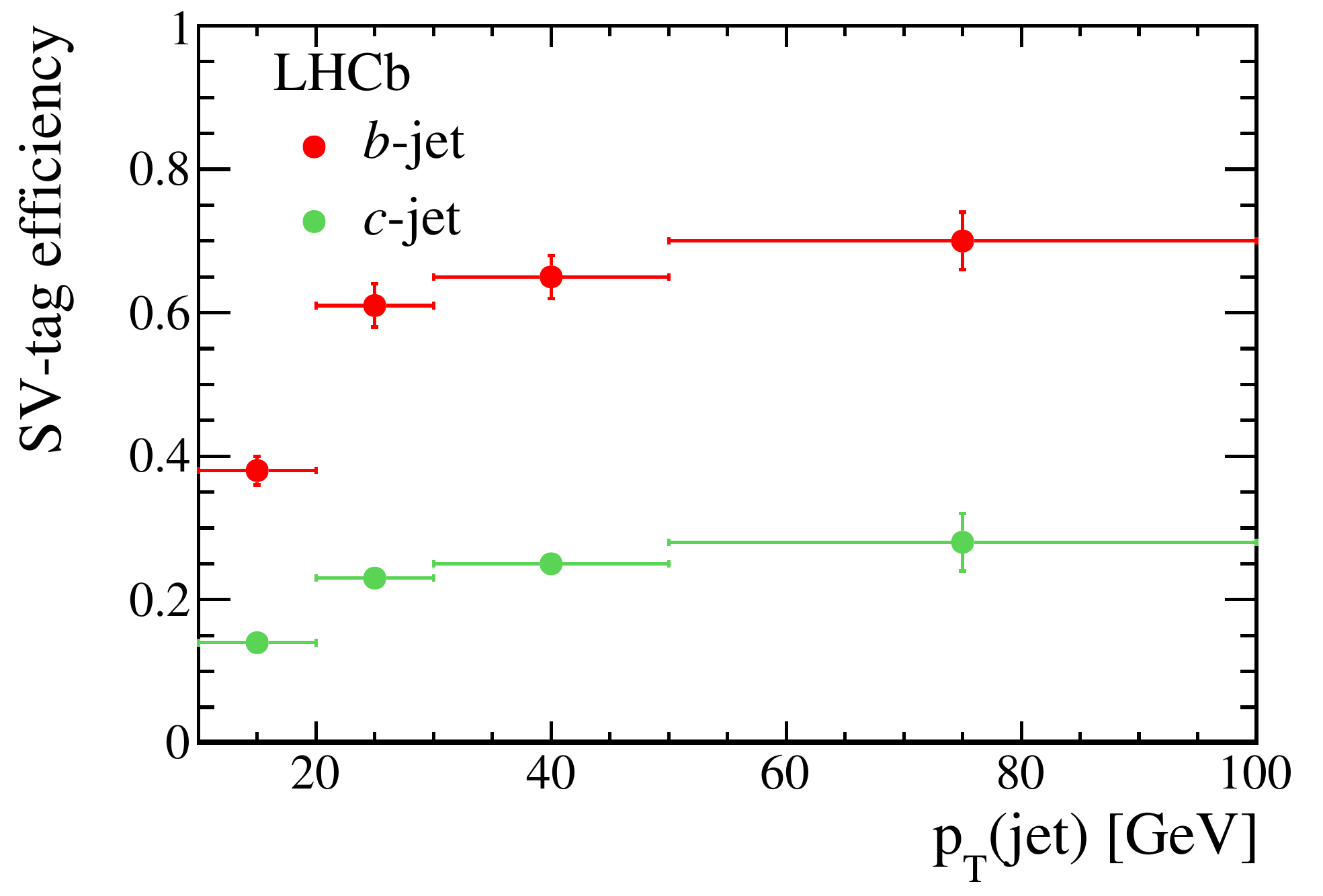}
 \caption{\label{fig:eff_vs_pt} Efficiencies of the SV-tagger algorithm measured in data relative to those obtained from simulation for $2.2 < \eta < 4.2$: (top left) results from the (closed markers) highest-\pt track and (open markers) muon-jet samples; (top right) the combined results assuming the scale factors are the same for semileptonic and inclusive $(b,c)$-hadron decays; and (bottom left) the combined results for $(b,c)$-jet using the highest-\pt-track approach assuming the scale factors are the same for $b$ and $c$ jets.
 The absolute efficiencies corresponding to the combined $(b,c)$-jet results (bottom right).
}
\end{figure}

\begin{figure}[] 
  \centering 
  \includegraphics[width=0.45\textwidth]{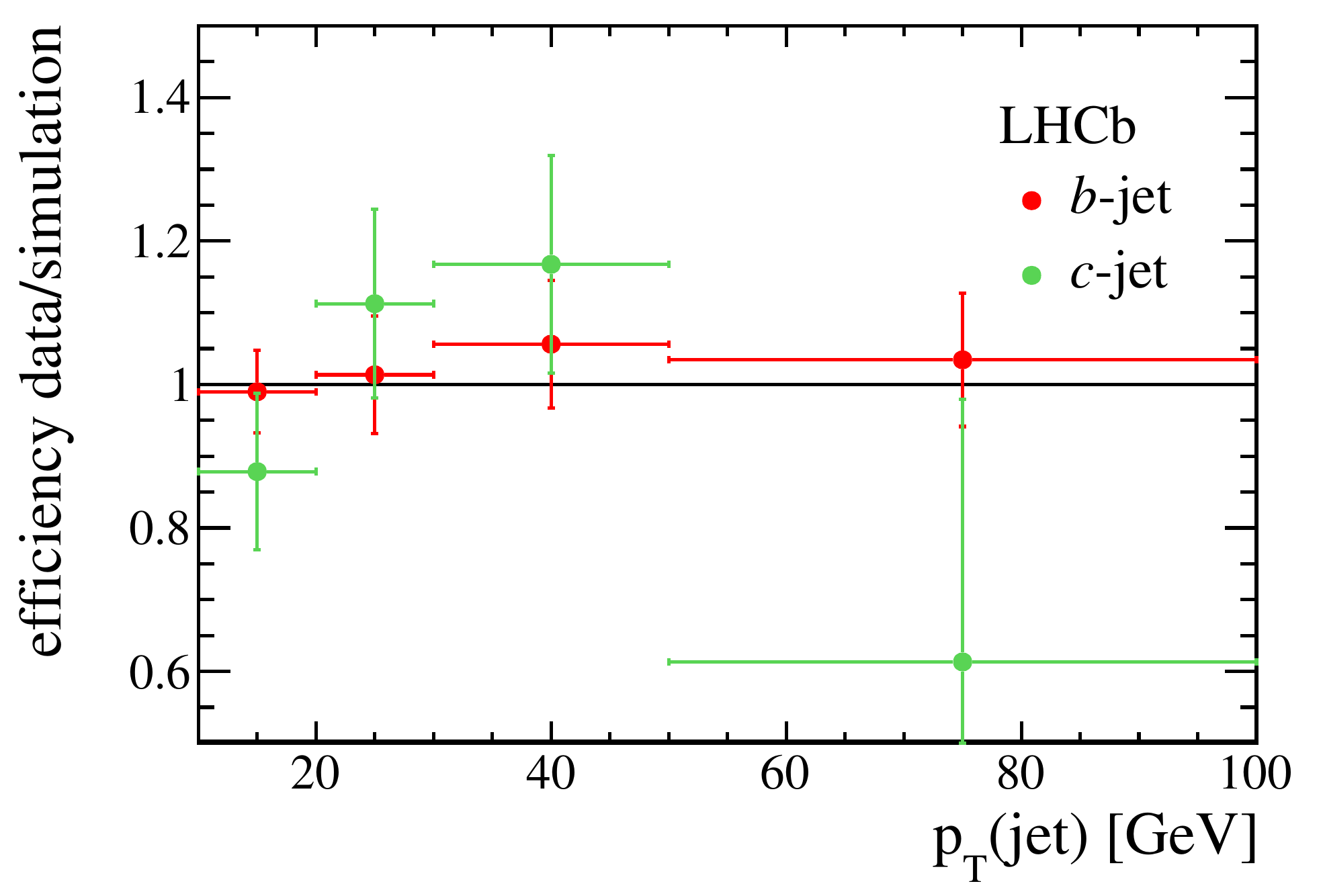}
  \caption{\label{fig:topoeffs_final} 
    TOPO algorithm $(b,c)$-tagging efficiencies, using the ``loose'' BDT requirement, in data relative to those obtained in simulation.}
\end{figure}

The scale factors for the SV-tagger algorithm are measured versus jet \pt in the region $2.2 < \eta < 4.2$, where the efficiencies are expected to be nearly uniform versus $\eta$, and in the region $2 < \eta < 2.2$ for jet $\pt > 20\gev$, where the efficiencies are nearly uniform versus jet \pt (there are not sufficient statistics to measure the efficiencies in the $\eta > 4.2$ region).
The results versus jet \pt are shown in Fig.~\ref{fig:eff_vs_pt} and are summarized as follows:
\begin{itemize}
\item The scale factors obtained from the highest-\pt track approach are all consistent with unity at the $\pm20\%$ level.  They show no trend in \pt for $b$ or $c$ jets.  
\item The scale factors for muon jets are found to be consistent, albeit with large uncertainties, with those obtained using the highest-\pt track approach.  The results are combined assuming that the scale factors are the same for semileptonic and inclusive $(b,c)$-hadron decays (see Fig.~\ref{fig:eff_vs_pt}) and are summarized in Table~\ref{tab:svtag_results}.  The scale factors are consistent with unity for jet $\pt > 20\gev$, but 10-20\% below unity for low-\pt jets.
\item  The scale-factor results obtained from the global fits are strongly anti-correlated between $b$ and $c$ jets. It is likely that the true scale factors are similar between $b$ and $c$ jets since many of the contributing factors, {\em e.g.} mismodeling of the SV position resolution, are expected to affect $b$ and $c$ jets in a similar manner.
  The highest-\pt track fits are repeated assuming that the scale factors are the same for $b$ and $c$ jets (see Fig.~\ref{fig:eff_vs_pt}) and summarized in Table~\ref{tab:svtag_results}.  
The results for jet $\pt > 20\gev$ are consistent with unity at about the 5\% level, while at low jet \pt the scale factor is again less than unity by about 10\%.  The muon jet results are not combined for $b$ and $c$ jets since the $b$-jet results are much more precise.
\end{itemize}
Neither of the assumptions made in the combinations has to be completely valid; however, they should each be a good approximation.  
Overall, the efficiencies measured in data are consistent with those in simulation for jet $\pt > 20\gev$ with a conservative systematic uncertainty estimate of 10\%. 
At low jet \pt the scale factors are about 0.9 for $b$ jets and 0.8 for $c$ jets.  Using the difference in central values obtained from the highest-\pt track, combined highest-\pt track and muon jet, and combined $b$ and $c$ jet results, produces a conservative systematic uncertainty estimate of 10\%. 
The absolute efficiencies measured assuming the scale factors are the same for $b$ and $c$ jets are given in Table~\ref{tab:svtag_results}.  For jet $\pt > 20\gev$ and $2.2 < \eta < 4.2$, the mean SV-tagging efficiency is about 65\% for $b$ jets and 25\% for $c$ jets.
Finally, the TOPO algorithm efficiencies are measured in data and found to be consistent with simulation to about 5\% for $b$ jets and 20\% for $c$ jets (see Fig.~\ref{fig:topoeffs_final}). 
The absolute efficiencies measured using the TOPO for $b$ jets are: $21\pm1\%$ for 10--20\gev; $44\pm4\%$ for 20--30\gev; $60\pm5\%$ for 30--50\gev; and $66\pm6\%$ for 50--100\gev.

\begin{table}[]
  \begin{center}
    \caption{\label{tab:svtag_results} SV-tagger algorithm $(b,c)$-tagging efficiencies measured in data compared to those obtained in simulation.  The $b$ and $c$ results are obtained by combining the highest-\pt track and muon-jet results under the assumption that the scale factors are the same for semileptonic and inclusive $(b,c)$-hadron decays.
The $(b,c)$ results are obtained by fitting the highest-\pt-track sample under the assumption that the scale factors are the same for $b$ and $c$ jets.  The absolute efficiencies observed in data are provided using the ``$(b,c)$ jets'' results.
}
    \begin{tabular}{cc|cc|c|cc}
      \toprule
      {} & {} & \multicolumn{3}{c}{$\epsilon($data$)/\epsilon($simulation$)$} & \multicolumn{2}{c}{$\epsilon($data$)$ (\%)} \\
      jet \pt (\gev) & jet $\eta$ & $b$ jets & $c$ jets & $(b,c)$ jets & $b$ jets & $c$ jets \\
      \midrule
      10--20\phantom{0} & 2.2--4.2 & $0.89\pm0.04$ & $0.81\pm0.09$ & $0.91\pm0.04$ & $38\pm2$ & $14\pm1$ \\
      20--30\phantom{0} & 2.2--4.2 & $0.92\pm0.07$ & $0.97\pm0.09$ & $0.97\pm0.04$ & $61\pm3$ & $23\pm1$ \\
      30--50\phantom{0} & 2.2--4.2 & $1.06\pm0.08$ & $1.04\pm0.09$ & $0.97\pm0.04$ & $65\pm3$ & $25\pm1$ \\
      50--100 & 2.2--4.2 & $1.10\pm0.09$ & $0.81\pm0.15$ & $1.05\pm0.06$ & $70\pm4$ & $28\pm4$ \\
      \midrule
      20--100 & \phantom{.2}2--2.2 & $1.00\pm0.07$ & $1.12\pm0.10$ & $1.05\pm0.03$ & $56\pm2$ & $20\pm1$ \\
      \bottomrule
    \end{tabular}
  \end{center}
\end{table}

\section{Light-parton jet misidentification}
\label{sec:light}

Light-parton jets contain SVs due to any of the following: (1) misreconstruction of prompt particles as displaced tracks; (2) decays of long-lived strange particles; or (3) interactions with material.  Type (1) can be studied in data using jets that contain an SV whose inverted direction of flight lies in the jet cone (referred to as a backward SV).  Types (2) and (3) can be studied using SVs for which the ratio of the SV flight distance divided by the SV momentum is too large for the decay of a $(b,c)$ hadron (referred to as a too-long-lived SV).  
The mistag probability for simulated \light jets using backward and too-long-lived SVs is consistent with the nominal mistag probability at the 20\% level (the nominal mistag probability is shown in Fig.~\ref{fig:mc}).  Furthermore, the SV BDT distributions obtained using backward and too-long-lived SVs are similar to the nominal \light-jet BDT distributions.  Therefore, the mistag probability of \light jets and SV properties can be studied in data using backward and too-long-lived SV-tagged jets.

\begin{figure}[] 
  \centering 
  \includegraphics[width=0.45\textwidth]{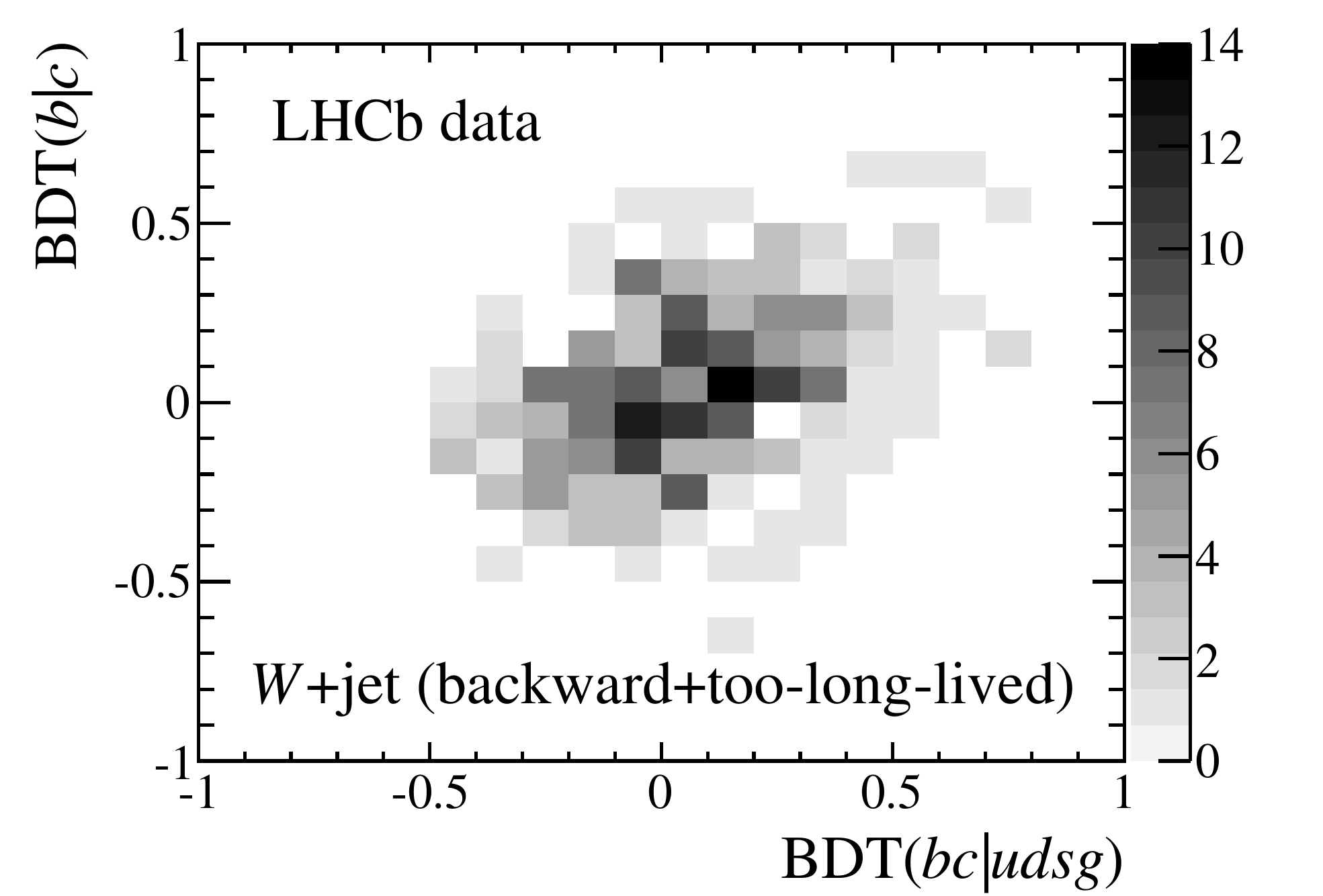}
  \includegraphics[width=0.45\textwidth]{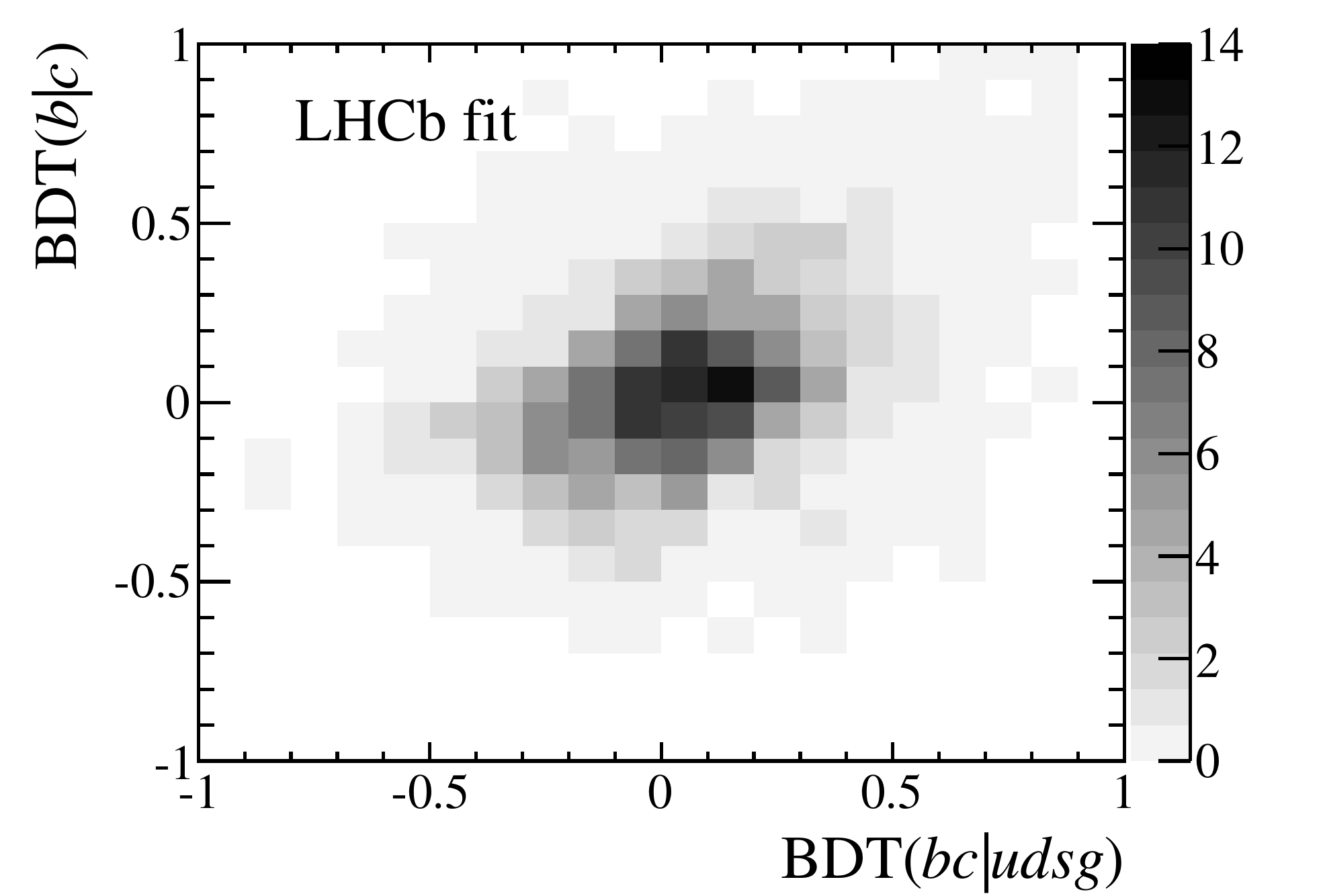}\\
  \includegraphics[width=0.45\textwidth]{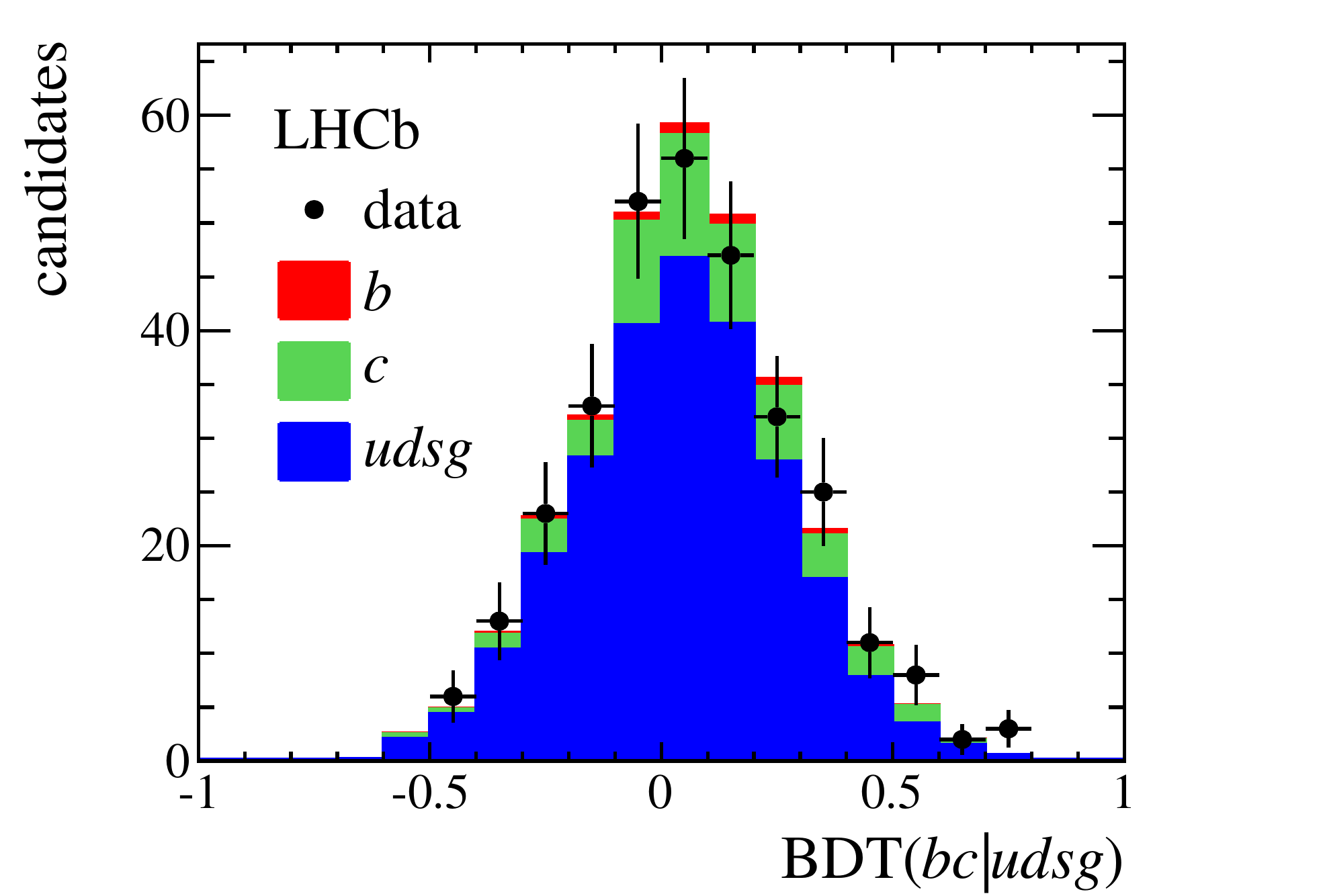}
  \includegraphics[width=0.45\textwidth]{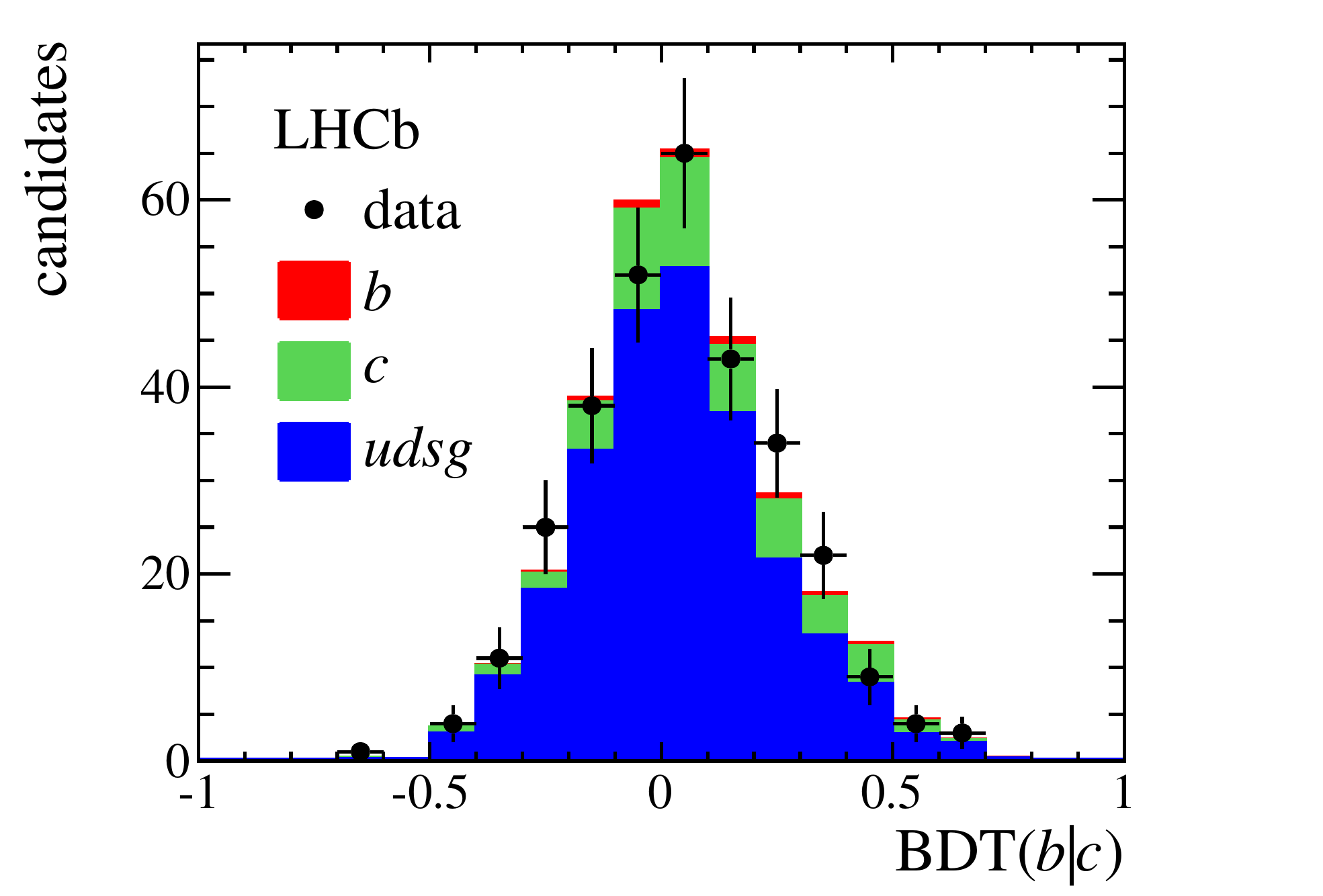}
\caption{\label{fig:bdtfit_wmu_bkll}  SV-tagger algorithm BDT distributions for backward and too-long-lived SVs in the $W\!+$jet data sample:  (top left) distribution in data; (top right) two-dimensional template-fit result; and (bottom) projections of the fit result with the $b$, $c$, and \light contributions shown as stacked histograms.
}
\end{figure}

Such a study is complicated by the fact that prompt tracks in $(b,c)$ jets can also be misreconstructed as displaced, and that $(b,c)$ jets also produce strange particles and material interactions.  
Therefore, both backward and too-long-lived SVs are also found in $(b,c)$ jets.
The $W\!+$jet data sample, which is dominantly composed of \light jets, is used to mitigate effects from mistagged $(b,c)$ jets.
Figure~\ref{fig:bdtfit_wmu_bkll} shows the BDT distributions from backward and too-long-lived SVs observed in data compared to simulation.  
The backward and too-long-lived BDT templates are similar for all jet types.  The $(b,c)$ yields here are fixed by fitting the nominal SV-tagged data to obtain the total $(b,c)$-jet content then taking the backward and too-long-lived SV-tag probabilities for $(b,c)$ jets from simulation.
The distributions in data and simulation are consistent, which demonstrates that the SV properties are well-modeled for \light jets.

\begin{figure}[] 
  \centering 
  \includegraphics[width=0.45\textwidth]{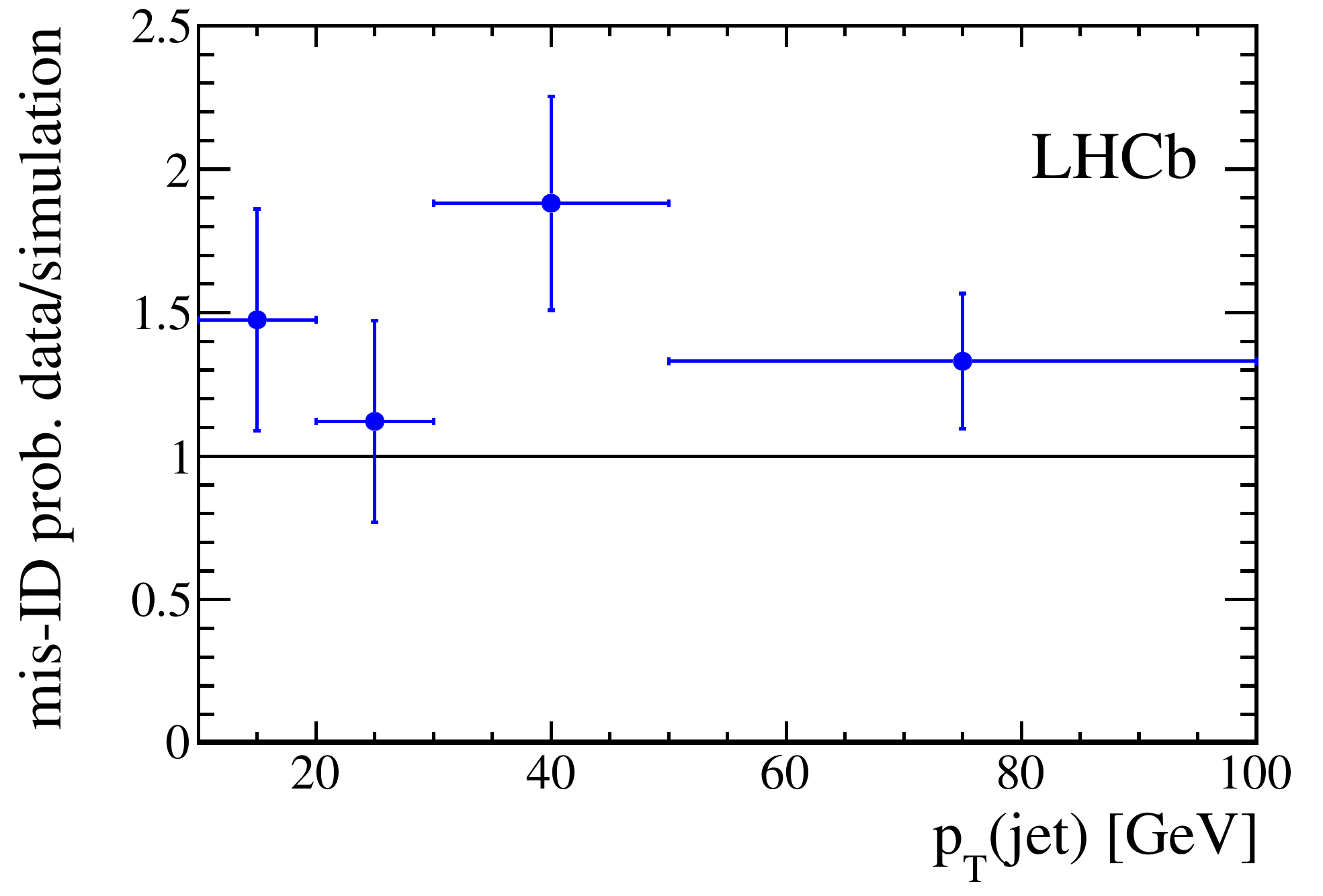}
  \includegraphics[width=0.45\textwidth]{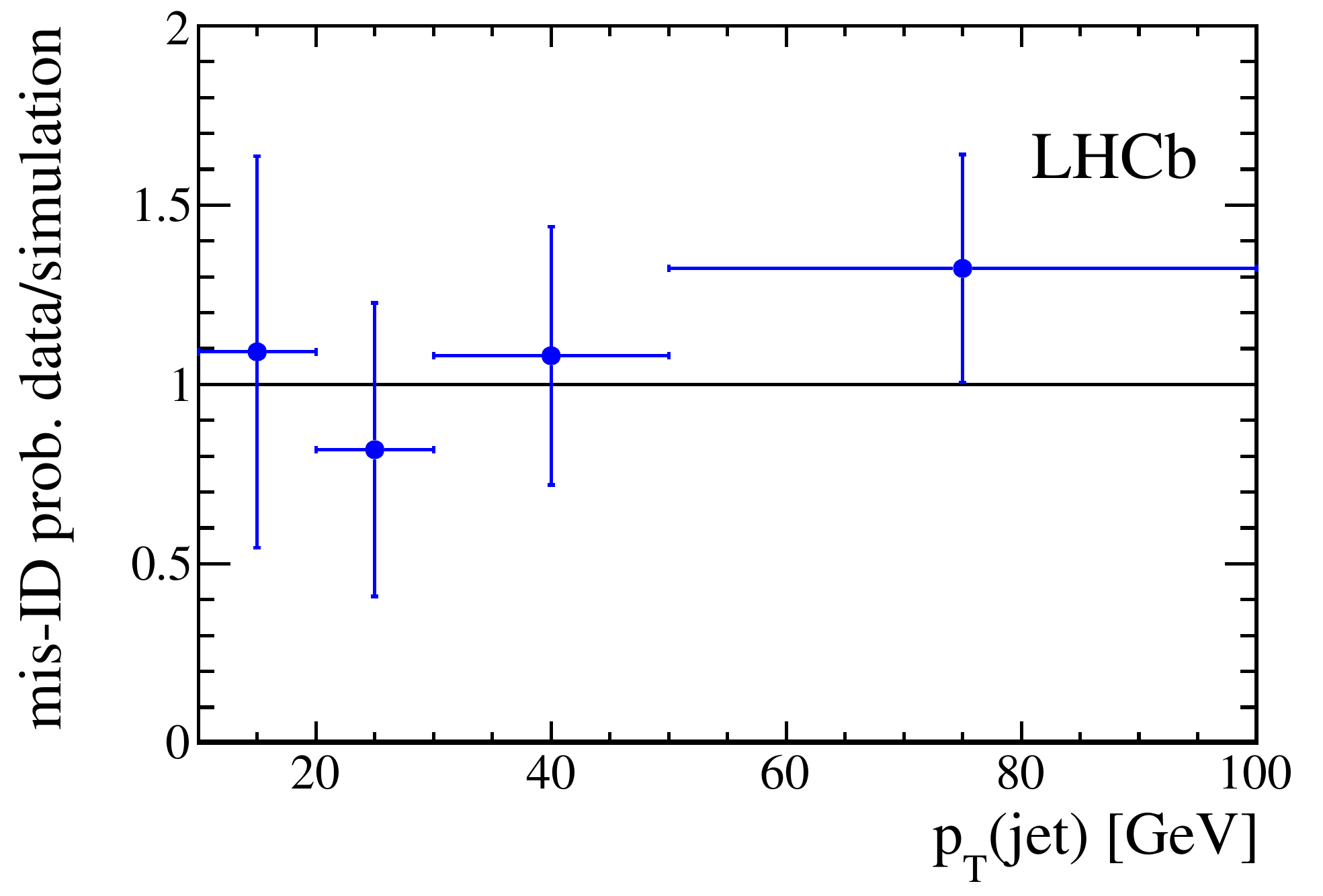}
  \caption{\label{fig:mistag} 
 Ratio of \light-jet mistag probabilities observed in data to those in simulation for the (left) SV-tagger and (right) TOPO algorithms.
}
\end{figure}

The total \light-jet composition of this sample, without applying any SV-tagging algorithm, is found to be 95\%, by fitting the nominal SV-tagged BDT distributions and applying the data-driven $(b,c)$-tagging efficiencies from the previous section. 
The mistag probability of \light jets is obtained as the ratio of the number of SV-tags for those jets (obtained by fitting the SV BDT distributions) to the total number of \light jets. 
The ratio of this probability in data to that in simulation is shown in Fig.~\ref{fig:mistag}; data and simulation agree at about the $\pm30\%$ level integrated over jet \pt.  
A detailed study of $W\!+$jet production in LHCb using the SV-tagger algorithm introduced in this paper, in which the jets are required to satisfy $\pt > 20\gev$ and $2.2 < \eta < 4.2$, finds that the nominal \light-jet mistag probability is 0.3\% which is consistent with simulation\cite{LHCb-PAPER-2015-021}.   The same ratio for the TOPO algorithm is also shown in Fig.~\ref{fig:mistag}.

The performance of any tagging algorithm on \light jets can be affected by the presence of $(b,c)$ jets in the event.  The misidentification probability of \light jets is studied in simulated di-$b$-jet events and compared to the performance obtained in simulated events that contain no $(b,c)$ jets.  
The absolute difference in the fraction of \light jets that are SV-tagged and have $\bdtbcl > 0.2$ is found to be at the per mille level for low-\pt jets, but increases to about 1\% for jet \pt of 50\gev and to about 2--3\% at 100\gev.  The BDT shapes are distorted relative to those obtained in events that contain no $(b,c)$ jets, but there is still significant discrimination between the \light and $(b,c)$ distributions.   The difference is largely due to particles originating from a $b$-hadron decay and produced with $\Delta R < 0.5$ relative to the \light-jet axis.  These tracks may then form SVs with misreconstructed prompt tracks in the \light jets.  

\section{Summary}

The LHCb collaboration has developed several algorithms that efficiently identify jets that arise from the hadronization of $b$ and $c$ quarks.  The performance of these algorithms has been studied in data and is found to agree with that in simulation at about the 10\% level for $(b,c)$ jets, and at the 30\% level for \light jets.  
The SV properties of all jet types are found to be well modeled by LHCb simulation.
The efficiency for identifying a $b(c)$ jet is about 65\%(25\%) with a probability of misidentifying a \light jet of 0.3\% for jets with transverse momentum $\pt > 20\gev$ and pseudorapidity $2.2 < \eta < 4.2$.

\section*{Acknowledgments}

\noindent We express our gratitude to our colleagues in the CERN
accelerator departments for the excellent performance of the LHC. We
thank the technical and administrative staff at the LHCb
institutes. We acknowledge support from CERN and from the national
agencies: CAPES, CNPq, FAPERJ and FINEP (Brazil); NSFC (China);
CNRS/IN2P3 (France); BMBF, DFG, HGF and MPG (Germany); INFN (Italy); 
FOM and NWO (The Netherlands); MNiSW and NCN (Poland); MEN/IFA (Romania); 
MinES and FANO (Russia); MinECo (Spain); SNSF and SER (Switzerland); 
NASU (Ukraine); STFC (United Kingdom); NSF (USA).
The Tier1 computing centres are supported by IN2P3 (France), KIT and BMBF 
(Germany), INFN (Italy), NWO and SURF (The Netherlands), PIC (Spain), GridPP 
(United Kingdom).
We are indebted to the communities behind the multiple open 
source software packages on which we depend. We are also thankful for the 
computing resources and the access to software R\&D tools provided by Yandex LLC (Russia).
Individual groups or members have received support from 
EPLANET, Marie Sk\l{}odowska-Curie Actions and ERC (European Union), 
Conseil g\'{e}n\'{e}ral de Haute-Savoie, Labex ENIGMASS and OCEVU, 
R\'{e}gion Auvergne (France), RFBR (Russia), XuntaGal and GENCAT (Spain), Royal Society and Royal
Commission for the Exhibition of 1851 (United Kingdom).

\addcontentsline{toc}{section}{References}
\setboolean{inbibliography}{true}
\bibliographystyle{LHCb}
\bibliography{svtag.bbl}

\clearpage
\centerline{\large\bf LHCb collaboration}
\begin{flushleft}
\small
R.~Aaij$^{38}$, 
B.~Adeva$^{37}$, 
M.~Adinolfi$^{46}$, 
A.~Affolder$^{52}$, 
Z.~Ajaltouni$^{5}$, 
S.~Akar$^{6}$, 
J.~Albrecht$^{9}$, 
F.~Alessio$^{38}$, 
M.~Alexander$^{51}$, 
S.~Ali$^{41}$
% ...

{\footnotesize \it
% ...
$ ^{5}$Clermont Universit\'{e}, Universit\'{e} Blaise Pascal, CNRS/IN2P3, LPC, Clermont-Ferrand, France\\
$ ^{6}$CPPM, Aix-Marseille Universit\'{e}, CNRS/IN2P3, Marseille, France\\
% ...
$ ^{9}$Fakult\"{a}t Physik, Technische Universit\"{a}t Dortmund, Dortmund, Germany\\
% ...
$ ^{37}$Universidad de Santiago de Compostela, Santiago de Compostela, Spain\\
$ ^{38}$European Organization for Nuclear Research (CERN), Geneva, Switzerland\\
% ...
$ ^{41}$Nikhef National Institute for Subatomic Physics, Amsterdam, The Netherlands\\
% ...
$ ^{46}$H.H. Wills Physics Laboratory, University of Bristol, Bristol, United Kingdom\\
% ...
$ ^{51}$School of Physics and Astronomy, University of Glasgow, Glasgow, United Kingdom\\
$ ^{52}$Oliver Lodge Laboratory, University of Liverpool, Liverpool, United Kingdom\\
% ...
}
\end{flushleft}

\end{document}